\documentclass[preprint, review,3p,times, 10pt]{elsarticle}

\usepackage{natbib}
\usepackage{amsmath,amssymb,amsfonts}
\usepackage{footnote}
\usepackage{comment}
\usepackage{threeparttable}
\usepackage{romannum}
\usepackage{longtable}
\usepackage{lscape}
\usepackage{soul}
\usepackage{array, makecell} 
\usepackage{multirow} 
\usepackage{caption} 
\usepackage{subcaption} 
\usepackage{graphicx} 
\usepackage{float} 

\usepackage{longtable} 

\usepackage[hyphens]{url}
\usepackage{hyperref}
\hypersetup{colorlinks=true,breaklinks=true}
 
\newcommand*\rot{\rotatebox{90}}

\journal{Elsevier}

\begin{document}

\begin{frontmatter}
\title{An Attention-Guided Deep Learning Approach for Classifying 39 Skin Lesion Types} 

\author[1]{Sauda Adiv Hanum}\ead{u1804063@student.cuet.ac.bd}

\author[1]{Ashim Dey}\ead{ashim@cuet.ac.bd}

\author[3]{Muhammad Ashad Kabir\corref{correspondingauthor}}\ead{akabir@csu.edu.au}

\cortext[correspondingauthor]{Corresponding author: Charles Sturt University, Panorama Ave, Bathurst, NSW 2795. Ph.+61263386259} 

\affiliation[1]{organization={Department of Computer Science and Engineering, Chittagong University of Engineering and Technology}, city={Raozan}, state={Chittagong}, postcode={4349}, country={Bangladesh}}

\affiliation[3]{organization={School of Computing, Mathematics and Engineering, Charles Sturt University}, city={Bathurst}, state={NSW}, postcode={2795}, country={Australia}}

\begin{abstract}
The skin, as the largest organ of the human body, is vulnerable to a diverse array of conditions collectively known as skin lesions, which encompass various dermatoses. Diagnosing these lesions presents significant challenges for medical practitioners due to the subtle visual differences that are often imperceptible to the naked eye. While not all skin lesions are life-threatening, certain types can act as early indicators of severe diseases, including skin cancers, underscoring the critical need for timely and accurate diagnostic methods.
Deep learning algorithms have demonstrated remarkable potential in facilitating the early detection and prognosis of skin lesions. This study advances the field by curating a comprehensive and diverse dataset comprising 39 categories of skin lesions, synthesized from five publicly available datasets. Using this dataset, the performance of five state-of-the-art deep learning models -- MobileNetV2, Xception, InceptionV3, EfficientNetB1, and Vision Transformer -- is rigorously evaluated.
To enhance the accuracy and robustness of these models, attention mechanisms such as the Efficient Channel Attention (ECA) and the Convolutional Block Attention Module (CBAM) are incorporated into their architectures. Comprehensive evaluation across multiple performance metrics reveals that the Vision Transformer model integrated with CBAM outperforms others, achieving an accuracy of 93.46\%, precision of 94\%, recall of 93\%, F1-score of 93\%, and specificity of 93.67\%.
These results underscore the significant potential of the proposed system in supporting medical professionals with accurate and efficient prognostic tools for diagnosing a broad spectrum of skin lesions.
The dataset and code used in this study can be found at \url{https://github.com/akabircs/Skin-Lesions-Classification}.

\end{abstract}

\end{frontmatter} 

\section{Introduction}
\label{sec:introduction}
The human body comprises numerous vital organs, among which the skin stands out as the largest, enveloping and protecting the entire body. Its extensive coverage makes it particularly susceptible to a wide range of viral, bacterial, and inflammatory conditions, leading to diverse health complications. Skin diseases encompass any disorders that negatively impact the skin, affecting its appearance, function, or structure \citep{Skin_Disorder}.
Certain abnormalities in the skin’s appearance, such as nodular or waxy growths on areas like the ear, neck, or face; scar-like patches; or rough, scaly lesions that may itch, bleed, or become crusty, can serve as early indicators of skin cancer. Importantly, early detection and treatment of these abnormalities significantly improve outcomes, with most cases of skin cancer being curable when identified and managed in their initial stages \citep{Abnormal_behaviour_of_skin}.
The global epidemiological burden of melanoma—a severe and potentially life-threatening form of skin cancer—is a growing concern. In 2020, it was estimated that 325,000 new cases of melanoma were diagnosed worldwide, with approximately 57,000 deaths attributed to this malignancy. Alarmingly, projections suggest that if current trends persist, the global burden of melanoma could rise to 510,000 new cases and 96,000 deaths annually by 2040 \citep{Statistics}. These statistics underscore the urgent need for effective diagnostic tools and strategies to combat the escalating prevalence of melanoma and other skin-related disorders.

If skin lesions are detected early, it is possible to prevent complications and mitigate the progression of disease effects, improving patient outcomes \citep{choudhary2022skin}. Traditional diagnosis of skin lesions heavily depends on the expertise and experience of dermatologists. However, factors such as variations in skin tone and the subtlety of visual indicators can increase the likelihood of misdiagnosis. Consequently, manual diagnosis is often time-consuming and prone to error. To address these limitations, computer-aided diagnostic (CAD) tools and automated systems can provide significant support to dermatologists by improving diagnostic accuracy and efficiency \citep{al2020multiple}.
Skin lesions can arise from various causes, including the entrapment of microbes in skin pores or hair follicles, parasitic infestations, the presence of microorganisms on the skin, or systemic illnesses involving organs such as the thyroid and kidneys. Additionally, a compromised immune system can predispose individuals to skin disorders. External factors, such as contact with allergens or infected skin, can also contribute. In some instances, genetic predisposition plays a critical role in the development of skin lesions \citep{Cause}.
Medical professionals diagnose skin lesions primarily through visual examination. In certain cases, laboratory methods such as biopsies or microscopic examination of skin samples are employed to confirm diagnoses. Despite their utility, these methods are not universally accurate and often require substantial time and resources \citep{debelee2023skin}. Given these challenges, this study proposes the development of a deep learning-based classification system to address a wide spectrum of skin lesions, encompassing 39 distinct types, thus enhancing diagnostic precision and reducing the time required for analysis \citep{arora2023comparative}.

Numerous studies have demonstrated that deep learning approaches for diagnosing skin lesion types from images of affected skin areas significantly enhance effectiveness and accuracy. While certain strategies, such as those outlined in \citep{kassem2021machine}, perform well on datasets with a few number of classes, they are prone to overfitting and exhibit inconsistent behavior when tested on datasets with a large number of classes. Techniques such as transfer learning, image generation via adversarial generative networks, and data augmentation offer potential solutions to mitigate the challenges of training models on limited datasets.
However, some researchers rely on non-public datasets, which presents reproducibility challenges. The unavailability of datasets hinders the validation and replication of findings, and the selection of web-sourced images may introduce biases. When datasets contain a large number of images per class, deep learning significantly outperforms traditional machine learning methods. Even in cases with limited image availability, deep learning models can overcome this limitation by employing data augmentation techniques. These methods enable models to make autonomous and intelligent decisions with higher accuracy rates.
Pre-trained deep learning models and state-of-the-art strategies have already demonstrated promising results for high-accuracy skin lesion classification. This research focuses on utilizing deep learning methodologies to classify 39 skin lesion types. Additionally, it aims to compare the performance of these methods to identify the most accurate model for handling a large number of skin lesion classes effectively.
The key contributions of this study are as follows:

\begin{itemize}
    \item A dataset comprising 39 classes is curated by integrating five publicly available datasets \citep{rafay2023efficientskindis}, \citep{Skinnnn}, \citep{Skin_Lesions_Classification_Dataset}, \citep{Dataset_23}, \citep{SkinDiseases}. This integration enhances the diversity and variation of skin lesions represented in the dataset, ensuring comprehensive coverage of 39 distinct types of skin lesions. 
    
    \item This study introduces attention modules into state-of-the-art deep learning models and explores their advantages on the curated dataset. The impact of attention mechanisms on model performance is evaluated using diverse metrics, demonstrating their potential to enhance the robustness and precision of deep learning models in skin lesion classification.
    
    \item A rigorous analysis is conducted to identify the best-performing model on the curated dataset. Subsequently, an extensive evaluation compares the performance of the top-performing model with state-of-the-art studies, further validating its effectiveness and superiority. 
\end{itemize}

The remainder of this paper is structured as follows. Section \ref{Review} provides an overview of related works, discussing recent advancements and methodologies in the field of skin lesion classification. Section \ref{method} describes the overall methodological framework, including a detailed explanation of the curated dataset, the deep learning models explored, and the integration of attention modules. Section \ref{reslt} presents a comprehensive experimental analysis of the performance of the various models, along with a comparative evaluation of the proposed approach against existing methods on standard datasets. Following the discussion presented in Section~\ref{sec:discussion}, the paper concludes with the key findings and implications outlined in Section~\ref{conclu}.

\section{Related work}\label{Review}
Numerous studies have utilized machine learning techniques to classify skin lesions, primarily relying on publicly available datasets. Table \ref{differ} provides a summary of key contributions from existing works, including the datasets used, the number of skin lesion classes considered, and the approaches employed. While these studies have made significant strides in advancing lesion classification, most are constrained by their focus on a limited number of lesion types and their dependence on pre-existing datasets. The methodologies often leverage traditional CNN-based architectures or the latest transformer models; however, they fall short of addressing the challenges inherent in large-scale skin lesion classification. In contrast, this study introduces a novel approach that incorporates advanced attention-guided techniques, enabling comprehensive and robust classification across a significantly larger and more diverse dataset.

\begin{table}[!ht]
    \centering
    \caption{A summary of the related work} \label{differ} 
\resizebox{\textwidth}{!}{
     \begin{tabular}{cp{7cm}p{6.5cm}c}  
    
    \hline
    Study & Model & Dataset & Class\\ 
    \hline 
    \hline
 \citep{10421742} & Deep Reinforcement Learning & ISIC 2017 \citep{codella2018skin} & 3\\

    \citep{CHOUDHARY2022104659} & Deep Neural Networks & ISIC 2017 \citep{codella2018skin} & 2\\

        \citep{arora2023comparative} & DenseNet201 & ISIC 2018 \citep{codella2019skin} & 7\\
    \citep{young2019deep} & Inception & HAM10000 \citep{tschandl2018ham10000} & 2\\

    \citep{chowdhury2021exploring} & Custom CNN & HAM10000 \citep{tschandl2018ham10000} & 7\\
        \citep{srinivasu2021classification} & MobileNetV2 with LSTM & HAM10000 \citep{tschandl2018ham10000} & 7\\
    \citep{bioengineering10121430} & Custom CNN model & ISIC 2018 \citep{codella2019skin} and HAM10000 \citep{tschandl2018ham10000} & 7\\

    \citep{ayas2023multiclass} & Swin Transformer + CNN & ISIC 2019 \citep{tschandl2018ham10000, combalia2019bcn20000} & 8\\

    \citep{desale2024efficient} & Chimp Optimization with Vision Transformer & ISIC 2019 \citep{tschandl2018ham10000, combalia2019bcn20000} & 8\\

    \citep{saha2024yotransvit} & YOLOv3 and Vision transformer & ISIC 2019 \citep{tschandl2018ham10000, combalia2019bcn20000} & 8\\

    \citep{9121248} & Custom GoogleNet & ISIC 2019 \citep{tschandl2018ham10000, combalia2019bcn20000} & 8\\

    \citep{alsahafi2023skin} & Residual Deep Convolution Neural Network & ISIC 2019 \citep{tschandl2018ham10000, combalia2019bcn20000} & 8\\

    \citep{zeng2024dsp} & DSP-KD & ISIC 2019 \citep{tschandl2018ham10000, combalia2019bcn20000} & 8\\

    \citep{nakai2022enhanced} & Deep Bottleneck Transformer & ISIC 2017 \citep{codella2018skin} and HAM10000 \citep{tschandl2018ham10000} & 6 and 7\\
    
 \citep{RODRIGUES20208} & DenseNet201 + KNN & ISIC 2017 \citep{codella2018skin} and PH2 \citep{mendoncca2015ph2} & 3 and 2\\
 
   \citep{rezaee2024self} & Optimized ResNet + Transformer + Self-attention & ISIC 2019 \citep{tschandl2018ham10000, combalia2019bcn20000} and PH2 \citep{mendoncca2015ph2} & 8 and 3\\

    \citep{ahmad2024novel} & Vision Transformer & ISIC 2019~\citep{tschandl2018ham10000, combalia2019bcn20000}, ISIC 2020~\citep{rotemberg2021patient}, PH2 \citep{mendoncca2015ph2}, HAM10000 \citep{tschandl2018ham10000}  & 8\\

    \citep{nagadevi2024enhanced} & RAN + Inception + MobileNet & HAM10000 \citep{tschandl2018ham10000} and PH2 \citep{mendoncca2015ph2} & 4 and 3\\

    \citep{khan2024intelligent} & Custom CNN & HAM10000 \citep{tschandl2018ham10000}, ISBI 2016 \citep{gutman2016skin}, ISIC 2017 \citep{codella2018skin}, ISIC 2018 \citep{codella2019skin}, ISIC 2019~\citep{tschandl2018ham10000,combalia2019bcn20000} & 4 and 3\\

    \citep{rafay2023efficientskindis} & EfficientNet-B2 & Merged (Atlas Dermatology \citep{Atlas} (24 classes) and ISIC 2019 (7 classes)) & 31\\
    \citep{sadik2023depth} & Xception & Merged (Dermnet \citep{Dermnet} and HAM10000 \citep{tschandl2018ham10000}) & 5\\
\hline

    This study & ViT + CBAM & Curated from (ISIC 2019~\citep{tschandl2018ham10000, combalia2019bcn20000}, Atlas Dermatology~\citep{Atlas}, HAM10000 \citep{tschandl2018ham10000}, MSLD 2.0 \citep{ali2024web}, \citep{dermn} and other sources) & 39\\
    \hline 

    \hline
    \end{tabular}
    }   
\end{table} 

The ISIC 2017 dataset has been extensively used in skin lesion classification research, primarily focusing on binary or small multiclass problems. Prasanna Kumar et al. \citep{10421742} proposed an enhanced segmentation formula and a self-attention mechanism to extract crucial features from lesion images. Although effective, their work was limited to three lesion classes, reducing its applicability to broader classification tasks. Similarly, Choudhary et al. \citep{CHOUDHARY2022104659} explored feature extraction techniques such as GLCM and 2D DWT, coupled with deep learning models for binary classification. While these studies advanced segmentation and feature extraction, their restricted scope of lesion types limited their generalizability.

The ISIC 2018 dataset, synonymous with the HAM10000 dataset, has been widely adopted due to its inclusion of seven lesion types. Arora et al. \citep{arora2023comparative} conducted a comparative study of 14 deep learning networks, providing valuable insights into their capabilities but without addressing the challenges of scaling to larger datasets. Similarly, Ali et al. \citep{bioengineering10121430} explored binary and multiclass classification by leveraging transfer learning, but their work primarily focused on benign versus malignant distinctions. Other researchers, such as Srinivasu et al. \citep{srinivasu2021classification}, utilized MobileNetV2 and LSTM models to classify lesions but were constrained by the predefined lesion classes of HAM10000, limiting their generalizability to larger and more diverse datasets. Young et al. \citep{young2019deep} applied the Inception model on the dataset, focusing on binary classification (Melanoma vs. Naevus), but their restricted class scope reduced its applicability to multiclass problems. Chowdhury et al. \citep{chowdhury2021exploring} explored CNN architectures with self-attention to classify seven lesion types, offering a novel approach but without addressing the integration of more advanced attention mechanisms for feature enhancement.

The ISIC 2019 dataset offers an incremental increase in lesion types with eight classes and has been widely studied. Ayas et al. \citep{ayas2023multiclass} employed a hybrid Swin Transformer and CNN model, demonstrating the potential of transformer-based approaches. Desale and Patil \citep{desale2024efficient} integrated optimization techniques with Vision Transformer, while Saha et al. \citep{saha2024yotransvit} combined YOLOv3 segmentation with Vision Transformer. Alsahafi et al. \citep{alsahafi2023skin} introduced a custom Residual Deep Convolution Neural Network (Skin-Net), leveraging multilevel feature extraction and cross-channel correlation to address classification challenges. Kassem et al. \citep{9121248} applied a custom GoogleNet model, resizing the dataset to balance image distribution across classes, but the focus remained on eight lesion types. Zeng et al. \citep{zeng2024dsp} introduced multi-source knowledge fusion distillation techniques to enhance feature learning, although their framework did not address scaling to datasets with greater lesion diversity.

Several studies have explored multi-dataset approaches to improve generalizability and address data imbalances. Ahmad et al. \citep{ahmad2024novel} evaluated Vision Transformers and DeepLabv3+ across ISIC 2019, ISIC 2020, and PH2 datasets, highlighting the potential of cross-dataset analysis but facing challenges related to imbalanced data distributions. Nagadevi et al. \citep{nagadevi2024enhanced} employed ensemble learning techniques on HAM10000 and PH2, achieving improved performance but remaining limited in the number of classes considered. Sadik et al. \citep{sadik2023depth} implemented transfer learning models, including Xception, to classify skin lesions across HAM10000 and Dermnet datasets, providing a broader perspective but focusing on a limited number of lesion classes. Nakai et al. \citep{nakai2022enhanced} incorporated self-attention mechanisms into a deep bottleneck transformer model, working across ISIC 2017 and HAM10000 datasets. Although their approach improved feature extraction, the study was still limited to six and seven lesion classes, respectively.

Other studies explored integrating multiple datasets to create more diverse collections. Rezaee and Zadeh \citep{rezaee2024self} proposed a hybrid ResNet and Transformer-based framework to classify skin lesions using ISIC 2019 and PH2 datasets. Their method used cross-fusion techniques to integrate global and local features but was constrained by data imbalances and compatibility issues. Similarly, Rodrigues et al. \citep{RODRIGUES20208} leveraged DenseNet201 and KNN classifiers to extract features from ISIC 2017 and PH2 datasets, enhancing feature learning but focusing on a limited number of lesion types. While these studies addressed some challenges of working with multi-datasets, they often lacked advanced attention mechanisms or scalability to larger lesion classes.

Merged dataset studies aim to address the limitations of individual datasets by creating larger and more diverse collections. Rafay and Hussain \citep{rafay2023efficientskindis} merged ISIC with the Atlas Dermatology dataset to form a collection of 31 lesion classes and evaluated multiple transfer learning models. Although this approach improved dataset diversity, it did not incorporate attention-based mechanisms to refine feature extraction. Sadik et al. \citep{sadik2023depth} extended their analysis to merged datasets, including HAM10000 and Dermnet, employing transfer learning models such as Xception for classification. While their study broadened the scope by incorporating data from multiple sources, it still focused on a limited number of lesion classes, reducing its applicability to large-scale multiclass classification tasks.

Despite the progress made by these studies, significant gaps remain. Most works focus on datasets with a limited number of lesion types, relying on traditional CNN-based or transformer-based architectures without exploring the benefits of attention mechanisms. Additionally, while some studies merge datasets to enhance diversity, they fail to fully address the challenges of scalability and generalization in large-scale, multiclass classification scenarios.

This study addresses these gaps by curating a dataset comprising 39 skin lesion classes through the integration of five publicly available datasets. By incorporating advanced attention mechanisms such as CBAM into state-of-the-art deep learning models, this work enhances feature extraction and classification robustness. Moreover, it provides an extensive evaluation across a large and diverse dataset, distinguishing itself from previous studies and offering a comprehensive solution for large-scale skin lesion classification.

\section{Methodology}\label{method} 

\begin{figure}[!ht] 
\centering
\includegraphics[width=0.75\textwidth]{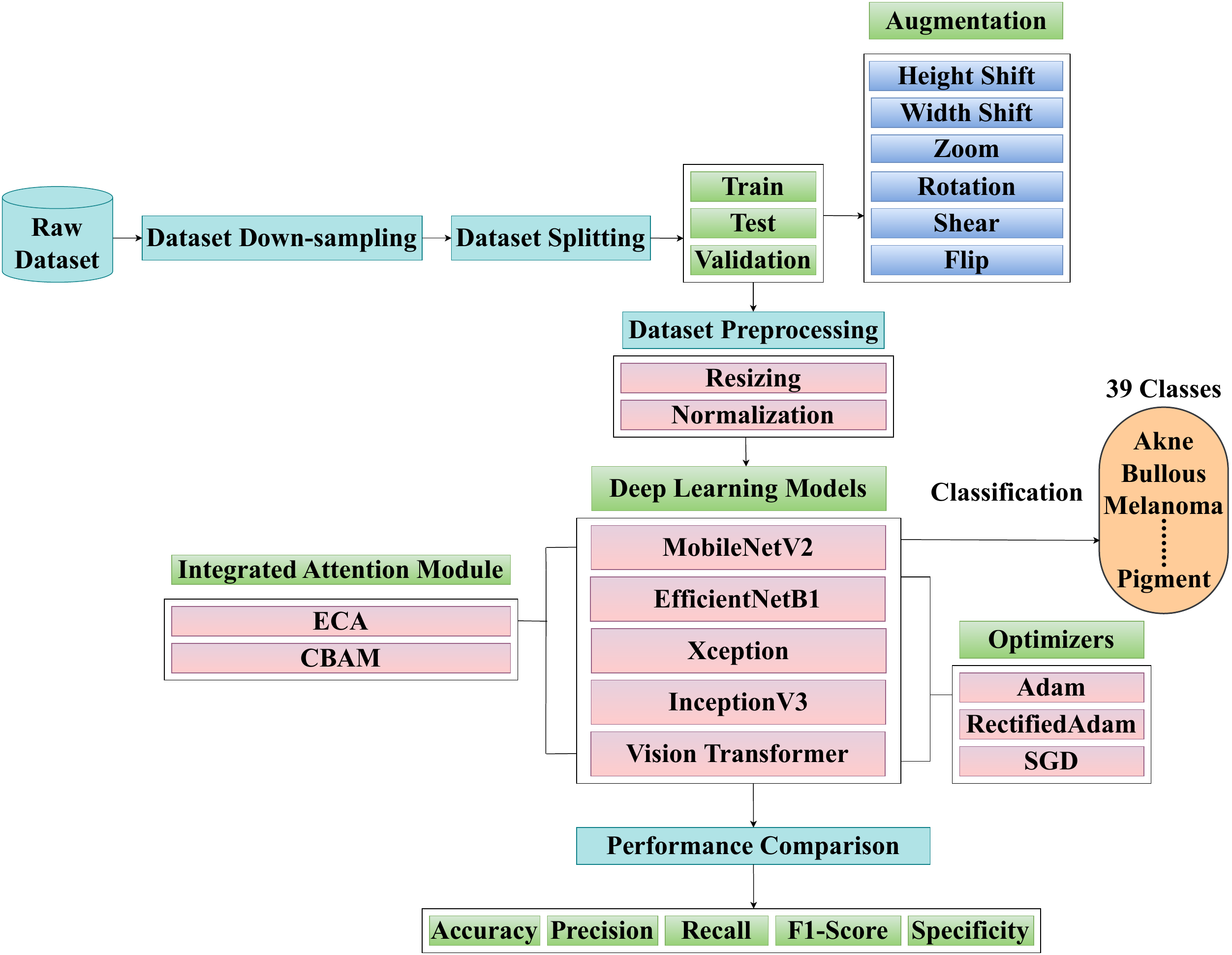} 
\caption{Methodological Flow of the Proposed Skin Disease Classification Approach} 
\label{general}
\end{figure}

The proposed methodology, illustrated in Figure \ref{general}, encompasses several critical stages designed to achieve robust and accurate skin lesion classification. These stages include dataset integration, where multiple publicly available datasets are merged to create a diverse and balanced dataset; data preprocessing, which involves normalization and augmentation techniques to improve data quality and variability; and image processing, aimed at refining input images for optimal model training. Furthermore, the methodology involves the exploration of deep learning models, both with and without the integration of advanced attention mechanisms, to evaluate their effectiveness in feature extraction and classification. Finally, performance evaluation is conducted using a comprehensive set of metrics to ensure a rigorous assessment of model performance.

\subsection{Dataset Curation}
In this study, a comprehensive dataset was curated by integrating five publicly available datasets, each contributing unique lesion types and features. The details of the datasets, including their sources, the number of classes, and the specific classes included, are outlined in Table \ref{data_list}.

\begin{table}[!ht] 
    \centering
    \caption{Sources of our curated dataset}
    \begin{tabular}{lcc}
    
    \hline
    Source & Number of classes & Classes taken\\ 
    \hline 
    \hline

    Merged dataset~\citep{rafay2023efficientskindis} (ISIC 2019~\citep{tschandl2018ham10000, combalia2019bcn20000} and Atlas Dermatology~\citep{Atlas}) & 31 & 12\\

    Skin Diseases 2 \citep{Skinnnn} 
    & 11 & 3\\

    Skin Lesion Classification Dataset \citep{Skin_Lesions_Classification_Dataset} (HAM10000 \citep{tschandl2018ham10000} and MSLD 2.0 \citep{ali2024web}) & 14 & 7\\
    
    dataset-23-skin \citep{Dataset_23} (Dermnet \citep{dermn}) & 23 & 16\\

    Skin Diseases \citep{SkinDiseases} (Atlas Dermatology~\citep{Atlas}) & 6 & 1\\
    \hline

    \hline
    \end{tabular}
\label{data_list}    
\end{table}

Integrating datasets from diverse sources presented several challenges. The datasets vary significantly in terms of class distributions, image resolutions, and quality. For example, while some datasets, such as ISIC 2019, contain detailed annotations and standardized imaging, others, such as Dermnet, include more general clinical images. These differences necessitated careful preprocessing and standardization to ensure uniformity in the curated dataset. Additionally, the initial integration revealed severe class imbalances, with some lesion types, such as Lichen Planus, represented by only 130 images, while others had thousands of samples. To address this, the dataset was balanced by capping each class at 130 images, ensuring equal representation and mitigating biases during model training. A sample image from each class in the curated dataset is presented in Table \ref{visu2}.
\begin{table}[!ht]
    \centering
    \caption{Sample images from all classes in the curated dataset} 
    \begin{tabular}{clc||clc} 
    
    \hline
    No. & Class Name & Image & No. & Class Name & Image\\
    \hline
    \hline

    1. & Akne 
    & \includegraphics[width=0.08\textwidth,height=0.03\textheight]{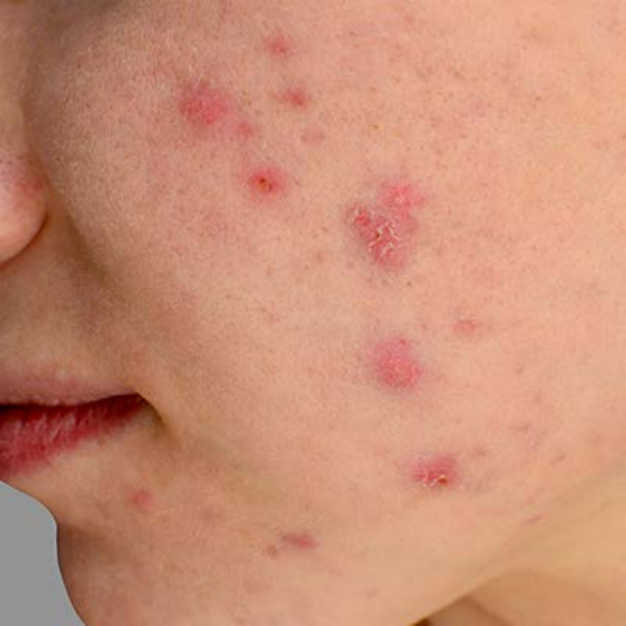}
     & 21. & Lupus 
    & \includegraphics[width=0.08\textwidth,height=0.03\textheight]{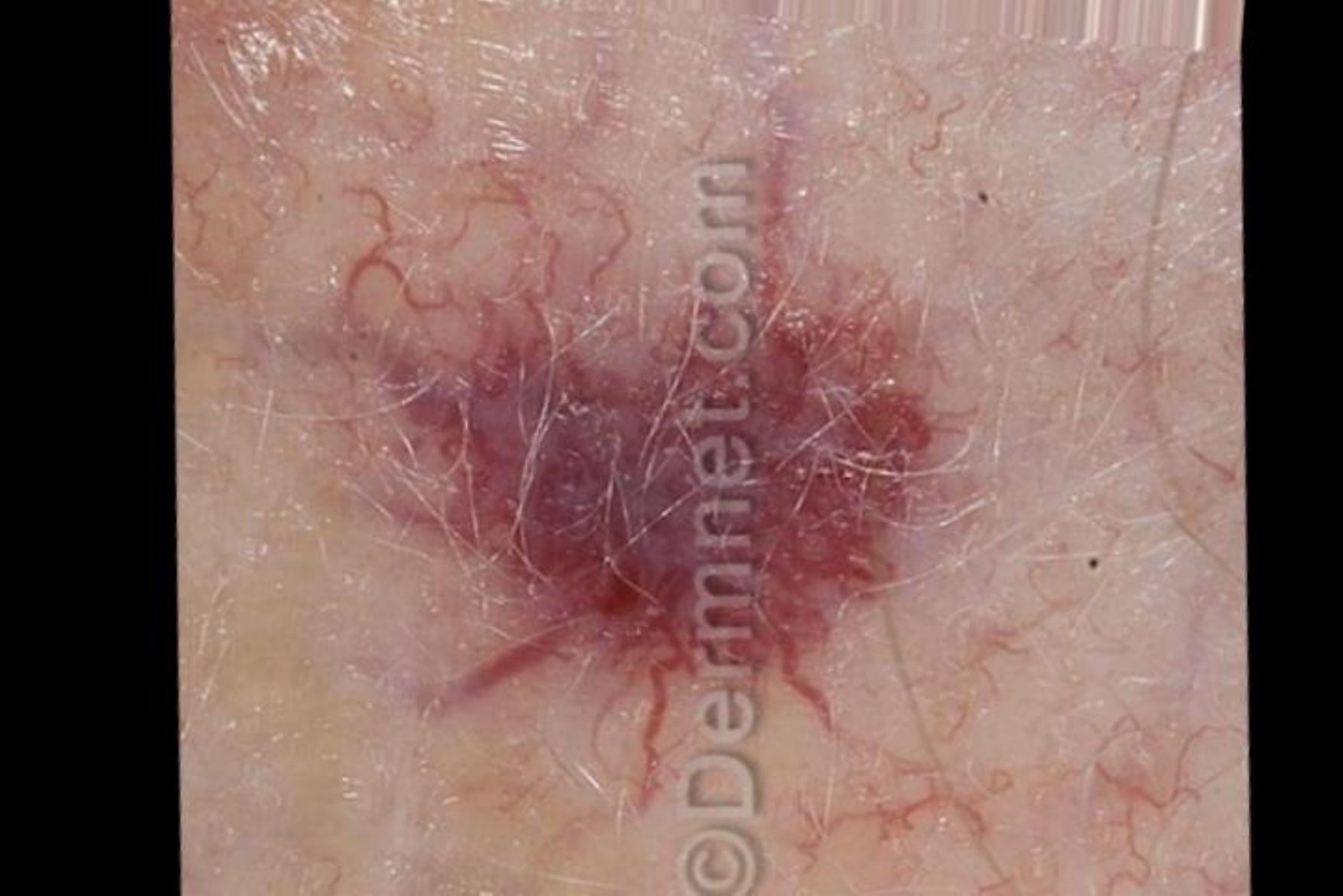}\\
       
    2. & Atopic Dermatitis 
    & \includegraphics[width=0.08\textwidth,height=0.03\textheight]{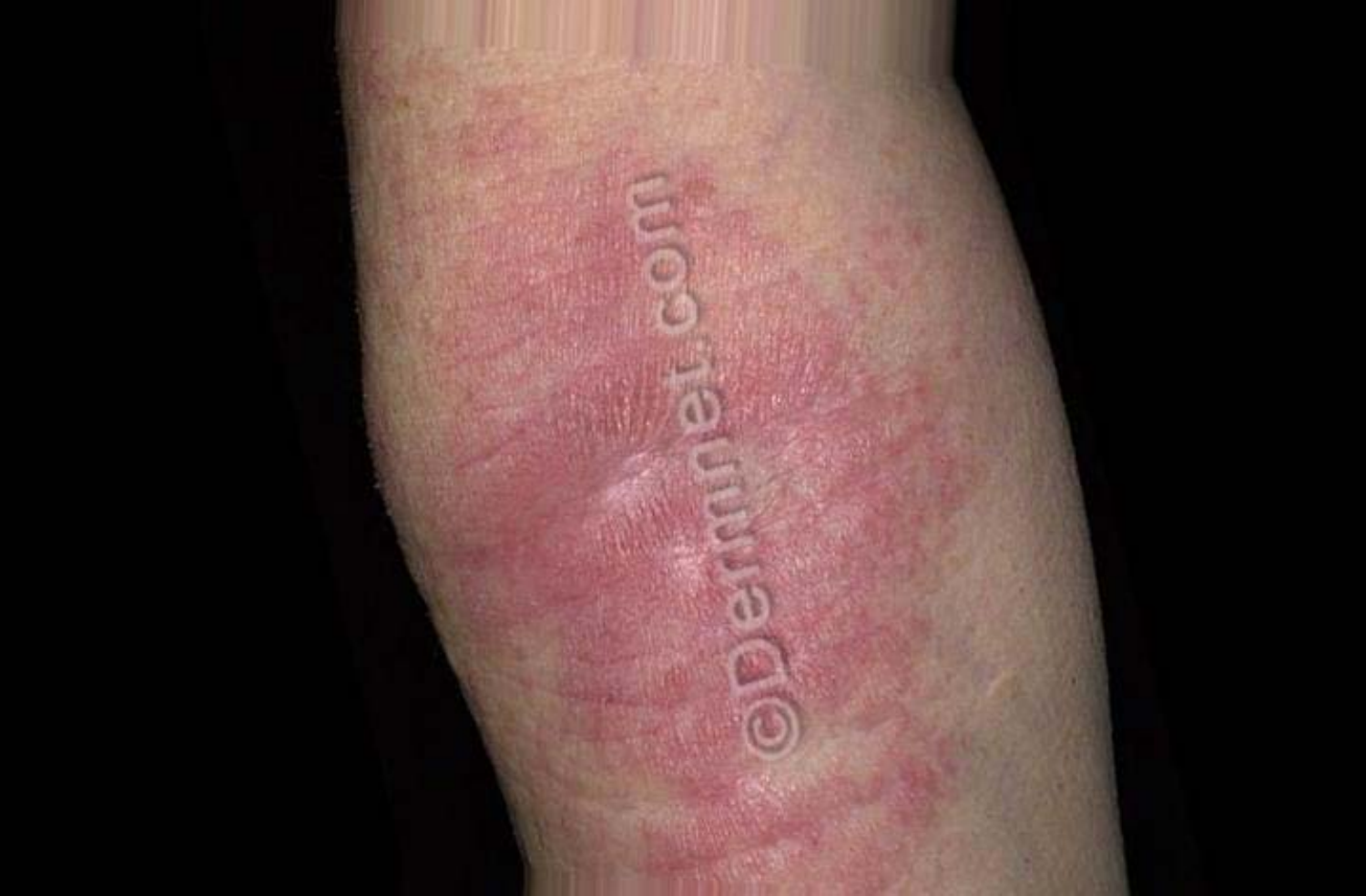}  
     & 22. & Measles 
    & \includegraphics[width=0.08\textwidth,height=0.03\textheight]{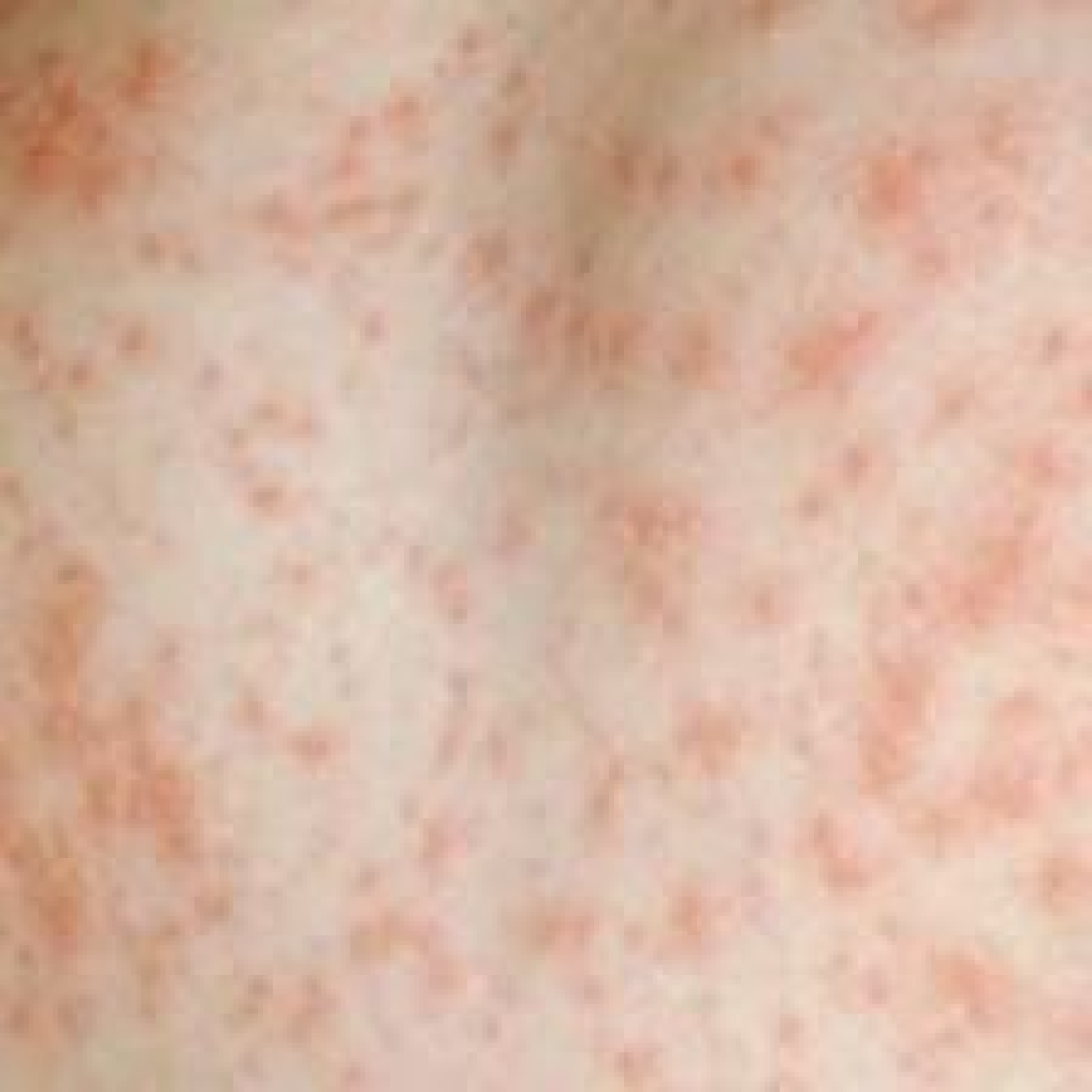}\\

    3. & Basal Cell Carcinoma 
    & \includegraphics[width=0.08\textwidth,height=0.03\textheight]{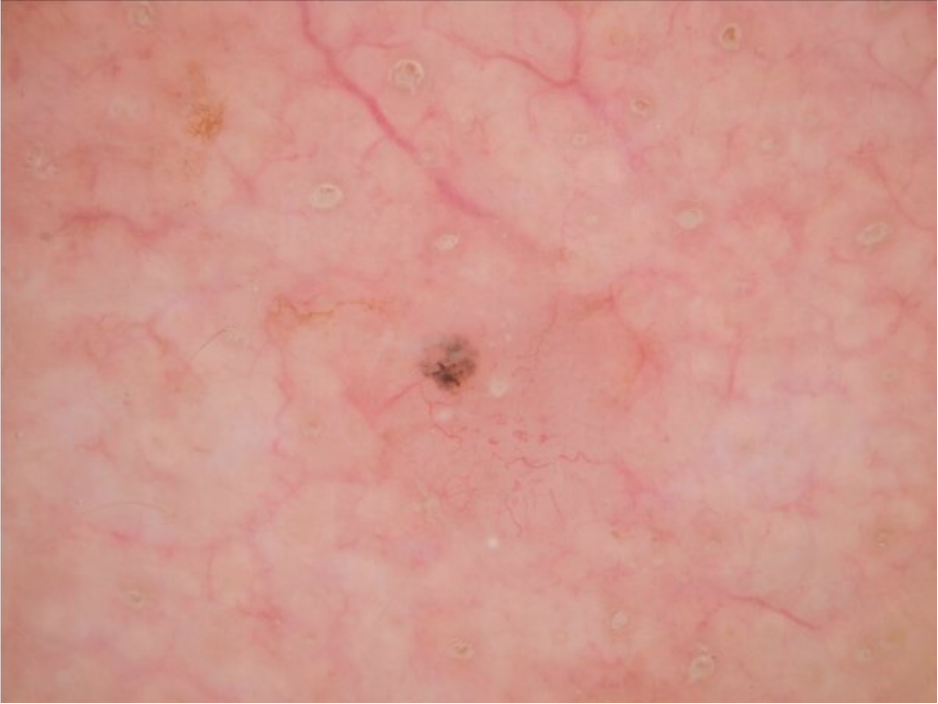} 
    & 23. & Melanocytic Nevi 
    & \includegraphics[width=0.08\textwidth,height=0.03\textheight]{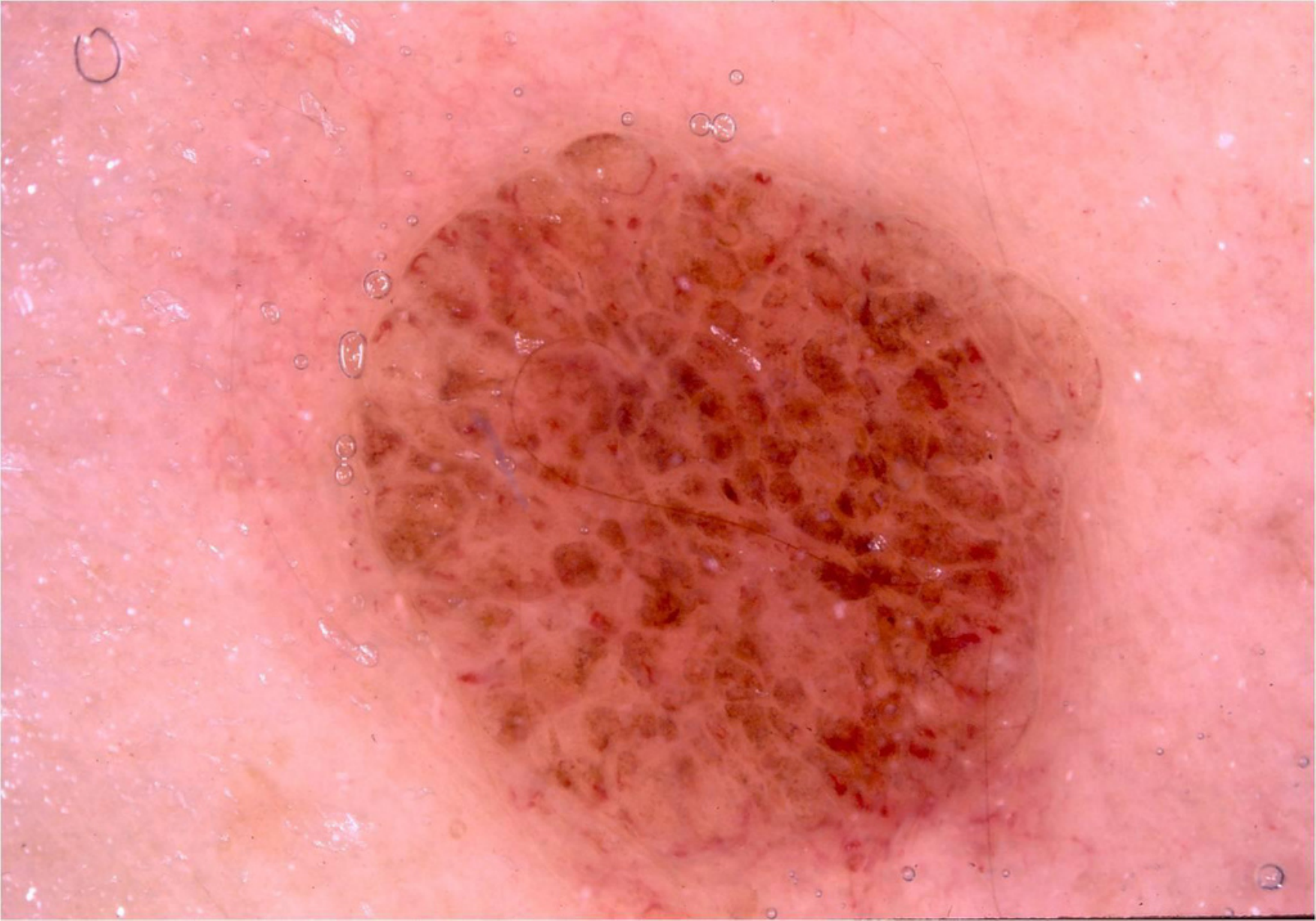}\\

    4. & Benign Keratosis 
    & \includegraphics[width=0.08\textwidth,height=0.03\textheight]{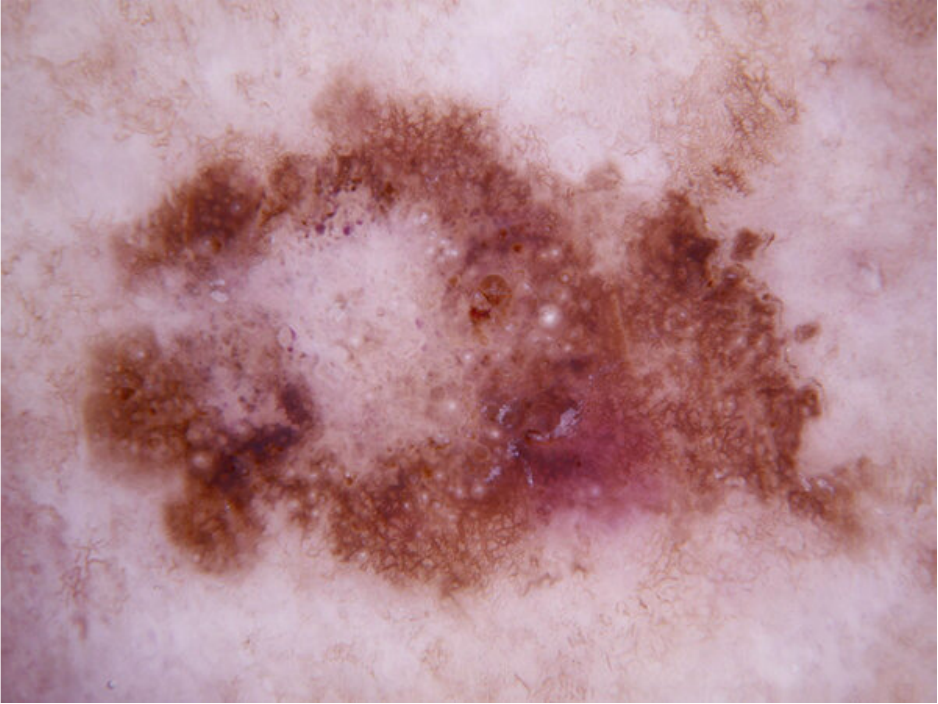} 
      & 24. & Melanoma
    & \includegraphics[width=0.08\textwidth,height=0.03\textheight]{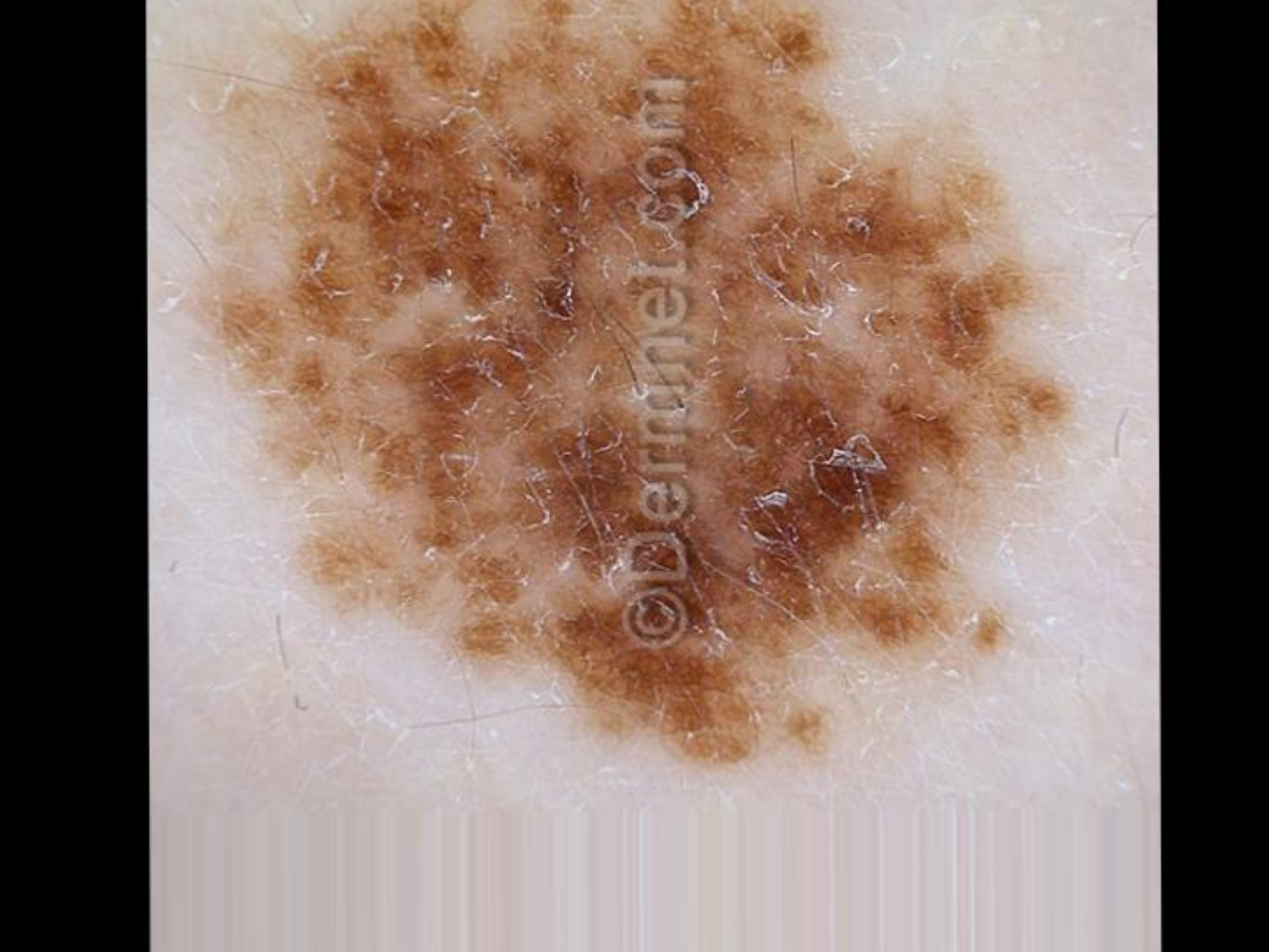}\\

    5. & Bullous 
    & \includegraphics[width=0.08\textwidth,height=0.03\textheight]{rotation3benign-familial-chronic-pemphigus-17.pdf} 
    & 25. & Molluscum Contagiosum 
    & \includegraphics[width=0.08\textwidth,height=0.03\textheight]{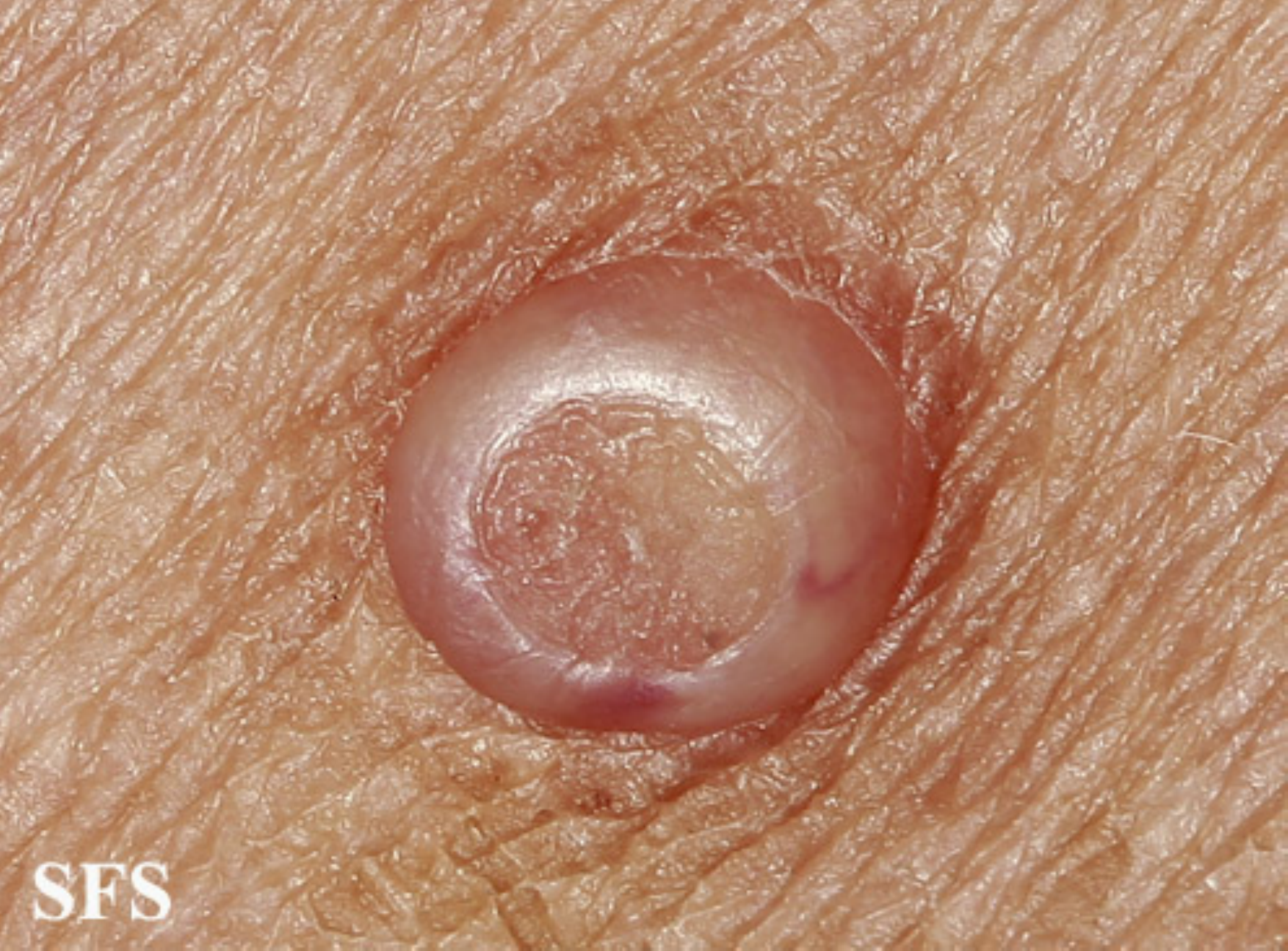}\\

    6. & Chickenpox 
    & \includegraphics[width=0.08\textwidth,height=0.03\textheight]{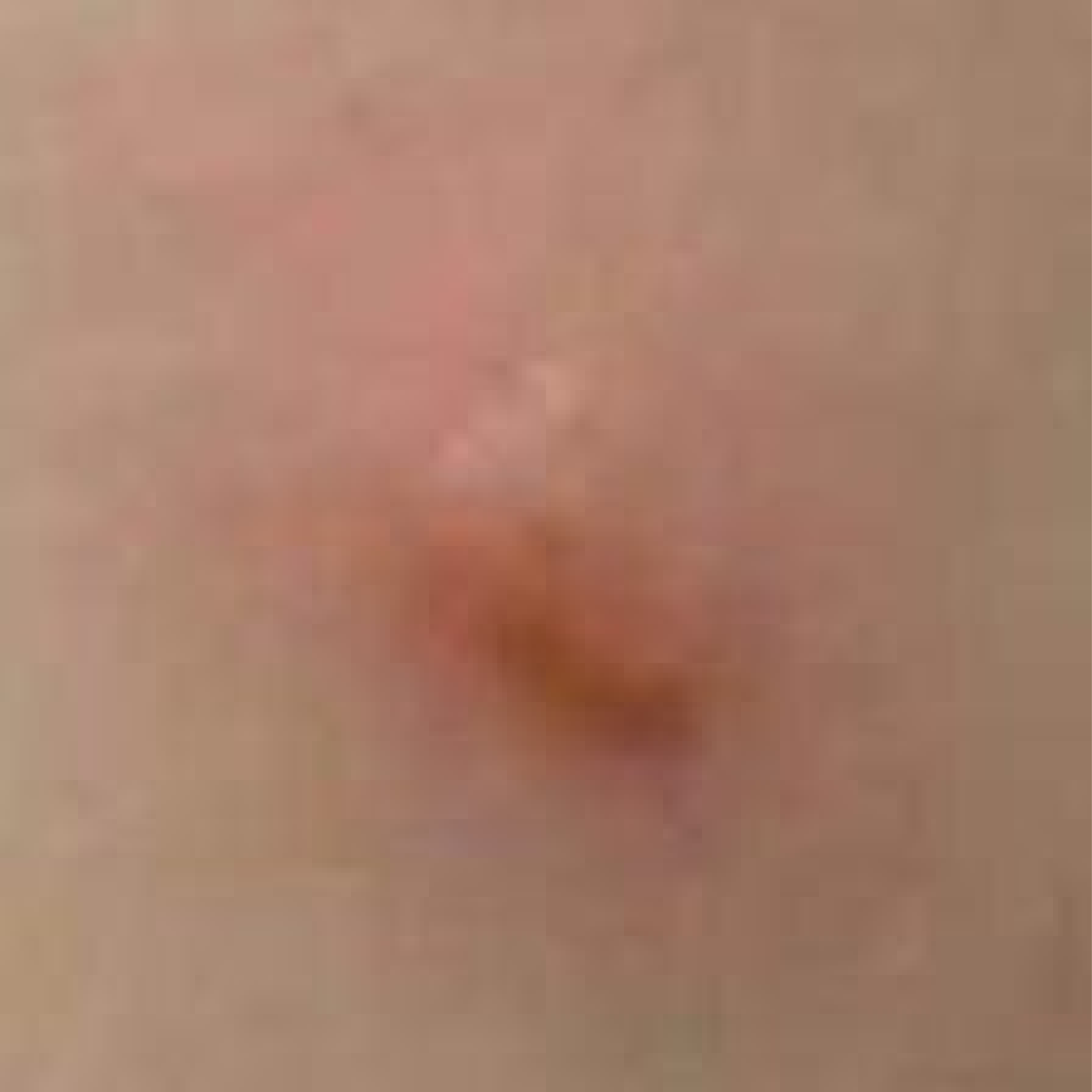} 
     & 26. & Monkeypox 
    & \includegraphics[width=0.08\textwidth,height=0.03\textheight]{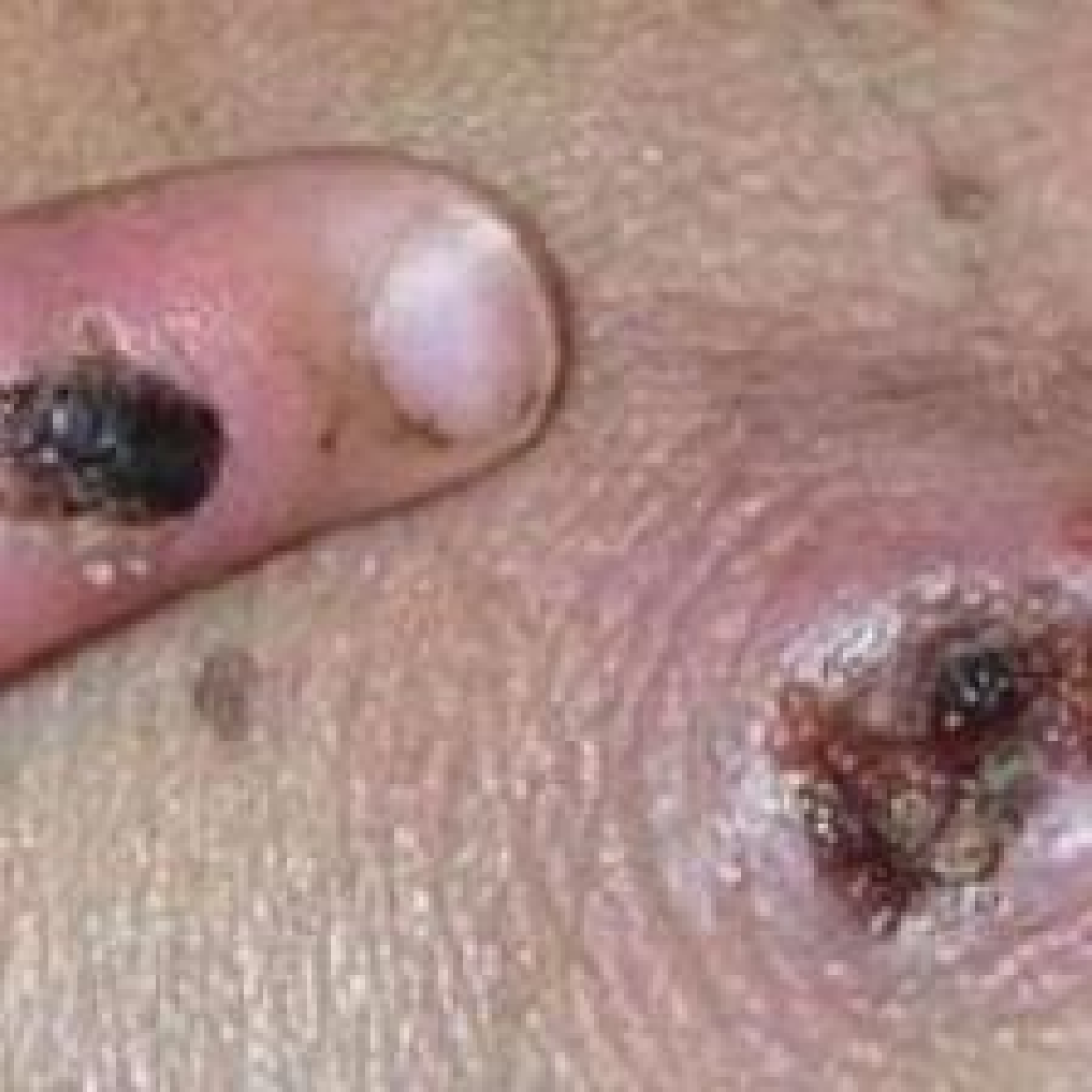}\\

    7. & Cowpox
    & \includegraphics[width=0.08\textwidth,height=0.03\textheight]{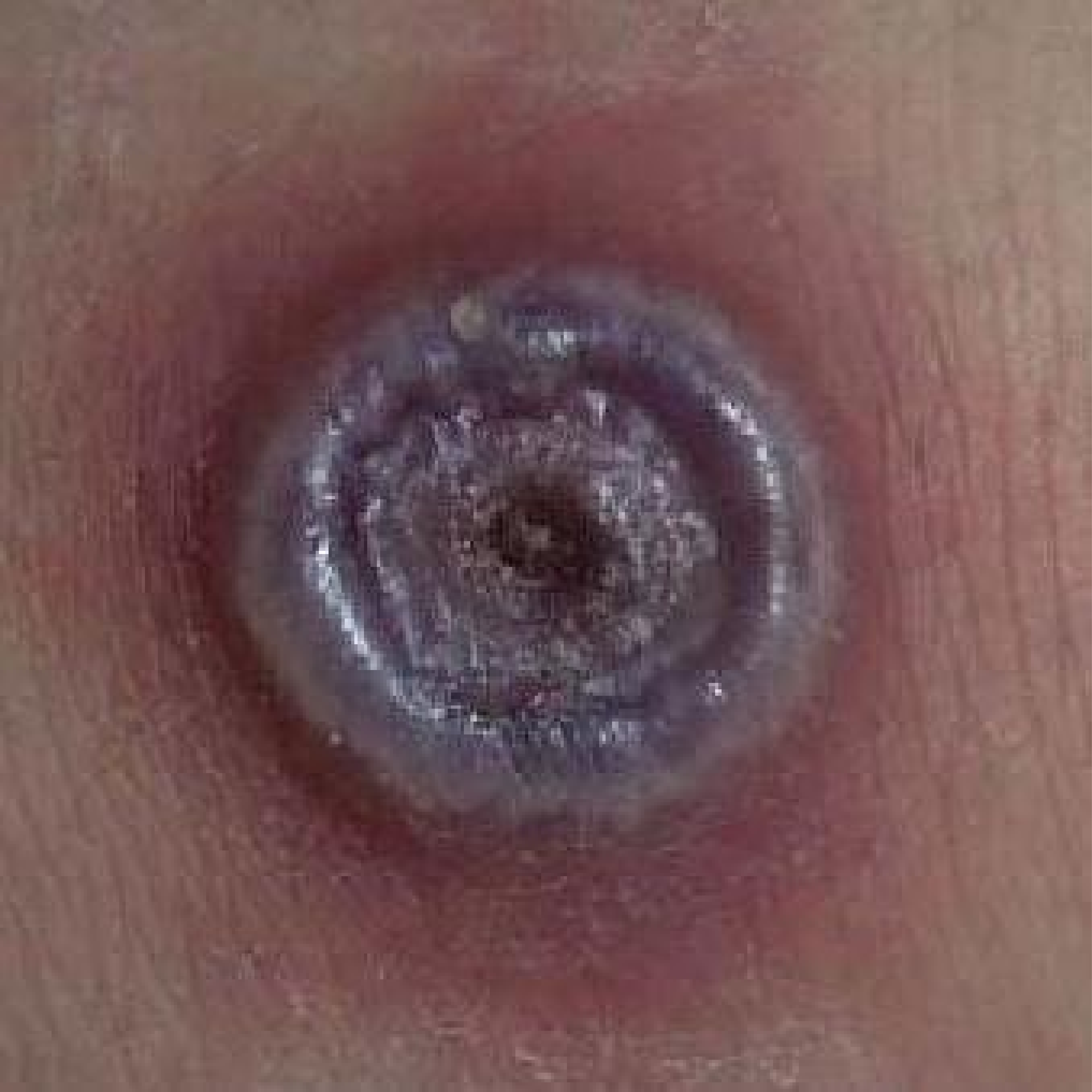}
    & 27. & Nail Fungus  
    & \includegraphics[width=0.05\textwidth,height=0.03\textheight]{164.pdf}\\
 
    8. & Dermatofibroma 
    & \includegraphics[width=0.08\textwidth,height=0.03\textheight]{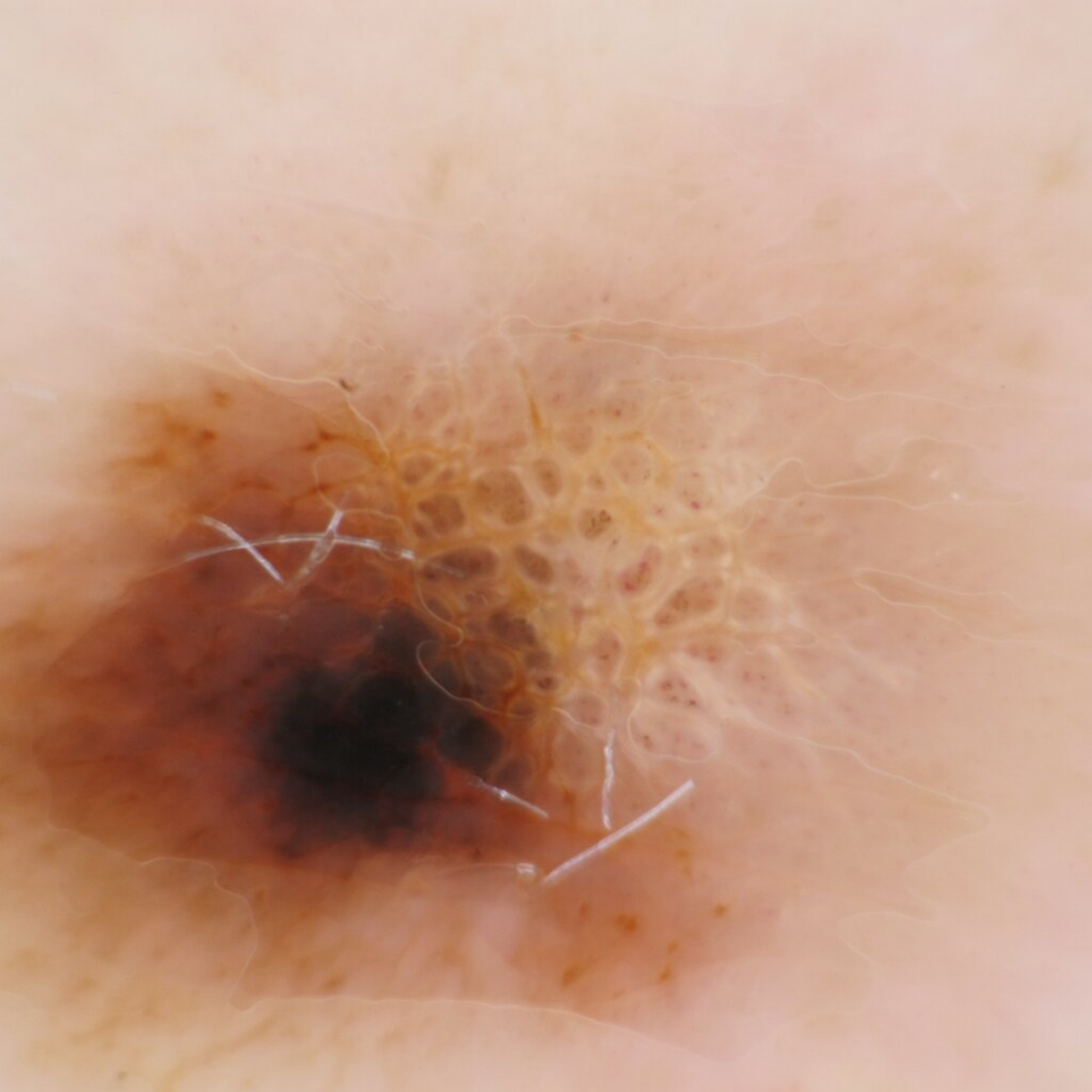} 
   & 28. & Pigment   
    & \includegraphics[width=0.08\textwidth,height=0.03\textheight]{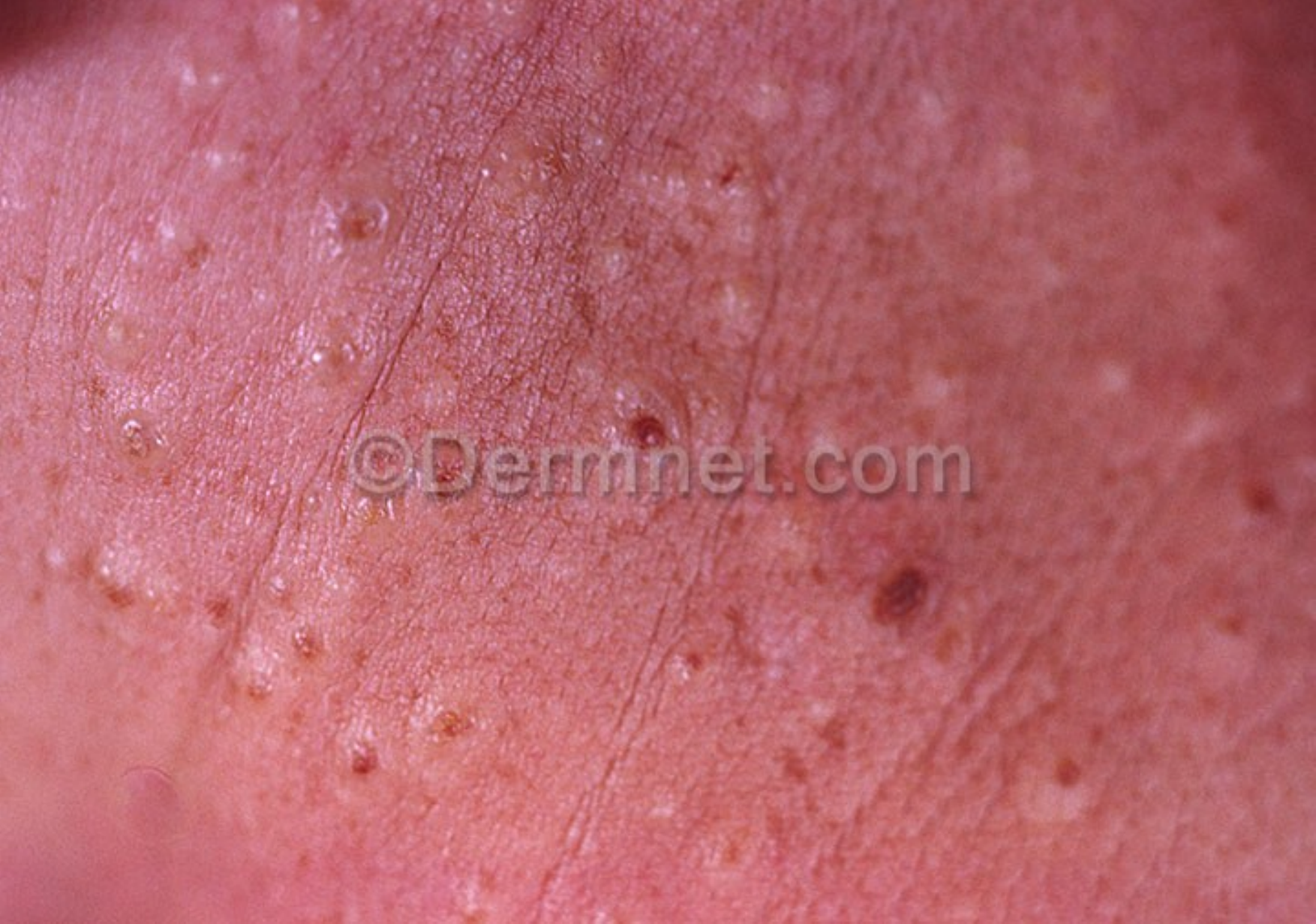}\\

    9. & Eczema 
    & \includegraphics[width=0.08\textwidth,height=0.03\textheight]{90.pdf} 
   & 29. & Pityriasis Rosea  
    & \includegraphics[width=0.08\textwidth,height=0.03\textheight]{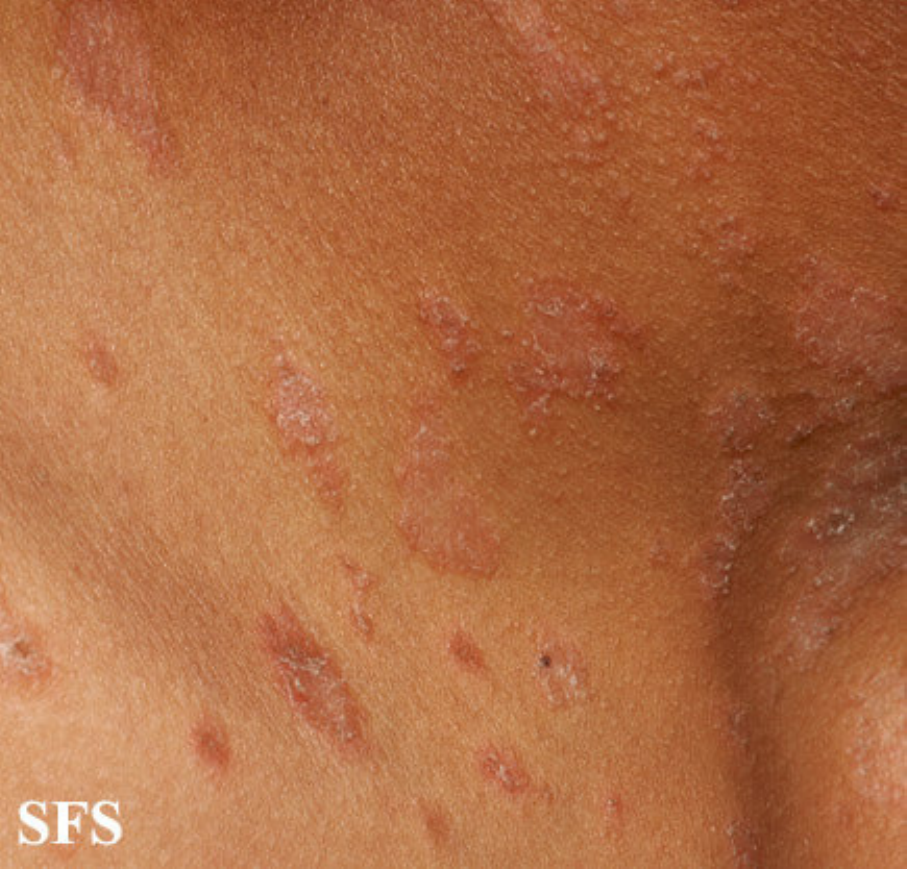} \\
 
    10. & Exanthems 
    & \includegraphics[width=0.08\textwidth,height=0.03\textheight]{193.pdf} 
    & 30. & Poison Ivy  
    & \includegraphics[width=0.08\textwidth,height=0.03\textheight]{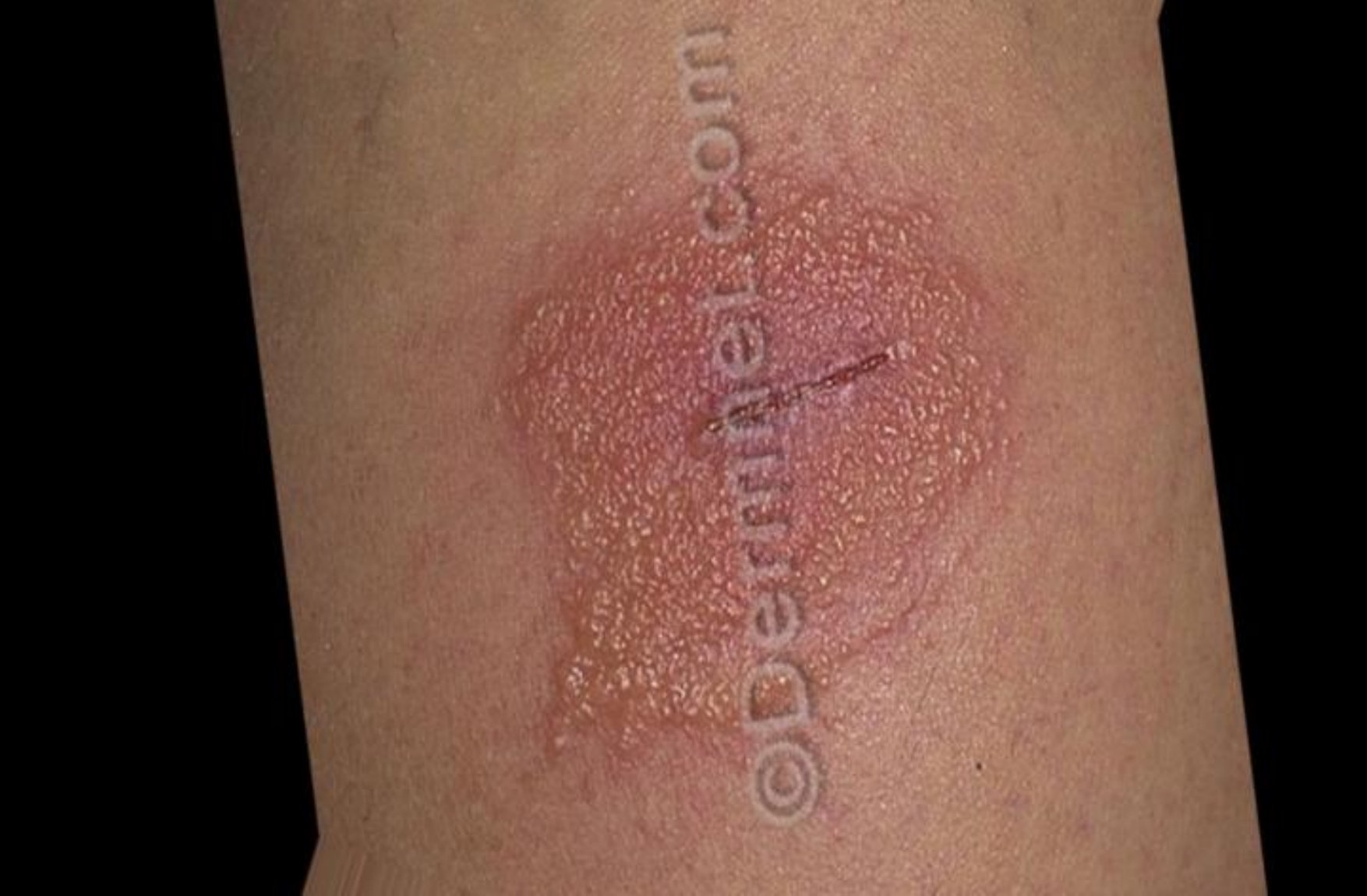}\\

     11. & Hand Foot Mouth Disease
    & \includegraphics[width=0.08\textwidth,height=0.03\textheight]{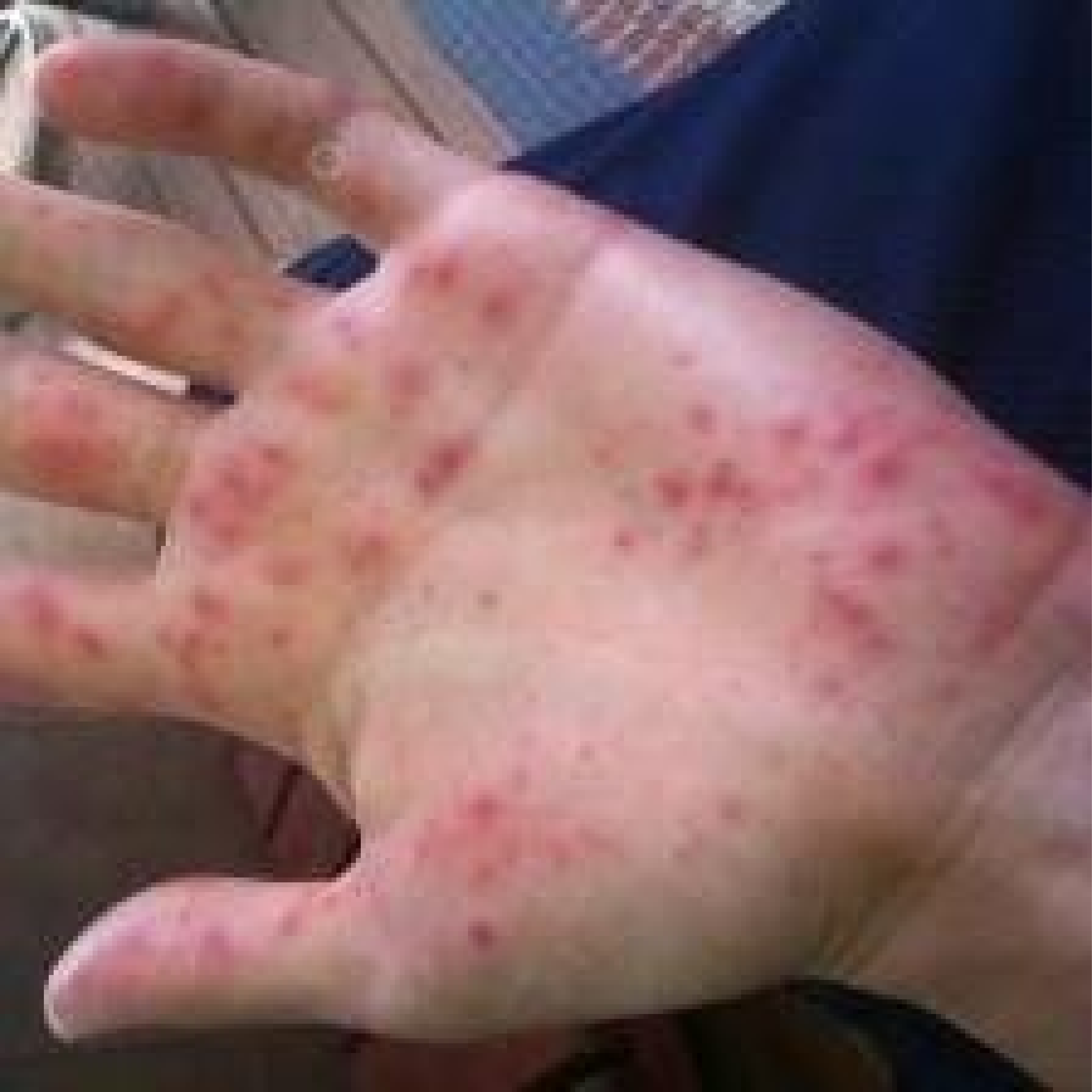}
   & 31. & Porokeratosis Actinic
    & \includegraphics[width=0.08\textwidth,height=0.03\textheight]{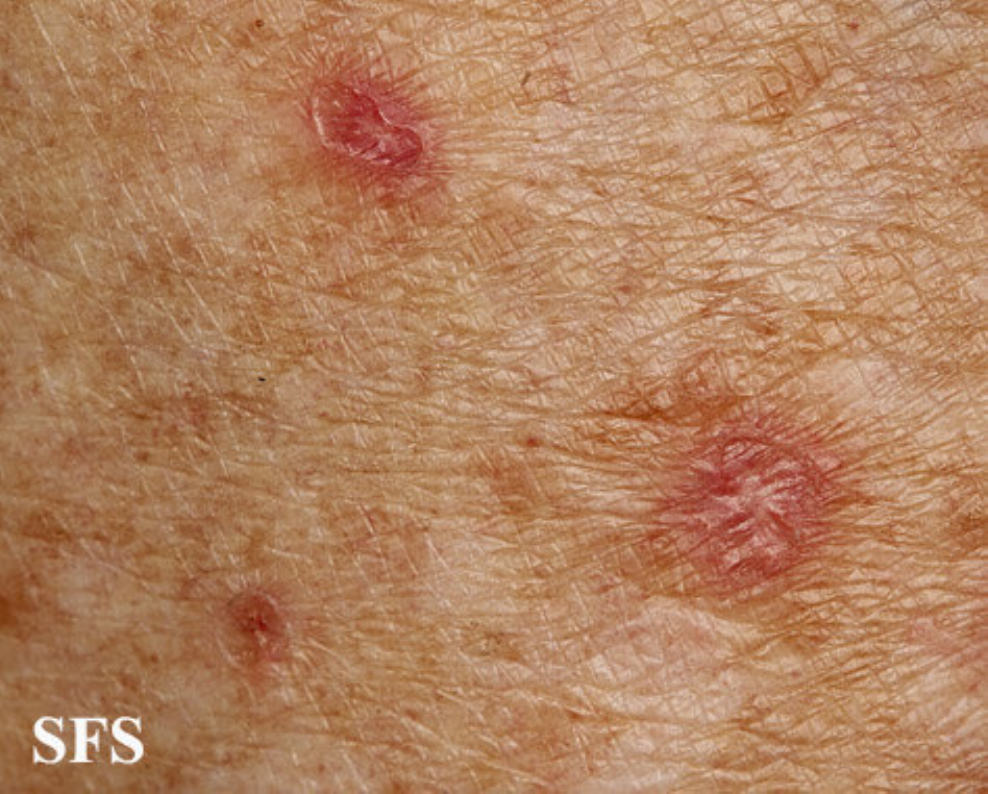} \\
    
    12. & Hailey-Hailey Disease
    & \includegraphics[width=0.08\textwidth,height=0.03\textheight]{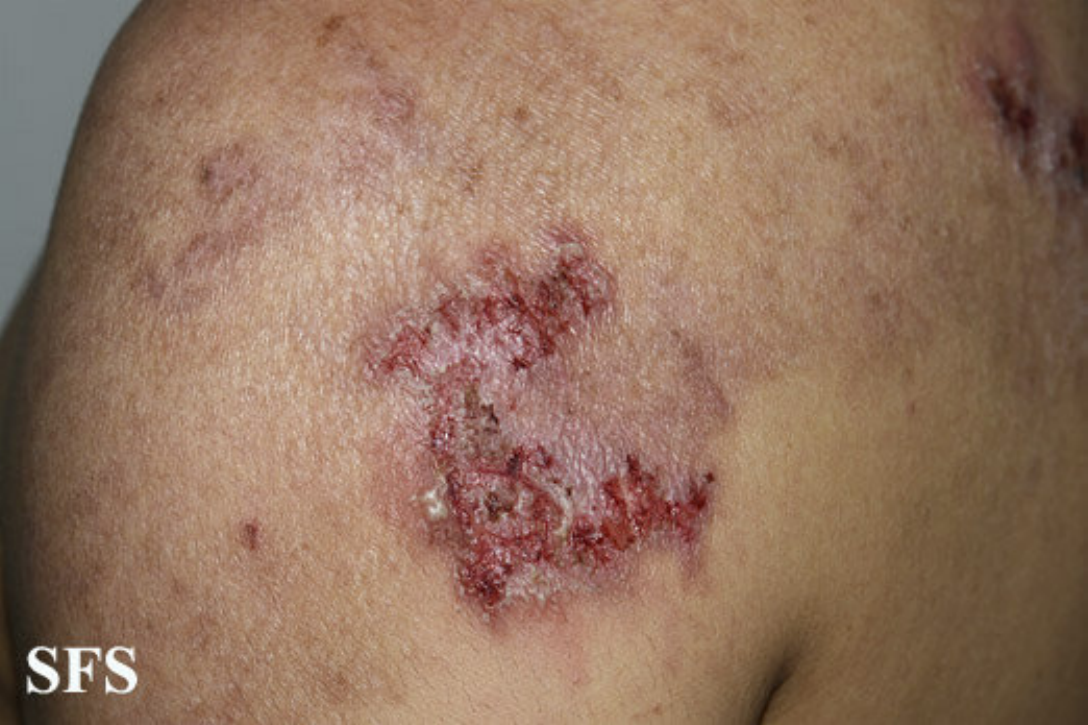} 
    & 34. & Psoriasis   
    & \includegraphics[width=0.08\textwidth,height=0.03\textheight]{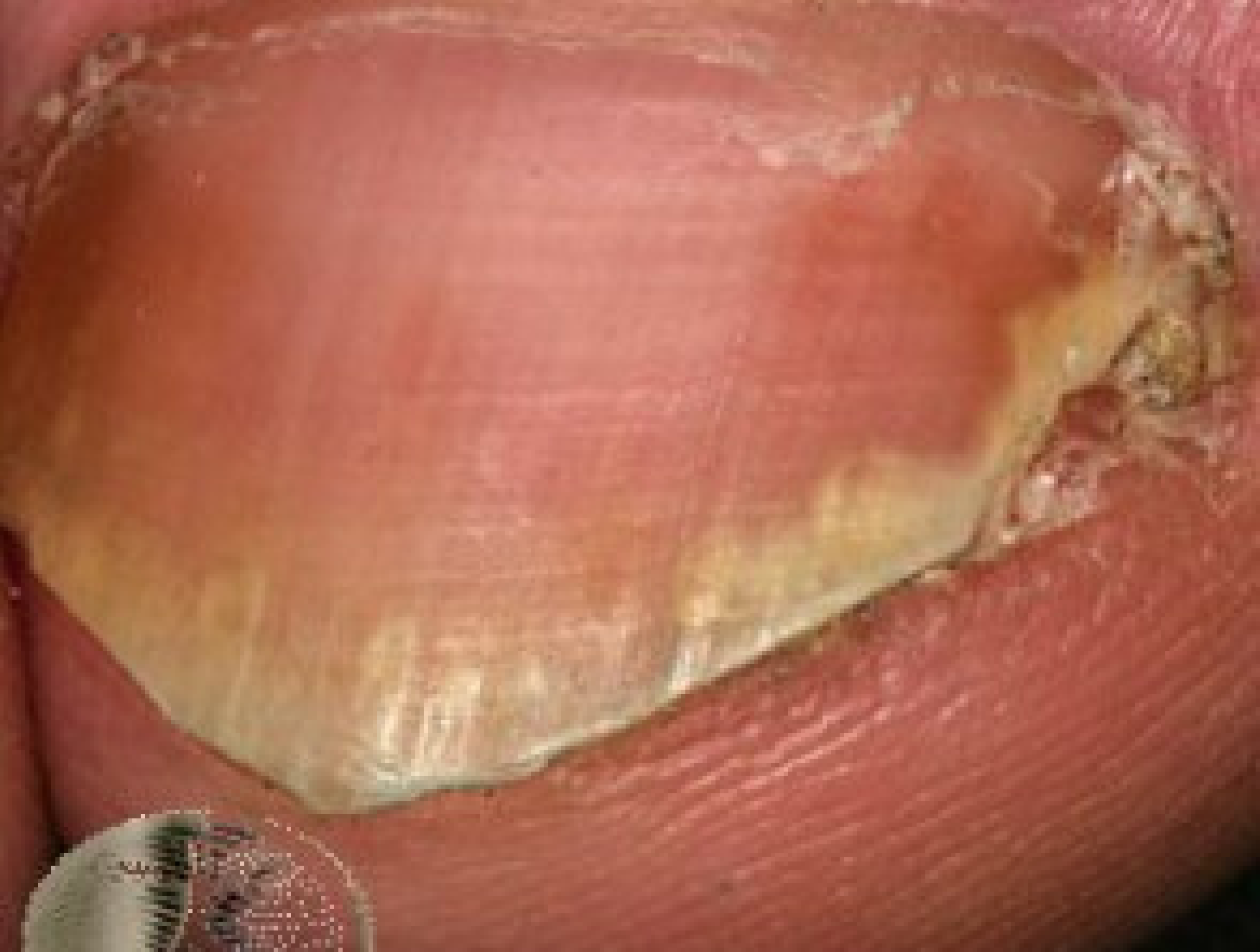}\\
    
    13. & Hair loss Alopecia
    & \includegraphics[width=0.08\textwidth,height=0.03\textheight]{194.pdf} 
    & 33. & Scabies Lyme Disease  
    & \includegraphics[width=0.08\textwidth,height=0.03\textheight]{caterpillar-dermatitis-4.pdf} \\

   14. & Impetigo 
    & \includegraphics[width=0.08\textwidth,height=0.03\textheight]{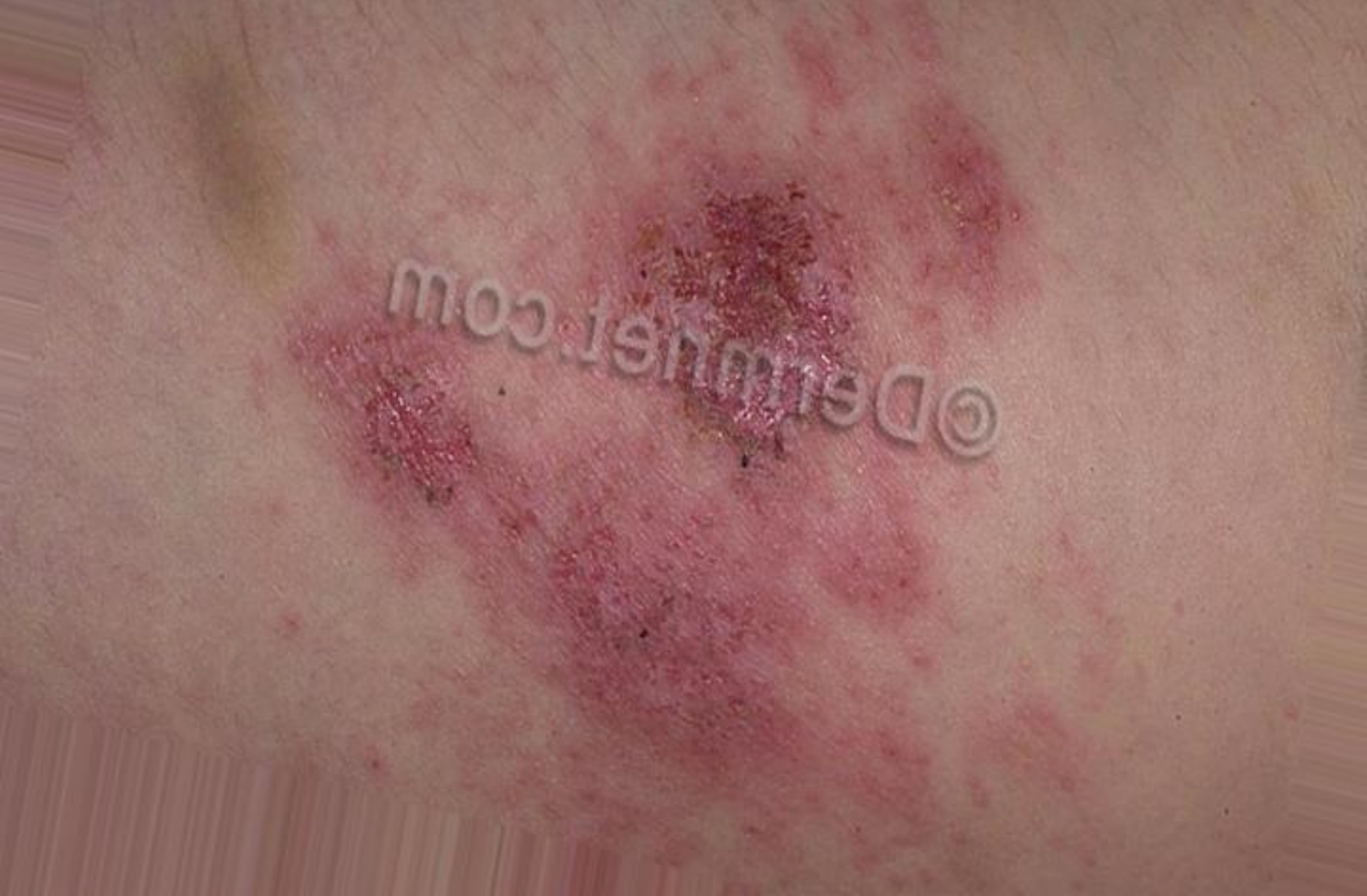}    
    & 34. & Seborheic Keratosis   
    & \includegraphics[width=0.08\textwidth,height=0.03\textheight]{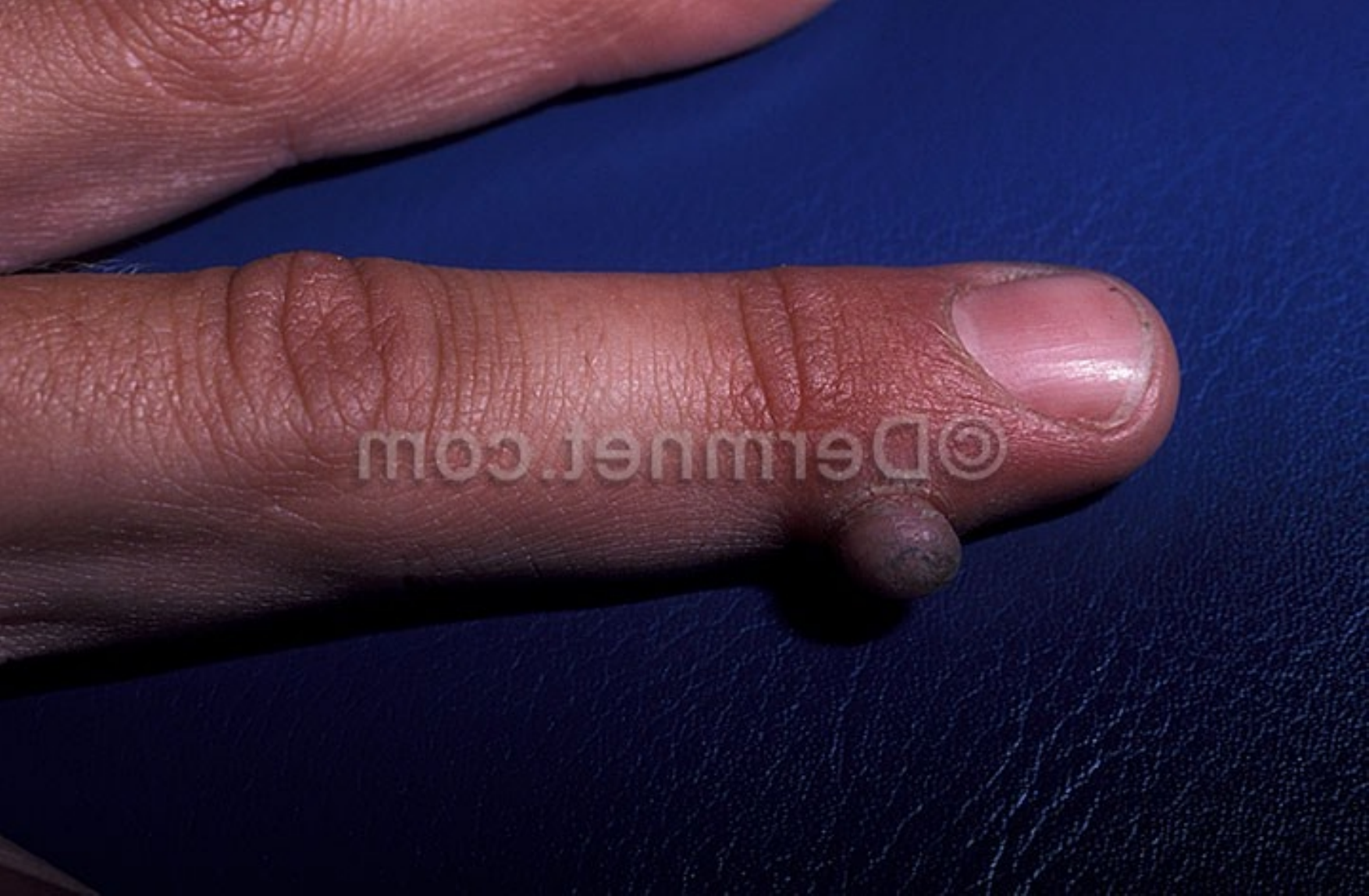}\\

    15. & Leprosy Borderline
    & \includegraphics[width=0.08\textwidth,height=0.03\textheight]{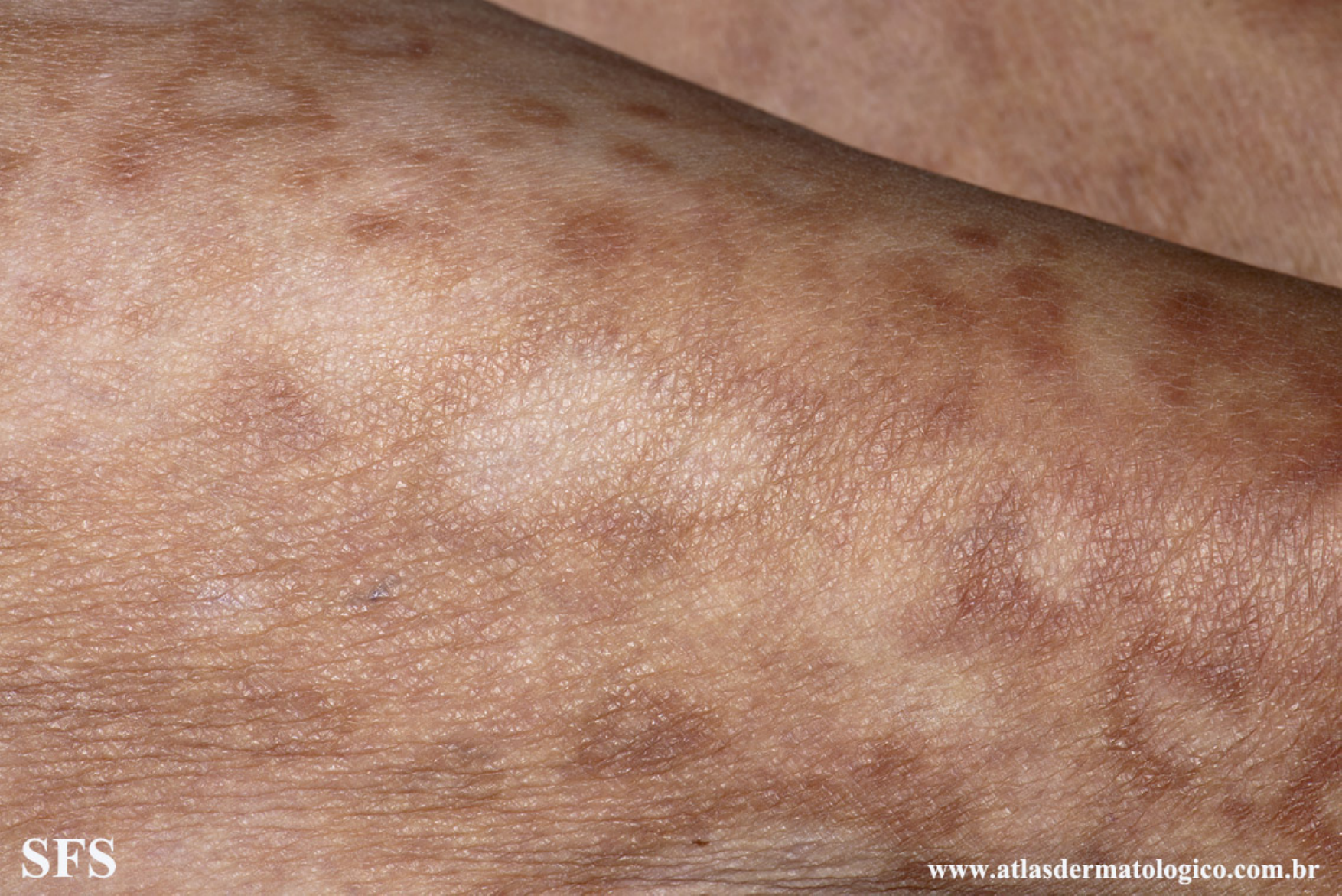}
    & 35. & Systemic Disease   
    & \includegraphics[width=0.08\textwidth,height=0.03\textheight]{imagelightamyloidosis-16.pdf}\\

    16. & Leprosy Lepromatous 
    & \includegraphics[width=0.08\textwidth,height=0.03\textheight]{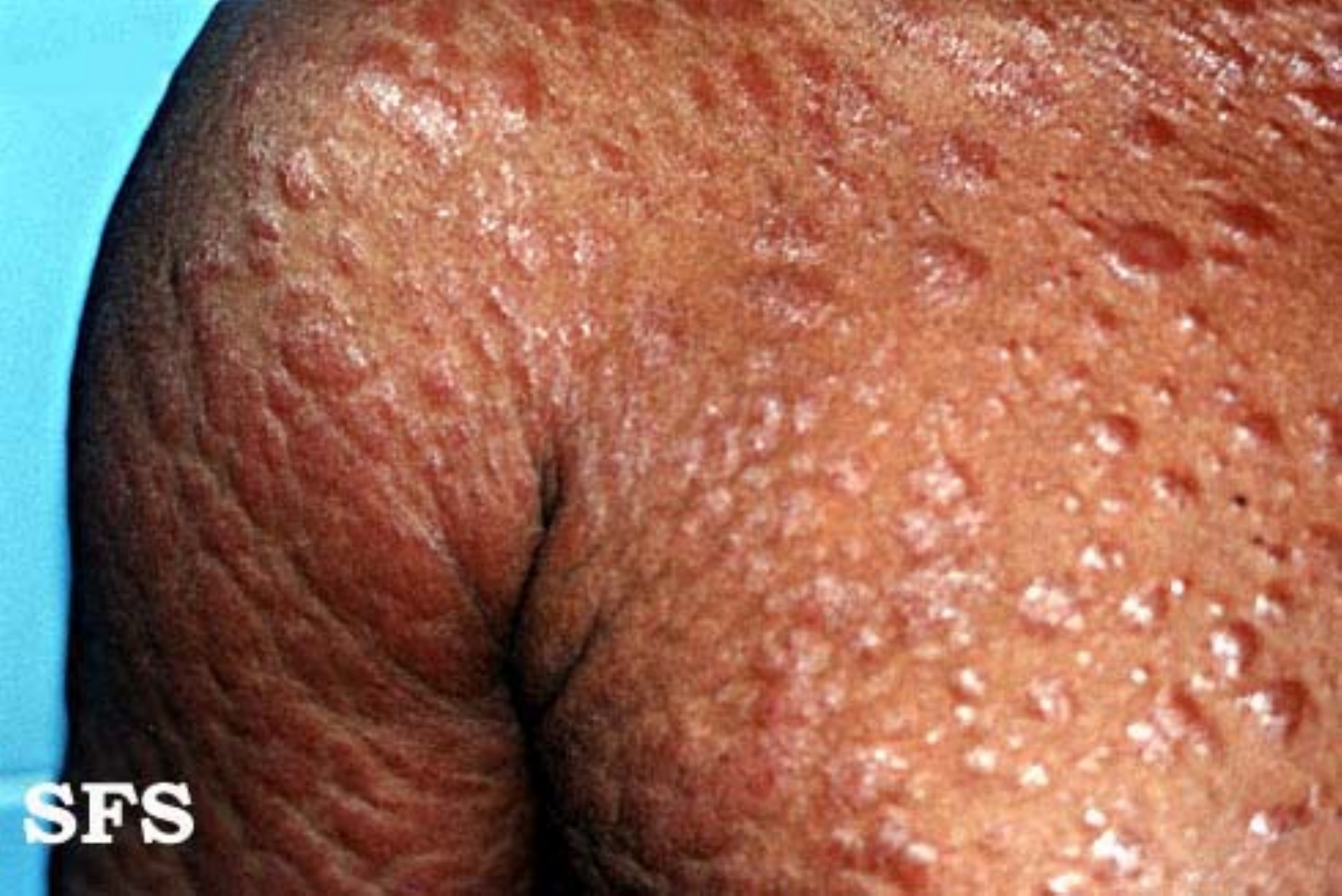}
    & 36. & Tinea Ringworm   
    & \includegraphics[width=0.08\textwidth,height=0.03\textheight]{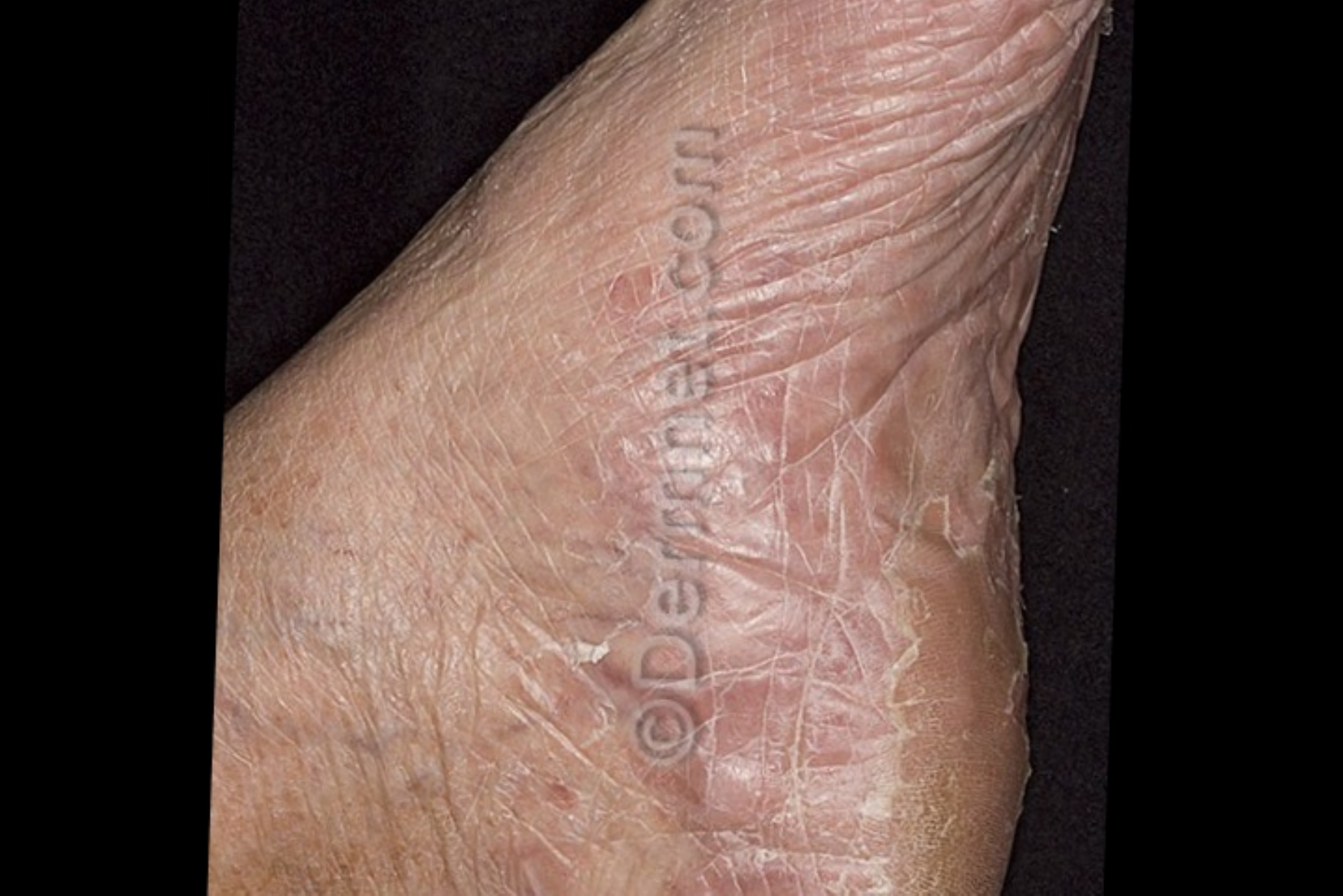}\\

    17. & Larva Migrans
    & \includegraphics[width=0.08\textwidth,height=0.03\textheight]{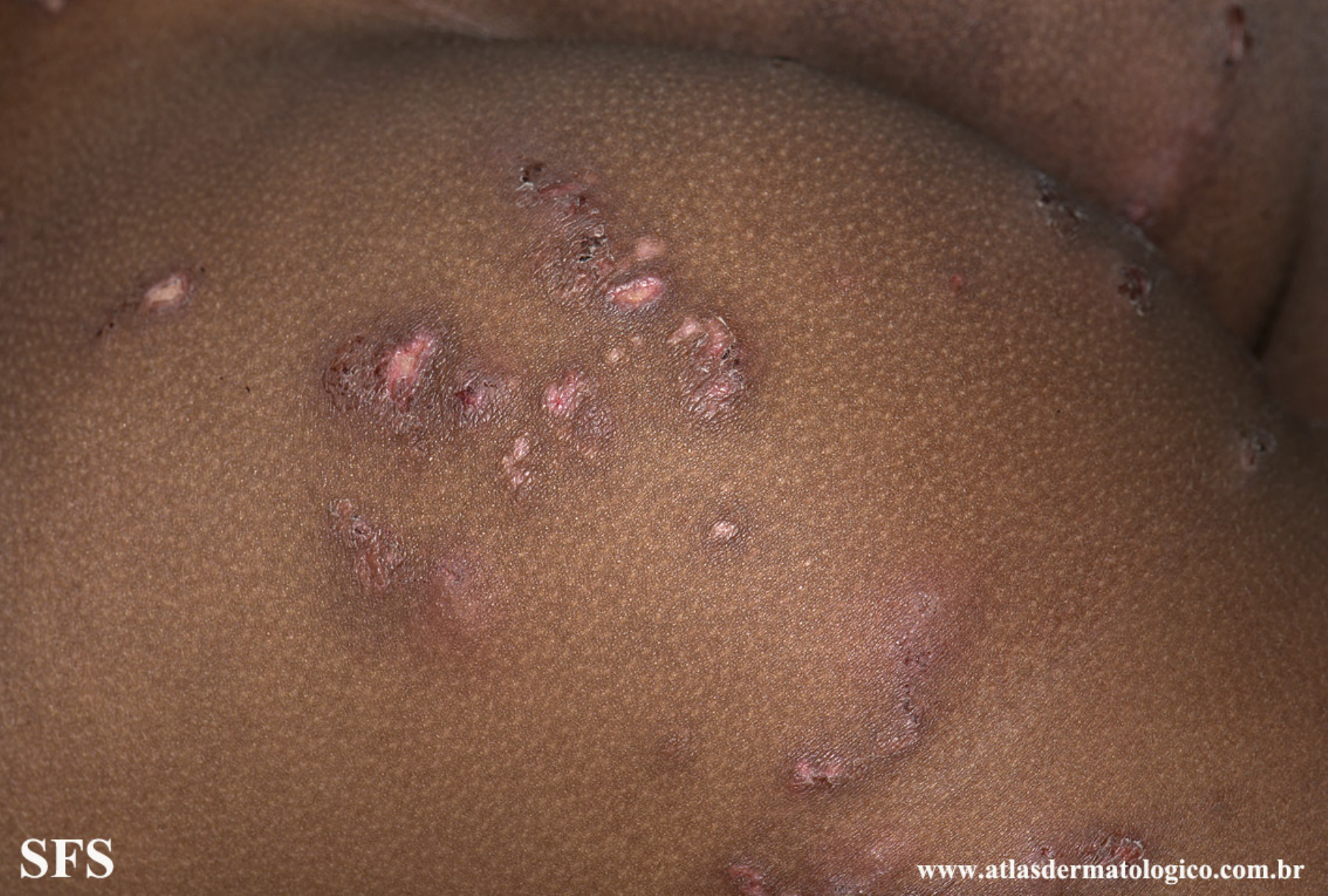}    
    & 37. & Tungiasis  
    & \includegraphics[width=0.08\textwidth,height=0.03\textheight]{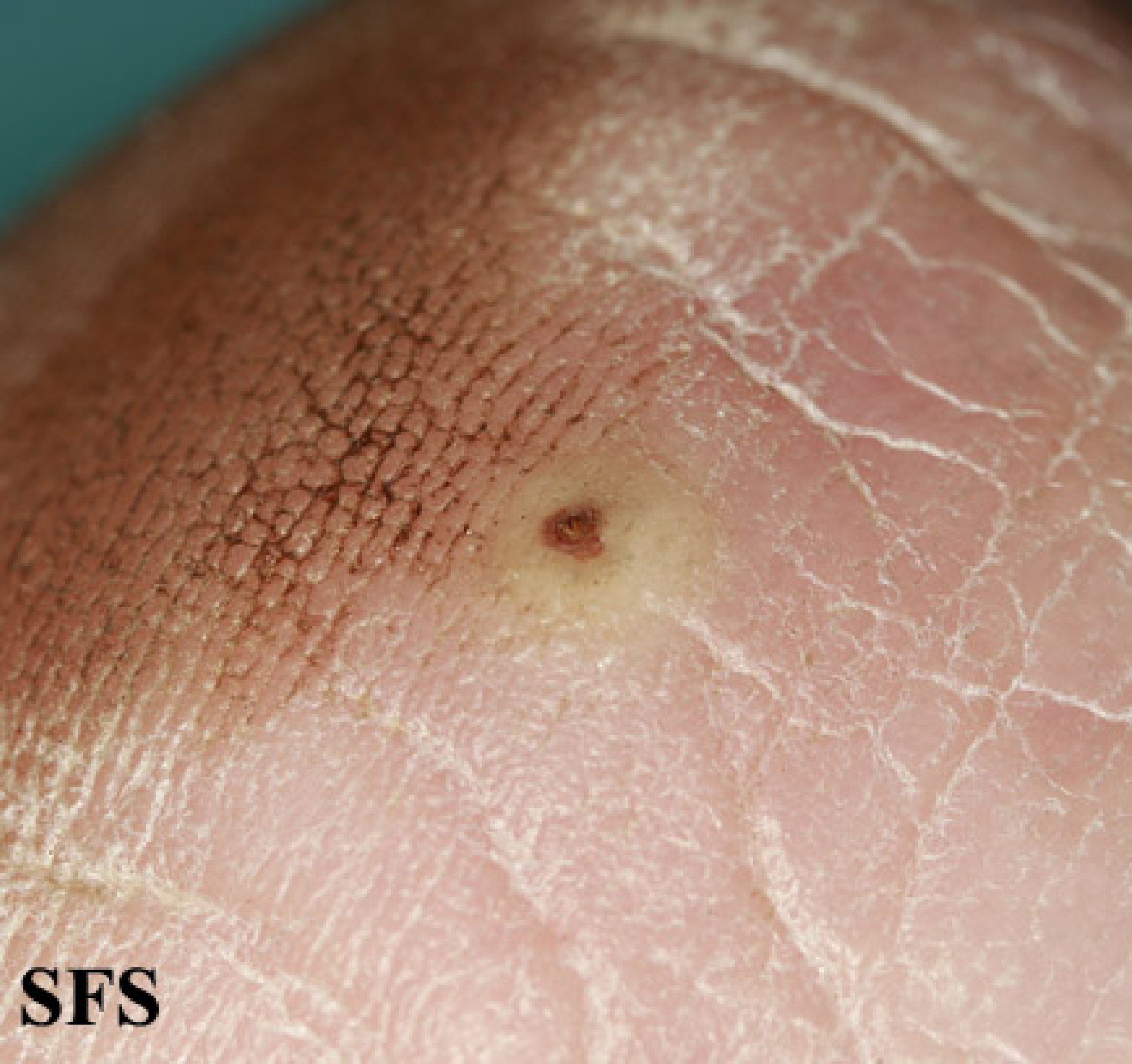}\\

    18. & Leprosy Tuberculoid 
    & \includegraphics[width=0.08\textwidth,height=0.03\textheight]{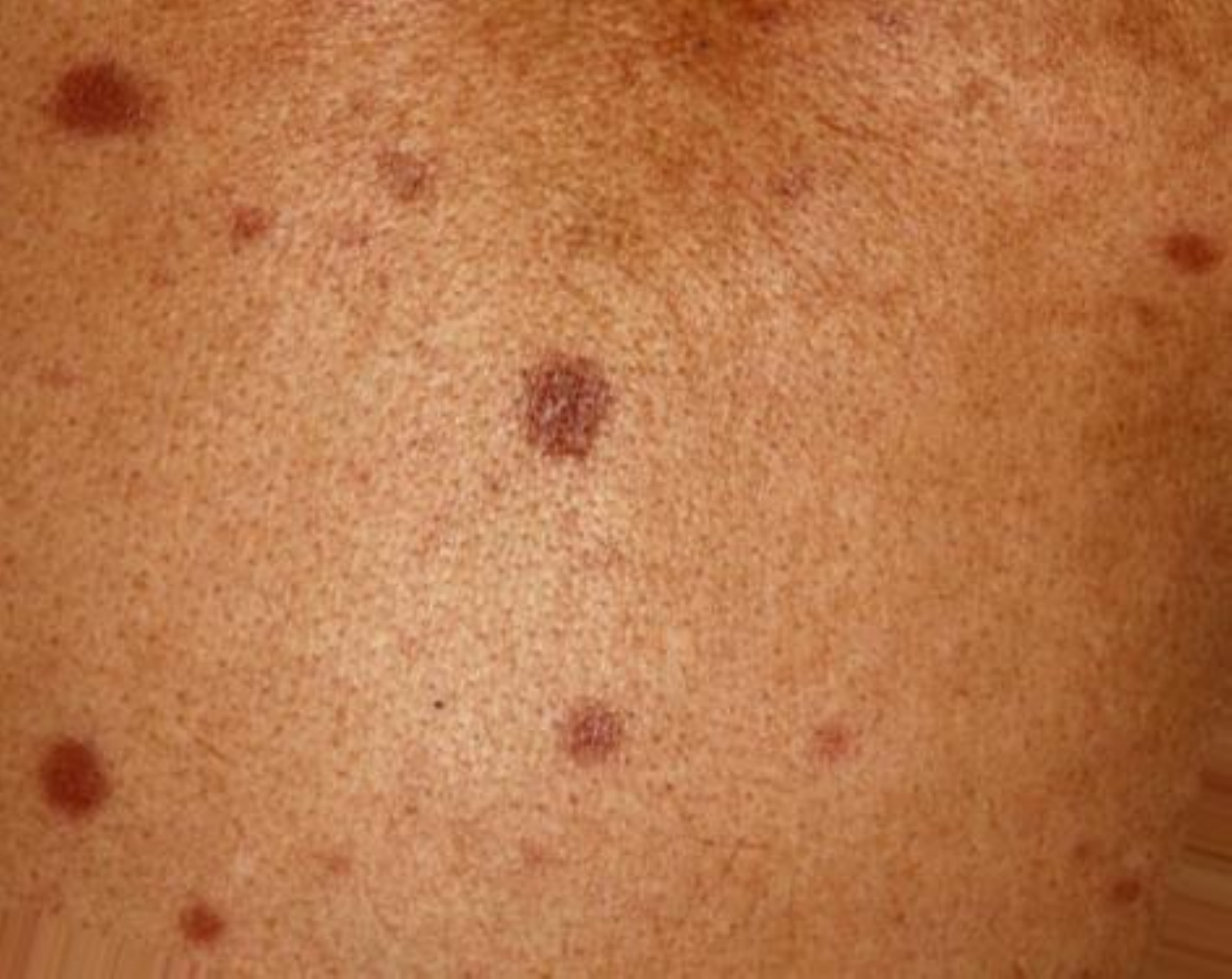}
    & 38. & Urticaria Hives   
    & \includegraphics[width=0.08\textwidth,height=0.03\textheight]{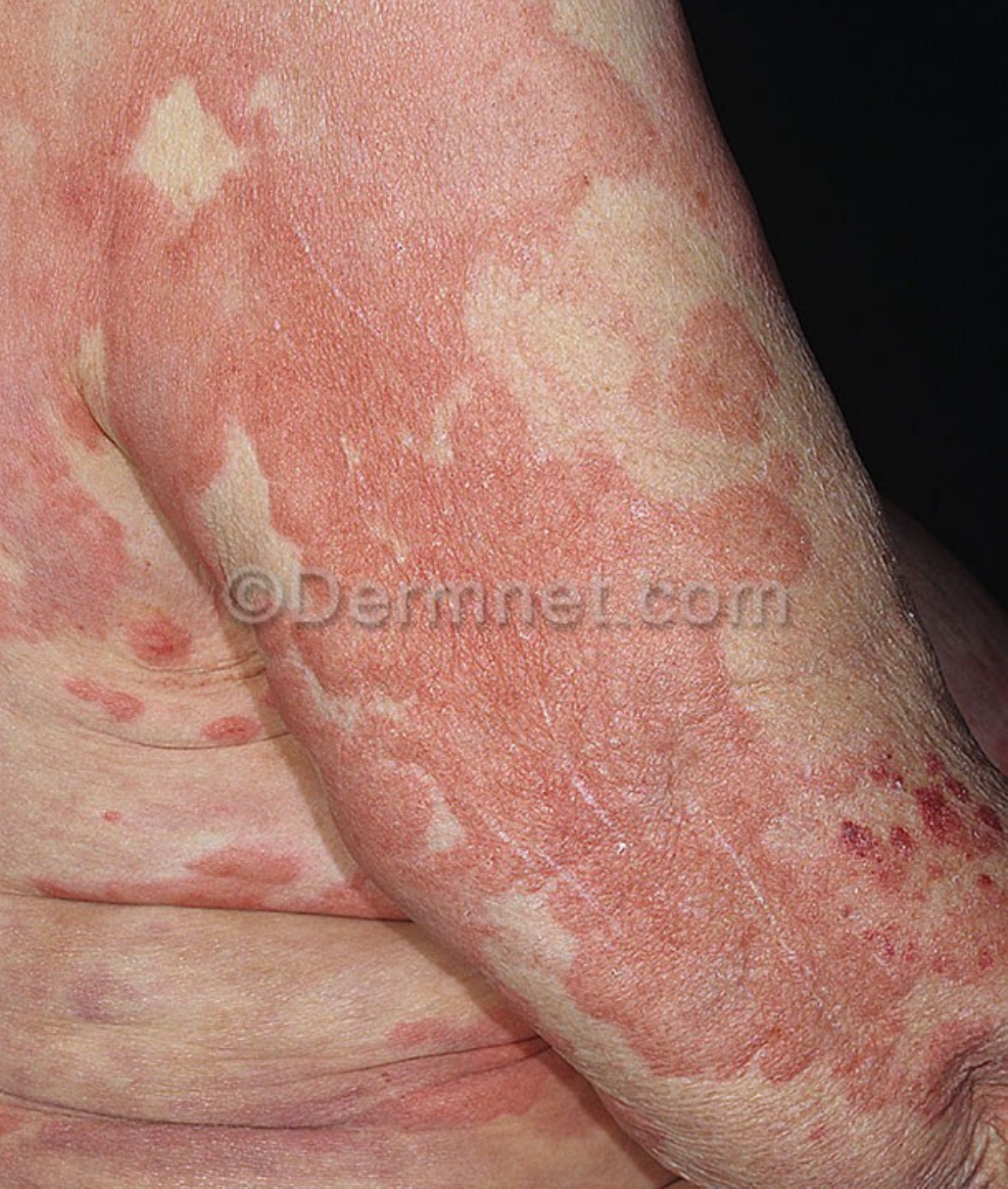}\\

    19. & Lichen Planus 
    & \includegraphics[width=0.08\textwidth,height=0.03\textheight]{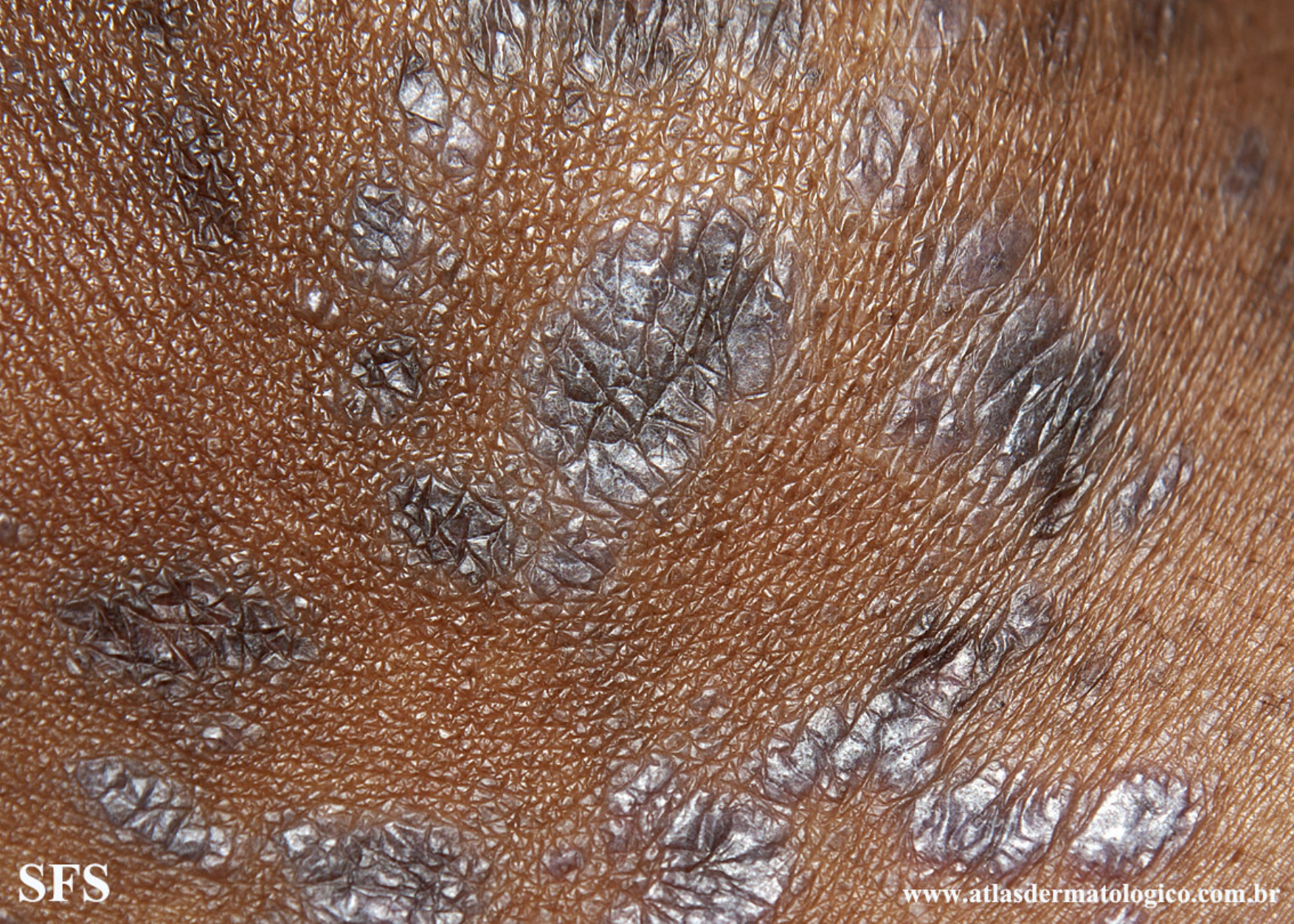}
    & 39. & Vasculitis  
    & \includegraphics[width=0.08\textwidth,height=0.03\textheight]{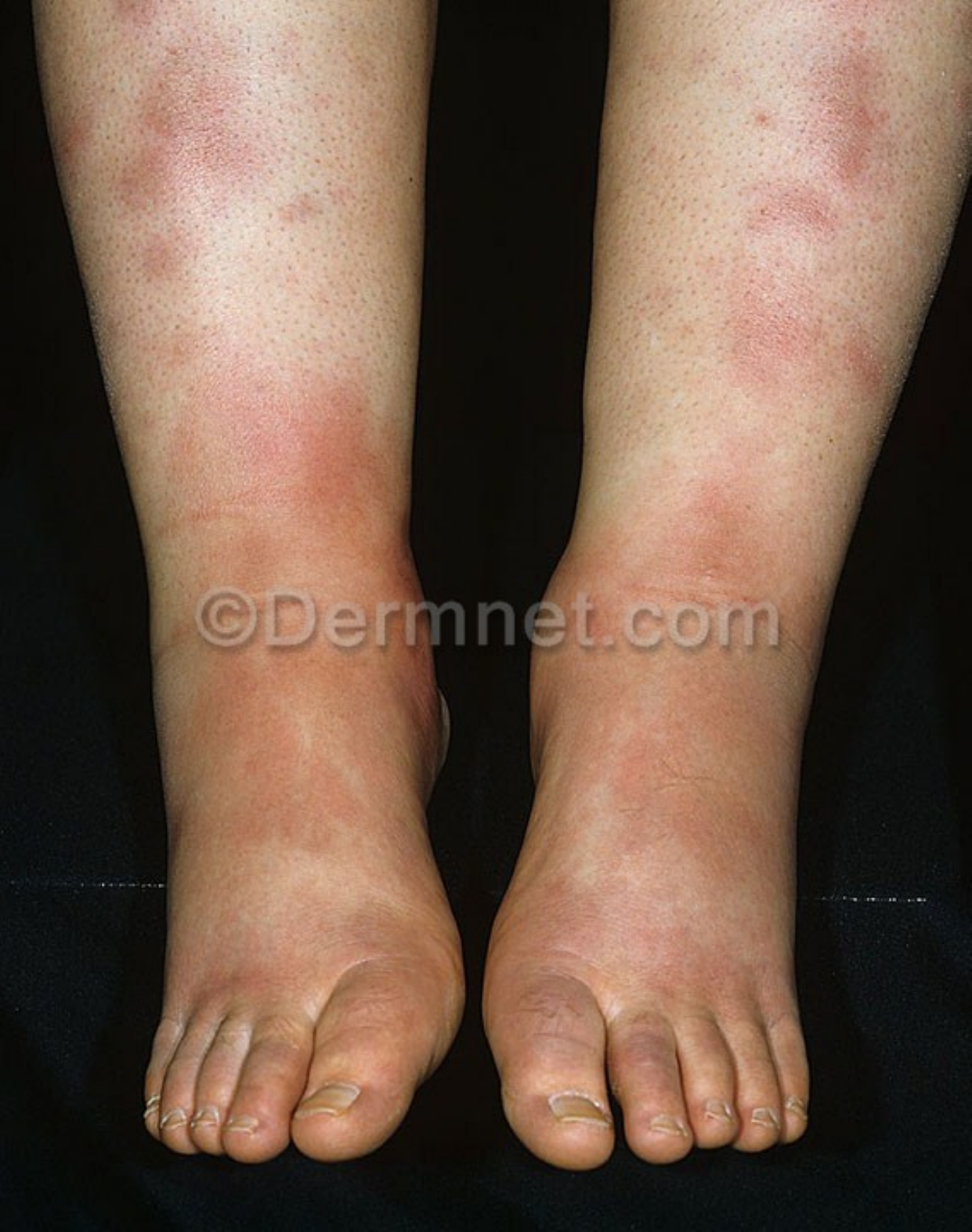}\\

    20. & Light Diseases 
    & \includegraphics[width=0.08\textwidth,height=0.03\textheight]{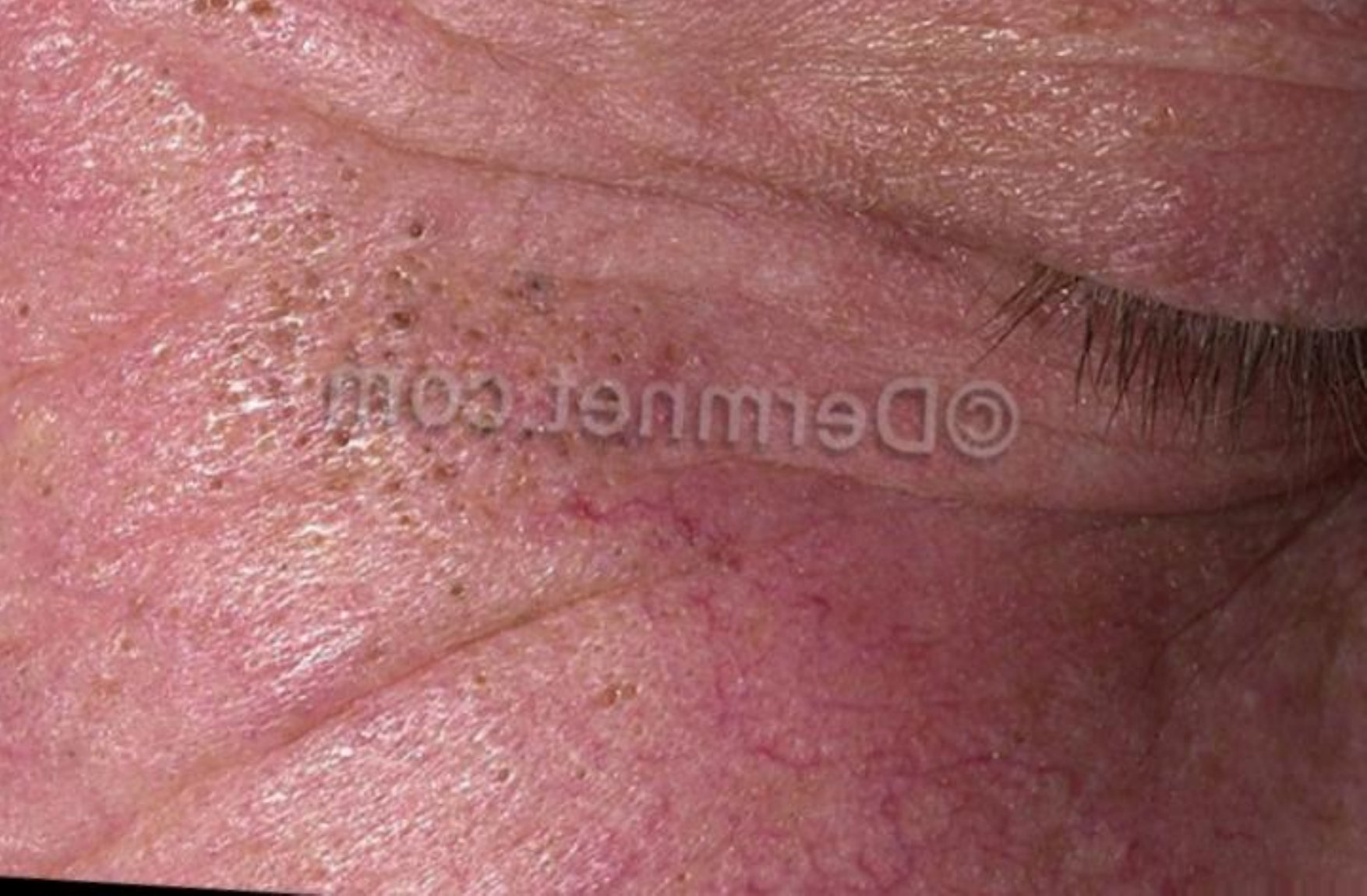} & & &\\
    
    \hline

    \hline
    \end{tabular}
\label{visu2}    
\end{table}  

The curated dataset offers several advantages over individual datasets. First, the integration of five datasets expands the diversity of skin lesion types, resulting in a collection of 39 distinct classes. This diversity is crucial for developing a model that is robust and capable of generalizing across a wide range of lesion types. Second, by standardizing and balancing the dataset, biases inherent in individual datasets are minimized, enabling fairer and more accurate evaluations of model performance. Third, the dataset encompasses lesions captured under varying conditions, including dermoscopic and clinical imaging, which enhances the model’s ability to handle real-world variations in input data.

To facilitate training, testing, and validation of the models, the curated dataset was split into three subsets: 70\% for training, 15\% for testing, and 15\% for validation. This resulted in 90 images per class for the training set and 20 images per class for both the testing and validation sets. This split ensures the model can generalize effectively across unseen data while providing sufficient data for both the training and evaluation phases.

\subsection{Dataset Preprocessing}
The initial curated dataset, while diverse, did not contain a sufficient number of images per class in the training set to enable effective model training and achieve robust classification accuracy. To address this limitation, data augmentation techniques were applied to increase the size of the training dataset. Augmentation not only enhances the dataset's size but also introduces variability, which helps models generalize better to unseen data by simulating real-world conditions.

The preprocessing techniques employed in this work include resizing, which ensures uniform image dimensions across the dataset, and normalization, which scales pixel values to a consistent range, facilitating faster and more stable model convergence. Additionally, geometric transformations such as height shift, width shift, rotation, zoom, shear, vertical flipping, and horizontal flipping were utilized. These transformations simulate variations in image orientation, scale, and perspective, thereby improving the model’s ability to handle diverse and complex inputs.

\subsection{Deep Learning Models}\label{model} 
In this study, five deep learning models have been used as baselines to evaluate their performance in classifying 39 types of skin lesions. A brief description of each model and its relevance to the study is provided below: 

\paragraph{MobileNetV2} It is a lightweight convolutional neural network (CNN) designed to balance model size and accuracy, making it particularly suitable for resource-constrained environments. As an enhancement of the original MobileNet model, MobileNetV2 introduces several key innovations, including depthwise separable convolutions, inverted residuals, and bottleneck designs. These architectural elements significantly reduce the number of parameters while maintaining the model’s ability to capture complex features. Furthermore, linear bottlenecks and squeeze-and-excitation (SE) blocks are employed to enhance feature extraction efficiency, enabling the model to process high-dimensional visual data with minimal computational overhead.

MobileNetV2 has proven effective across various computer vision tasks, such as image classification, object detection, and semantic segmentation, due to its ability to combine compactness and accuracy. In this study, MobileNetV2 was implemented using a learning rate of 0.001, a batch size of 8, and trained over 40 epochs. These hyperparameters were chosen to optimize the model’s performance on the curated skin lesion dataset while maintaining computational efficiency.

\paragraph{Xception} Xception, short for ``Extreme Inception," is a deep learning architecture that extends the principles of the Inception model by employing depthwise separable convolutions. This innovative design significantly reduces the number of parameters and computational requirements, while maintaining high accuracy in feature extraction and classification tasks. The architecture consists of 36 convolutional layers organized into a linear stack of depthwise separable convolutional layers. This design allows the model to effectively decouple spatial and channel-wise feature extraction, enabling more efficient processing of input data.

Xception's deep and efficient design has demonstrated excellent performance across a wide range of computer vision tasks, including image classification, object detection, and image segmentation. By leveraging depthwise separable convolutions, Xception achieves a balanced trade-off between accuracy and computational efficiency, making it suitable for applications requiring high performance with constrained resources.

In this study, the Xception model was implemented with a learning rate of 0.001, a batch size of 8, and trained over 8 epochs. These hyperparameters were selected to optimize the model's performance on the curated skin lesion dataset while ensuring computational feasibility. The use of Xception allows for effective extraction of intricate patterns in skin lesion images, contributing to robust classification results.

\paragraph{InceptionV3} It is a deep learning model that extends the Inception architecture, comprising 42 layers designed to improve both accuracy and computational efficiency. As a successor to earlier Inception models, InceptionV3 achieves a lower error rate while maintaining a streamlined architecture. Its design incorporates a variety of symmetric and asymmetric building blocks, such as convolutional layers, average pooling layers, max-pooling layers, concatenation layers, dropout layers, and fully connected layers. This diverse set of components enables the model to capture intricate patterns and details across a wide range of image features.

A key feature of InceptionV3 is the use of batch normalization on activation inputs, which stabilizes the training process and enhances convergence. This design also emphasizes efficiency by incorporating techniques that reduce the number of parameters and computations, making it suitable for resource-constrained environments. InceptionV3 has demonstrated strong performance in terms of accuracy, efficiency, and transfer learning capabilities, making it widely applicable across various image classification tasks.

In this study, the InceptionV3 model was implemented with a learning rate of 0.001, a batch size of 8, 8 epochs, and a dropout rate of 0.5. These hyperparameters were chosen to optimize performance on the curated skin lesion dataset while ensuring effective generalization. The model's ability to capture a broad range of image details contributes to its robustness in classifying diverse skin lesion types, further validating its suitability for this study..

\paragraph{EfficientNetB1} It is part of the EfficientNet family of models, which introduces a unique and systematic approach to scaling neural network architectures. Unlike traditional methods that scale depth (number of layers), width (number of channels), or resolution independently, EfficientNet employs a compound scaling method. This approach expands all three dimensions—depth, width, and resolution—simultaneously and proportionally, optimizing efficiency while maintaining high accuracy. By leveraging this innovative scaling strategy, EfficientNetB1 achieves exceptional performance without becoming computationally prohibitive.

EfficientNet offers several variants, ranging from B0 to B7, each with a different level of depth and width scaling. This flexibility allows users to select a model variant that best fits their resource constraints and accuracy requirements. EfficientNetB1, in particular, balances computational efficiency and accuracy, making it suitable for applications where resources are limited but high performance is essential.

The core strength of EfficientNetB1 lies in its ability to maintain a compelling balance between accuracy and efficiency, making it an adaptable and powerful architecture for image classification tasks. Its lightweight design, coupled with its ability to capture complex patterns in data, has made it widely applicable in domains requiring scalable and efficient deep learning solutions.

In this study, EfficientNetB1 was implemented with a learning rate of 0.001, a batch size of 8, and trained over 70 epochs. These hyperparameters were selected to optimize the model’s learning process and performance on the curated skin lesion dataset. The EfficientNetB1 architecture’s efficiency and scalability make it a valuable addition to the baseline models explored in this research.

\paragraph{Vision Transformer} The Vision Transformer (ViT) introduces a paradigm shift in image processing by representing images as sequences, akin to natural language processing (NLP) tasks. Unlike convolutional neural networks (CNNs), which rely on spatial hierarchies and convolutional operations, the Vision Transformer divides images into patches and treats them as sequential data. Each image is split into smaller patches, flattened, and linearly projected into a vector space, enabling the model to learn image structure autonomously. The ViT architecture builds on the foundation of transformer models, leveraging their ability to process sequential data effectively. Below is a step-by-step breakdown of its architecture:
\begin{enumerate}[i.]
    \item Image splitting into patches: The input image is divided into smaller, fixed-size patches, each representing a specific region of the image. This step transforms the 2D spatial structure of the image into discrete, independent patches that the model can process sequentially.
    \item Flattening the patches: Each patch's pixel values are flattened into a single vector by concatenating the channels of all pixels within the patch. This transformation allows the model to treat image patches as sequential data, analogous to words in a sentence for NLP tasks.
    \item Producing lower-dimensional linear embeddings: Flattened patch vectors are projected into a lower-dimensional space using trainable linear transformations. This step reduces the dimensionality of the data while preserving key features, ensuring computational efficiency and facilitating feature extraction. 
    \item Adding positional encodings: Positional encodings are added to the patch embeddings to retain information about the spatial arrangement of the patches. These encodings enable the model to understand the relative positions of patches, a crucial aspect for maintaining the spatial coherence of images.
    \item Feeding the sequence into a transformer encoder: The sequence of patch embeddings, enriched with positional encodings, is passed through a transformer encoder. The encoder consists of multiple layers, each featuring multi-head self-attention mechanisms (MSPs) and multi-layer perceptron (MLP) blocks. MSPs compute attention weights to prioritize relevant elements in the input sequence, enabling the model to focus on important features during classification. MLP blocks are fully connected layers that further process the sequence, transforming the learned features into more abstract representations. Each layer incorporates layer normalization before the attention and MLP blocks to ensure stable and efficient training.
    \item Final classification head: The output of the transformer encoder is passed to a classification head, consisting of a Gaussian Error Linear Unit (GELU) activation function and a final MLP block. The output of the MLP is processed through a softmax function to generate class probabilities for the input image.
\end{enumerate}
By relying on this step-by-step architecture, Vision Transformer eliminates the need for convolutional operations while leveraging the power of self-attention to model global dependencies in the input. This makes ViT particularly effective for image classification tasks, where understanding the relationships between distant image regions is essential.

In this study, the Vision Transformer is employed to classify 39 types of skin lesions. Its capability to learn high-dimensional, global features from images makes it a suitable choice for capturing the intricate and diverse patterns present in the curated dataset.

\subsection{Integrated Attention Module}\label{attention} 
Attention mechanisms play a critical role in enhancing deep learning models by enabling them to focus on the most relevant regions of input data. In the context of image classification, attention modules help in extracting features more precisely by emphasizing significant portions of images while ignoring less relevant details. To leverage this advantage, all five baseline models employed in this study are interrogated with well-known attention modules. A detailed discussion of these modules is provided below.

\subsubsection{Efficient Channel Attention (ECA)}
ECA is an architectural unit based on the squeeze-and-excitation (SE) block, designed to enhance feature representation with minimal computational overhead. Unlike traditional SE blocks, ECA avoids dimensionality reduction, maintaining the richness of feature maps while reducing model complexity. This is achieved by capturing local cross-channel interactions, which focus on the dependencies between different feature channels.

The ECA mechanism operates by performing channel-wise global average pooling on the feature maps, followed by a fast 1D convolution of size $k$. This convolution efficiently captures the interaction between each channel and its $k$-nearest neighbors. The kernel size $k$ determines the range of local interactions and is adaptively chosen to ensure an optimal balance between efficiency and performance \citep{wang2020eca}. By using this method, ECA ensures that attention prediction for each channel incorporates information from its neighboring channels without increasing the dimensionality of the feature maps.

\begin{figure}[!ht] 
\centering
\includegraphics[width=0.9\textwidth]{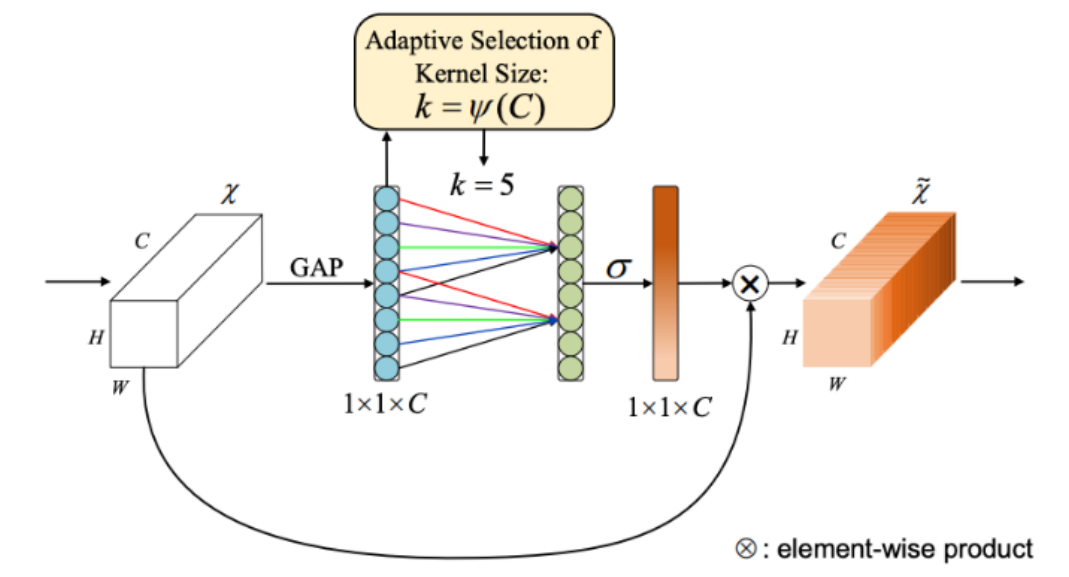} 
\caption{Efficient channel attention module \citep{wang2020eca}} 
\label{ecaa}
\end{figure}

The strength of ECA lies in its ability to enhance feature selectivity, allowing the model to focus on critical image details while maintaining computational efficiency. Its lightweight design makes it particularly suitable for integration into complex architectures, where computational constraints are a concern. Figure \ref{ecaa} illustrates the architecture of the Efficient Channel Attention module and its integration into the baseline models.

\subsubsection{Convolutional Block Attention Module (CBAM)} 
\begin{figure}[!ht]
\centering
\includegraphics[width=.9\textwidth]{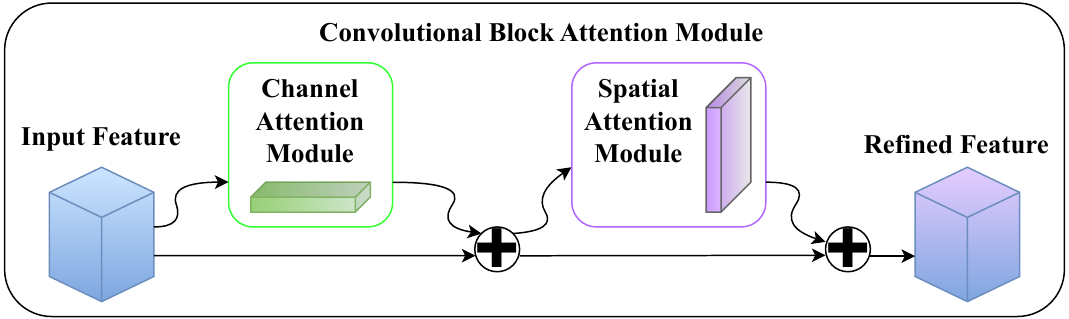} 
\caption{Convolutional block attention module} 
\label{cbamm}
\end{figure}

CBAM is a lightweight yet powerful attention mechanism designed to enhance feature extraction in deep learning models \citep{woo2018cbam}. CBAM combines channel attention and spatial attention modules in a sequential manner to refine feature representations adaptively, as illustrated in Figure \ref{cbamm}. By applying attention along both the channel and spatial dimensions, CBAM effectively focuses on the most relevant parts of an image while suppressing irrelevant information. Given an intermediate feature map $\mathbf{F} \in \mathbb{R}^{C \times H \times W}$ given as input, CBAM computes 1D channel attention map $\mathbf{M}_c \in \mathbb{R}^{C \times 1 \times 1}$ and a 2D spatial attention map $\mathbf{M}_s \in \mathbb{R}^{1 \times H \times W}$. The refined output $\mathbf{F}''$ is computed as follows:
\begin{align}
\mathbf{F}' &= \mathbf{M}_c(\mathbf{F}) \otimes \mathbf{F},  \\
\mathbf{F}'' &= \mathbf{M}_s(\mathbf{F}') \otimes \mathbf{F}'
\end{align}
where $\otimes$ denotes element-wise multiplication.

\begin{figure}[!ht] 
\centering
\includegraphics[width=.9\textwidth]{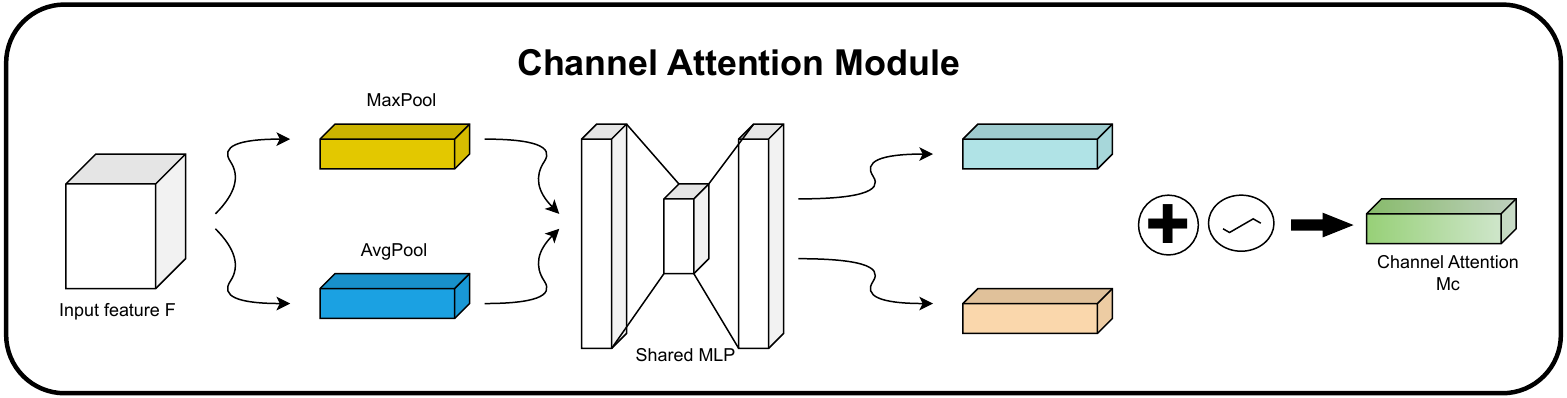} 
\caption{Channel attention module, redrawn based on } 
\label{channel}
\end{figure}

\paragraph{Channel Attention Module}
The channel attention module, illustrated in Figure \ref{channel}, emphasizes important feature channels by exploiting inter-channel dependencies~\citep{woo2018cbam}. This module determines ``what" is meaningful in an image. By squeezing the spatial dimensions, channel attention aggregates information across the input feature map's spatial domain. Both average-pooling and max-pooling operations are used to generate global descriptors, which are then passed through a shared network consisting of a multi-layer perceptron (MLP) with one hidden layer. The outputs of these two pooling operations are merged to produce the final channel attention map:

\begin{equation}
\mathbf{M}_c(\mathbf{F}) = \sigma \left( \text{MLP}(\text{AvgPool}(\mathbf{F})) + \text{MLP}(\text{MaxPool}(\mathbf{F})) \right)
= \sigma \left( \mathbf{W}_1 \left( \mathbf{W}_0 (\mathbf{F}^{c}_{\text{avg}}) \right) + \mathbf{W}_1 \left( \mathbf{W}_0 (\mathbf{F}^{c}_{\text{max}}) \right) \right),
\end{equation}
where $\mathbf{W}_0$ and $\mathbf{W}_1$ are the weights of the MLP layers.

\begin{figure}[!ht] 
\centering
\includegraphics[width=0.7\textwidth]{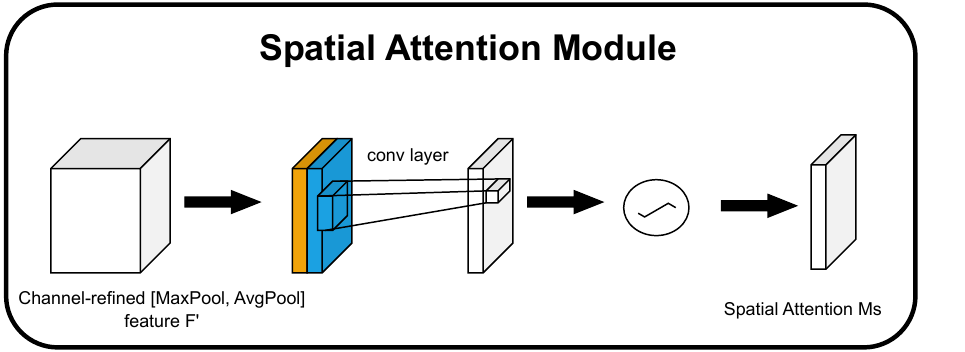} 
\caption{Spatial Attention Module, redrawn based on \citep{woo2018cbam}}  
\label{spatial}
\end{figure}

\paragraph{Spatial Attention Module} 
The spatial attention module, illustrated in Figure \ref{spatial}, focuses on ``where" the informative regions are located within an image. By analyzing inter-spatial relationships, the spatial attention module emphasizes critical regions while suppressing less relevant ones. Spatial attention is computed by applying average-pooling and max-pooling along the channel axis, generating two 2D feature maps. These maps are concatenated and passed through a convolutional layer with a $7\times7$ filter to produce the final spatial attention map:
\begin{equation}
F_{avg}^{s} = AvgPool(F) \\
\end{equation}
\begin{equation}
F_{max}^{s} = MaxPool(F) \\
\end{equation}
\begin{equation}
M_c(F) = \sigma(\text{f}^{7\times7}([F_{avg}^{s}; F_{max}^{s}])) \\
\end{equation}
where $\begin{aligned} \sigma \end{aligned}$ denotes the sigmoid function and $\begin{aligned} \text{f}^{7\times7} \end{aligned} $ represents a convolutional operation with a $7\times7$ filter.

\subsection{Vision Transformer with Convolutional Block Attention Module} 
In the proposed approach, the convolutional block attention module (CBAM) is integrated with a vision transformer (ViT) to enhance its feature extraction capabilities. Vision Transformers inherently excel at capturing global contextual information by processing images as sequences of patches, with each patch contributing to an embedded output set that encapsulates the global dependencies of the image. However, transformers can sometimes lack the ability to precisely capture localized, fine-grained details, such as the boundaries of specific regions, which are crucial for accurate classification.

By integrating CBAM into the vision transformer architecture, these shortcomings are addressed. After the vision transformer processes the input image, the global feature representations obtained from the transformer are passed to the channel attention module of CBAM. This module weights the feature maps produced by the vision transformer to emphasize the most informative and relevant features while suppressing less significant ones. The refined features from the channel attention module are then passed to the spatial attention module of CBAM. This module identifies specific spatial regions of the image that are most pertinent, ensuring that the model focuses on areas with the highest diagnostic value.

Through this integration, the vision transformer's global contextual understanding is combined with CBAM's ability to refine features both globally and locally. This synergy enhances the model's ability to extract highly informative features, resulting in more precise classification. The combined architecture not only improves the detection of subtle details within an image but also mitigates the limitations of the Vision Transformer in accurately identifying boundaries of altered or pathological regions~\citep{rs15092406}.

\begin{figure}[!ht] 
\centering
\includegraphics[width=1\textwidth]{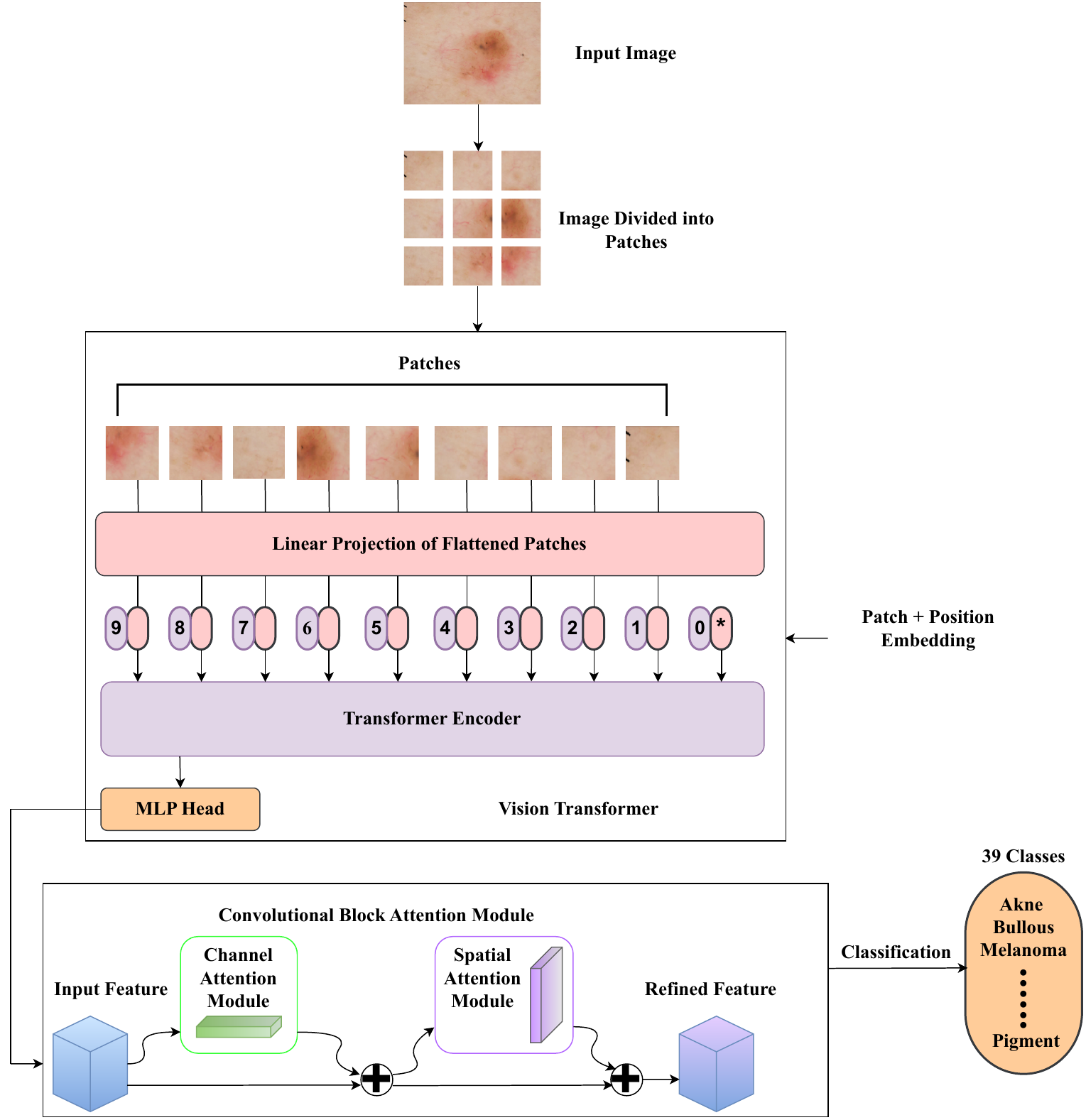} 
\caption{Vision transformer with CBAM} 
\label{vitwithcbam}
\end{figure}

Figure \ref{vitwithcbam} illustrates the architecture of the vision transformer with the CBAM attention module for the proposed approach. This hybrid design enables the model to capture features at multiple levels of granularity, leading to improved performance on complex image classification tasks. By leveraging this integration, the proposed approach becomes more efficient and accurate in identifying relevant patterns and structures in the input data, particularly for the challenging task of skin lesion classification.
\label{prop_app}

\section{Experimental Evaluation}\label{reslt} 
\subsection{Experiment Setup} 
The experiments conducted in this study were implemented using the Python programming language and executed on Kaggle's notebook environment. Kaggle provides access to powerful computational resources, including an NVIDIA Tesla P100 GPU and 16 GB of RAM, which were utilized to accelerate model training and evaluation. These resources were instrumental in handling the computationally intensive tasks associated with training deep learning models on the curated dataset.

The dataset was sourced from Kaggle and prepared for experimentation using Google Drive, which facilitated seamless data storage and accessibility. Additionally, parts of the preprocessing and experimentation phases were performed locally on an HP Core i5 laptop, ensuring flexibility and continuity of the workflow across multiple platforms. This hybrid setup allowed the experiments to leverage both cloud-based and local computational resources, providing a practical and cost-effective solution for developing and testing the proposed methodologies. 

\subsection{Evaluation Metrics} 
To evaluate the performance of the models developed in this study, several evaluation metrics were employed, providing a comprehensive analysis of their effectiveness and enabling meaningful comparisons. The foundation for these metrics is the confusion matrix, which categorizes predictions into four components: True Positive (TP), True Negative (TN), False Positive (FP), and False Negative (FN). These components are instrumental in calculating various performance metrics, such as Accuracy, Precision, Recall, F1-Score, and Specificity, each offering unique insights into the strengths and limitations of the models. Definitions of those evaluation metrics are depicted below:
\begin{equation}
\text{Accuracy} = \frac{TP + TN}{TP + TN + FP + FN}
\end{equation} 
\begin{equation}
\text{Precision} = \frac{TP}{TP + FP}
\end{equation}
\begin{equation}
\text{Recall} = \frac{TP}{TP + FN}
\end{equation}
\begin{equation}
\text{F1-Score} = \frac{2 * Precision * Recall}{Precision + Recall}
\end{equation}
\begin{equation}
\text{Specificity} = \frac{TN}{TN + FP} 
\end{equation} 

\subsection{Performance Evaluation and Analysis}
In this study, a range of pre-trained deep learning models -- Xception, InceptionV3, EfficientNetB1, MobileNetV2, and Vision Transformer (ViT) -- were evaluated on the curated dataset to determine the most effective model among the baseline methods. The baseline performance of these models is presented in Table \ref{table:deep}. Among them, the ViT achieved the highest accuracy of 91.79\%, demonstrating superior performance compared to other models in the baseline setup. 

\begin{table}[!ht]
    \centering
    \caption{Performance comparison with baseline methods}
    \label{table:deep}
    \begin{tabular}{lll ccccc}
    \hline
      \multicolumn{3}{c}{Method} & Accuracy (\%) & Precision (\%) & Recall (\%) & F1-Score (\%) & Specificity (\%)\\
        \hline\hline
    \multicolumn{2}{c}{\multirow{5}{*}{\rot{Baseline}}} & MobileNetV2 & 80.03 & 84 & 80 & 80 & 83.85\\
   
    &&EfficientNetB1 & 82.44 & 81 & 81 & 80 & 81\\
    
    &&InceptionV3 & 83.85 & 85 & 84 & 84 & 85.18\\
    
    &&Xception & 87.18 & 88 & 87 & 87 & 87.97\\
    
    &&Vision Transformer (ViT) & \textbf{91.79} & \textbf{92} & \textbf{92} & \textbf{92} & \textbf{91.86}\\
    \hline
    \multirow{10}{*}{\rot{Attention guided (this study)}} & \multirow{5}{*}{\rot{Baseline+ECA}} &
     MobileNetV2 + ECA & 85.13 & 88 & 85 & 85 & 87.93\\
    
    &&EfficientNetB1 + ECA & 90.38 & 91 & 90 & 90 & 90.93\\
    
    &&InceptionV3 + ECA & 87.69 & 89 & 88 & 88 & 88.52\\
    
    &&Xception + ECA & 87.69 & 88 & 88 & 87 & 88.24 \\
    
    &&ViT + ECA & \textbf{92.31} & \textbf{92} & \textbf{92} & \textbf{92} & \textbf{92.39}\\
    \cline{2-8}
    & \multirow{5}{*}{\rot{Baseline+CBAM}} &
    MobileNetV2 + CBAM & 86.79 & 88 & 87 & 87 & 87.96\\
    
    &&EfficientNetB1 + CBAM & 90.51 & 91 & 91 & 90 & 90.89\\
    
    &&InceptionV3 + CBAM & 86.92 & 88 & 87 & 87 & 88.33\\
    
    &&Xception + CBAM & 88.85 & 90 & 89 & 89 & 89.61 \\
    
    &&ViT + CBAM (Proposed) & \textbf{93.46} & \textbf{94} & \textbf{93} & \textbf{93} & \textbf{93.67}\\
    \hline

    \hline
    \end{tabular}
\end{table}

The table also illustrates the performance of the models when the efficient channel attention (ECA) module is integrated. The integration of ECA led to improved performance across all models. Notably, the ViT with ECA achieved an accuracy of 92.31\%, outperforming other ECA-augmented models. EfficientNetB1 with ECA also delivered impressive results, achieving an accuracy of 90.38\%, indicating that attention-guided mechanisms significantly enhance the baseline models' performance.

Furthermore, the table shows the results after integrating the convolutional block attention module (CBAM) into the models. Similar to the ECA results, the ViT and EfficientNetB1 stood out, with both models achieving accuracies above 90\%. Among all configurations, the ViT with CBAM achieved the highest accuracy of 93.46\%. This result highlights the effectiveness of the CBAM module in refining global and local features, which is particularly beneficial for handling the diverse and complex features of skin lesion images.

The comprehensive analysis presented in Table \ref{table:deep} demonstrates that the ViT integrated with CBAM (proposed approach) outperforms all other models and configurations. The integration of CBAM allows for the extraction of highly refined features, improving the model's ability to capture intricate patterns and dependencies in the dataset. 

These findings underscore the significance of using attention-guided mechanisms to enhance feature extraction in deep learning models, particularly when addressing complex multiclass classification tasks. The Vision Transformer with CBAM represents the most effective configuration, achieving the highest metrics across all evaluation parameters, including accuracy, precision, recall, F1-score, and specificity. 

\section{Discussion}\label{sec:discussion}
\subsection{Misclassification Analysis}
\begin{figure}[!ht]
\centering
\includegraphics[width=1\textwidth]{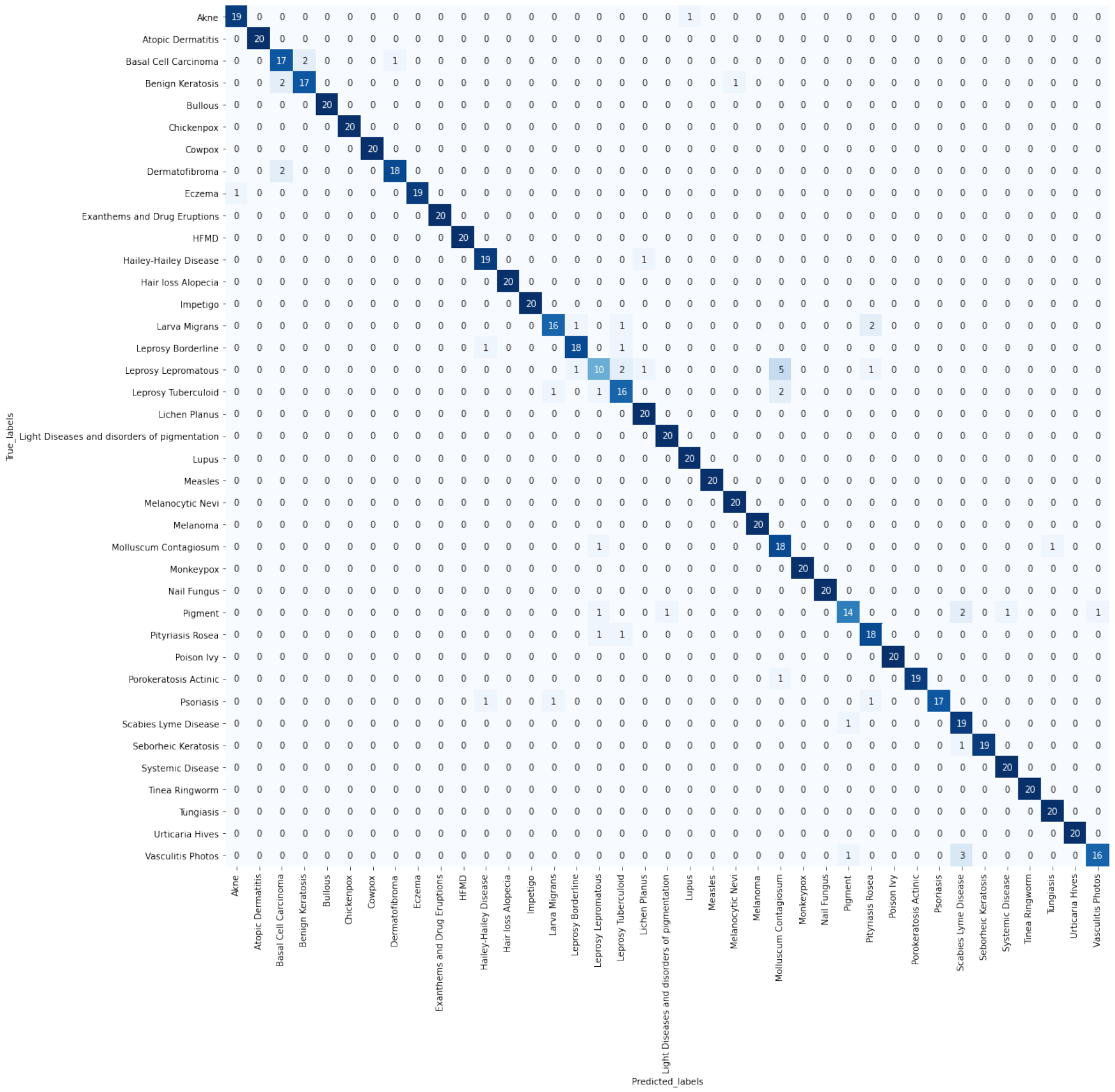}  
\caption{Confusion Matrix of proposed model} 
\label{confuse}
\end{figure}

\begin{figure}[!ht]
    \centering
    \includegraphics[width=.6\textwidth]{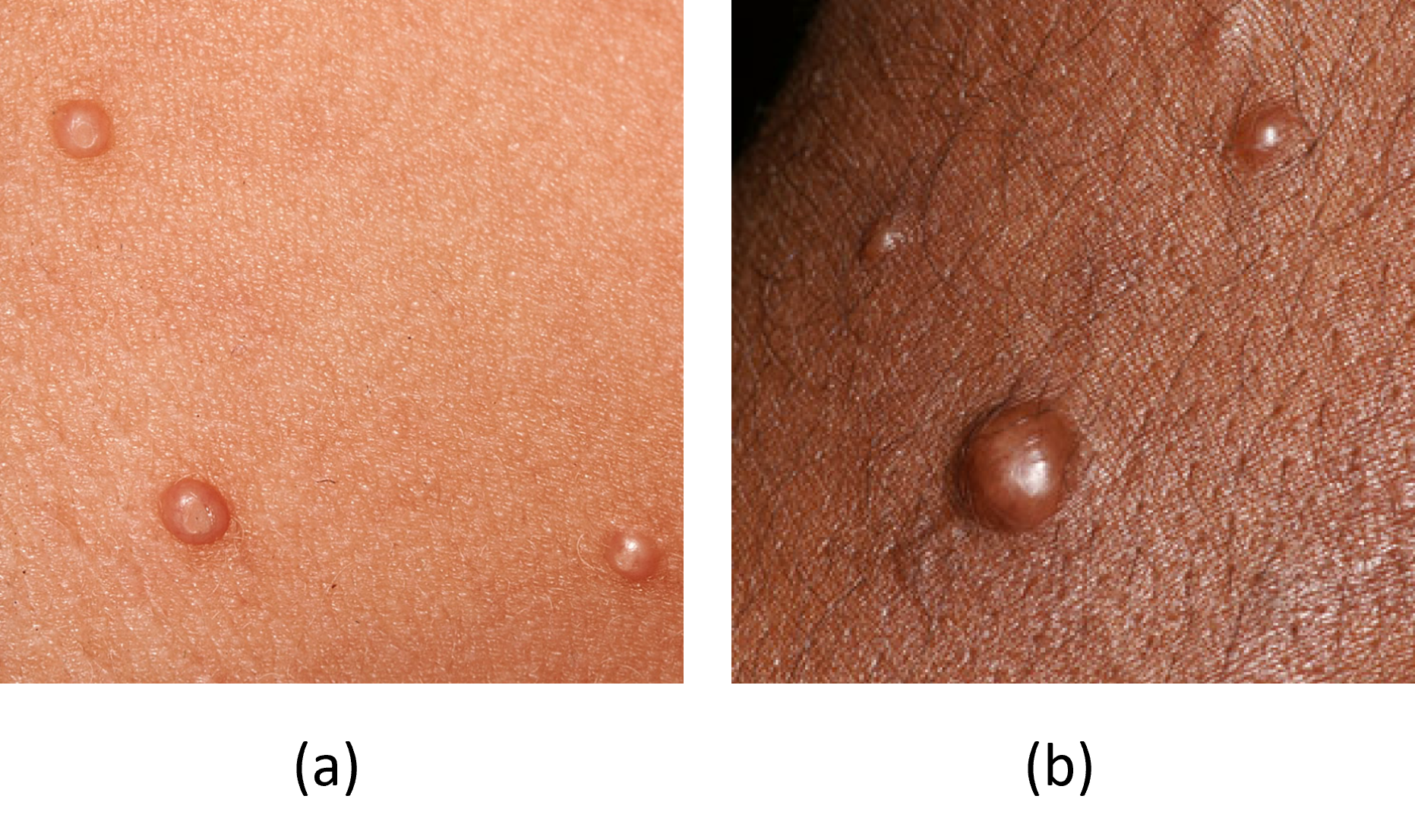}
    \caption{Visual similarity between (a) Molluscum Contagiosum and (b) Leprosy Lepromatous.} 
    \label{lepro}
\end{figure}

The class-wise performance analysis presented in Table \ref{table:class_wise} highlights the effectiveness of the Vision Transformer with CBAM attention module, achieving high accuracy across most classes, with notable successes such as 100\% accuracy in distinguishing 18 classes -- Atopic Dermatitis, Bullous, Chickenpox, Cowpox, Exanthems, Hand Foot Mouth Disease, Hair loss Alopecia, Impetigo, Lichen Planus, Light Diseases, Lupus, Measles, Melanocytic Nevi, Melanoma, Poison Ivy, Systemic Disease, Tungiasis, and Urticaria Hives.
\begin{table}[!htb]
    \centering
    \caption{Performance of all classes using the proposed method (ViT+CBAM)}
    \begin{tabular}{clc||clc} 
    \hline
    No. & Class name & Accuracy (\%) & No. & Class Name & Accuracy (\%)\\
    \hline
    \hline
    1. & Akne & 95 & 21. & Lupus & 100\\
    
    2. & Atopic Dermatitis & 100 & 22. & Measles & 100\\
    
    3. & Basal Cell Carcinoma & 85 & 23. & Melanocytic Nevi & 100\\
    
    4. & Benign Keratosis & 85 & 24. & Melanoma & 100\\
    
    5. & Bullous & 100 & 25. & Molluscum Contagiosum & 90\\

    6. & Chickenpox & 100 & 26. & Monkeypox & 90\\

    7. & Cowpox & 100 & 27. & Nail Fungus & 93\\
    
    8. & Dermatofibroma & 90 & 28. & Pigment & 70\\
    
    9. & Eczema & 95 & 29. & Pityriasis Rosea & 90\\
    
    10. & Exanthems & 100 & 30. & Poison Ivy & 100\\
    
    11. & Hand Foot Mouth Disease & 100 & 31. & Porokeratosis Actinic & 95\\
    
    12. & Hailey-Hailey Disease & 95 & 32. & Psoriasis & 85\\
    
    13. & Hair loss Alopecia & 100 & 33. & Scabies Lyme Disease & 95\\
    
    14. & Impetigo & 100 & 34. & Seborheic Keratosis & 95\\
    
    15. & Leprosy Borderline & 90 & 35. & Systemic Disease & 100\\
    
    16. & Leprosy Lepromatous & 50 & 36. & Tinea Ringworm & 93\\
    
    17. & Lerva Migrans & 80 & 37. & Tungiasis & 100\\
    
    18. & Leprosy Tuberculoid & 80 & 38. & Urticaria Hives & 100\\
    
    19. & Lichen Planus & 100 & 39. & Vasculitis & 80\\
    
    20. & Light Diseases & 100 &  &  &\\
    \hline
    
    \hline
    \end{tabular}
\label{table:class_wise}    
\end{table}
However, notable exceptions, particularly the 50\% accuracy for Leprosy Lepromatous, reveal areas where the model struggles to distinguish between visually similar conditions. The confusion matrix in Figure~\ref{confuse} highlights this misclassification issue, showing that Leprosy Lepromatous is frequently misclassified as Molluscum Contagiosum, contributing to its low class-wise accuracy of 50\%. 
This misclassification can be attributed to the striking visual similarity between the two conditions, illustrated in Figure~\ref{lepro}, both of which exhibit nodular lesions that are challenging to distinguish based solely on image features. Such challenges are common in dermatological image classification, where subtle morphological differences often require additional contextual or clinical information for accurate diagnosis.

One possible contributing factor to this misclassification is the dataset composition. While the dataset had a balanced representation across most classes, it is likely that the diversity of training samples for Leprosy Lepromatous is insufficient to capture its unique features. This lack of diversity may hinder the model’s ability to generalize effectively, especially when confronted with conditions that share overlapping visual characteristics. Furthermore, while the CBAM module enhances feature refinement by focusing on spatial and channel-level details, it may still struggle to adequately differentiate between subtle textural and morphological patterns inherent to these conditions.

\begin{figure}[!ht] 
\centering
\includegraphics[width=1\textwidth]{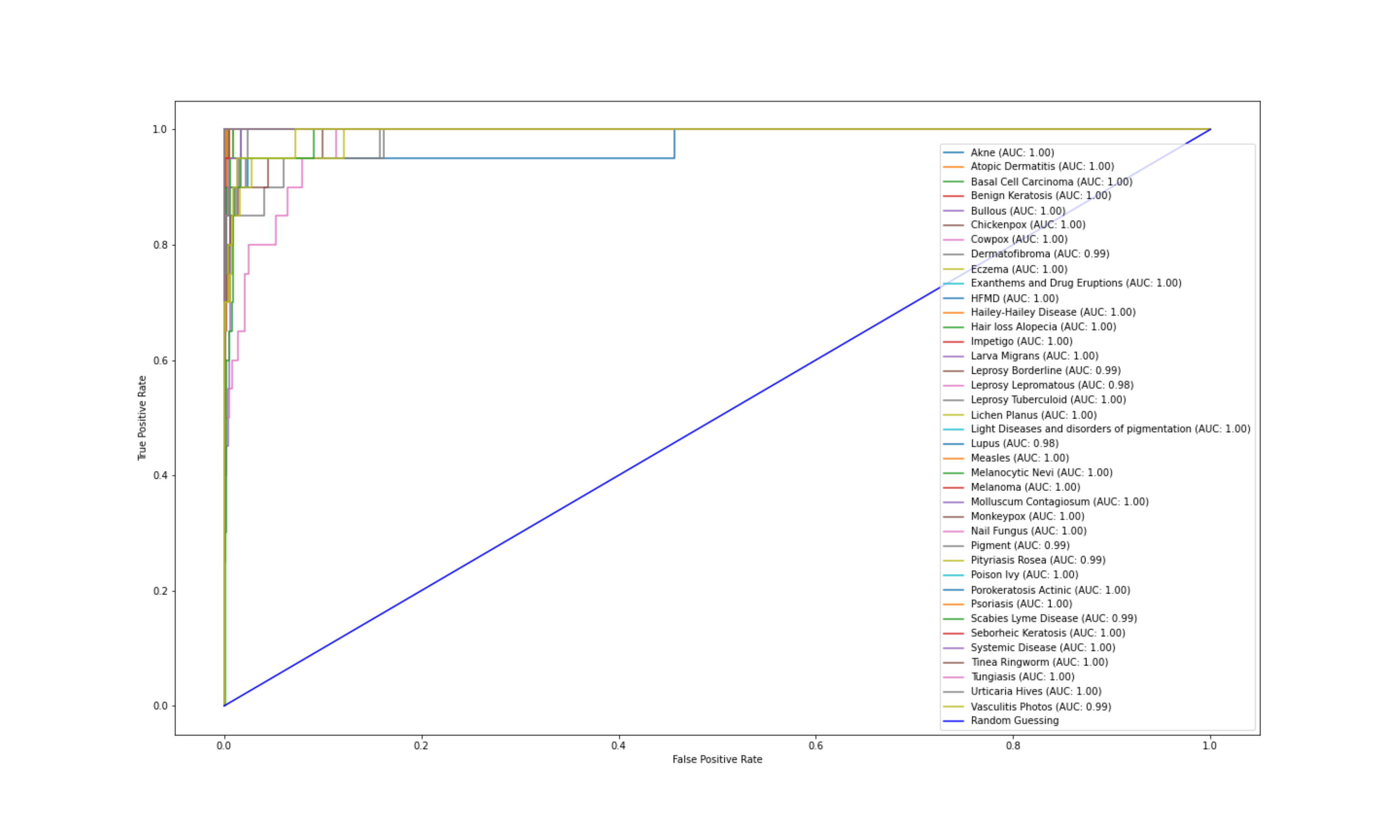}  
\caption{ROC curves for 39 classes} 
\label{roc} 
\end{figure}

Despite these challenges, the model's high specificity, as evidenced by the receiver operating characteristic (ROC) Curve in Figure~\ref{roc}, highlights its robustness in accurately distinguishing negative cases across the majority of classes. The overall area under the curve (AUC) score of 0.99 further underscores the model’s exceptional performance, reflecting its effectiveness in handling complex multi-class classification tasks. This performance reflects the efficacy of the attention-guided vision transformer architecture in capturing both global and local contextual features of skin lesions. However, the misclassification of visually similar conditions underscores the importance of integrating multimodal data, including clinical and histopathological inputs, to enhance the robustness and clinical applicability of such models.

\subsection{Limitation and Future Work} \label{limit}
While this study has made significant progress in multi-class skin lesion classification, several limitations remain, highlighting opportunities for future improvements. One key challenge encountered during the research was the misclassification of visually similar conditions, such as Leprosy Lepromatous and Molluscum Contagiosum. This issue underscores the limitations of relying solely on image-based features for classification. Similar findings have been reported in studies where overlapping visual characteristics posed challenges for accurate classification \citep{woo2018cbam, wang2020eca}. To address this, future work should explore incorporating additional clinical parameters, such as Bacterial Index (BI) values, which could serve as decisive features for distinguishing these conditions. For example, a BI threshold of 6+ could provide a critical differentiator, reducing the dependence on visual cues alone \citep{van2017field}. Furthermore, integrating histopathological data into the classification pipeline could significantly enhance the model’s ability to identify subtle patterns that are otherwise indistinguishable in dermoscopic images \citep{alshahrani2024analysis}.

Another notable limitation of this study was the lack of sufficient and diverse images for certain classes, which likely constrained the model’s generalization capabilities. Expanding the dataset with more diverse and representative samples is crucial. Collaborations with medical professionals to collect and annotate patient data can help address this limitation. Similar strategies have been employed in previous studies to improve dataset representation and model performance \citep{tschandl2018ham10000, combalia2019bcn20000}. Additionally, synthetic data generation techniques, such as Generative Adversarial Networks (GANs), have been shown to effectively augment datasets and introduce greater variability in lesion appearance \citep{mekala2024synthetic}.

While this study explored several baseline and attention-guided deep learning models, future research could focus on enhancing classification performance through ensemble learning techniques. Combining multiple models, such as Vision Transformers, CNNs, and hybrid architectures, could capitalize on their complementary strengths, leading to more robust and accurate predictions \citep{yunusa2024exploring}.
Moreover, this work primarily concentrated on image-based features for classification. A multimodal approach, integrating clinical, contextual, and patient-specific data, could provide a more holistic understanding of skin lesions. This approach could mimic the diagnostic process used by dermatologists, improving the model's reliability and clinical applicability \citep{luo2023artificial,yan2024multimodal}.

In addition to improving accuracy, future studies could explore the use of explainable AI (XAI) techniques to enhance the interpretability of model predictions. Providing insights into the decision-making process of the model would increase trust among clinicians and facilitate the adoption of AI systems in real-world medical practice \citep{van2022explainable, tjoa2020survey,ghnemat2023explainable}.
Lastly, optimizing computational efficiency is an essential area for future work, especially for deployment in resource-constrained environments. Techniques such as model pruning, quantization, and knowledge distillation have been demonstrated to reduce the computational and memory footprint of models while maintaining high performance, making them suitable for rural clinics or mobile health platforms \citep{tmamna2024pruning, mishra2023transforming}.


\section{Conclusion}\label{conclu}
This study has successfully addressed the complex task of classifying skin lesions into 39 distinct classes using advanced deep learning methodologies. The proposed Vision Transformer integrated with the Convolutional Block Attention Module (CBAM) achieved a remarkable accuracy of 93.46\%, demonstrating its robustness and effectiveness in handling multi-class skin lesion classification. By leveraging the strengths of attention-guided mechanisms, the model was able to refine global and local image features, thereby enhancing its ability to identify subtle patterns in medical images. These findings represent a significant contribution to the field of medical image analysis, illustrating the transformative potential of artificial intelligence in improving diagnostic accuracy. Such a system can assist medical professionals in early disease detection, personalized treatment planning, and improving patient outcomes, ultimately contributing to more efficient and effective healthcare delivery.

Despite its successes, this study acknowledges the need for further improvements, such as the inclusion of more diverse datasets, integration of clinical parameters, and exploration of multimodal approaches. These future developments could expand the research scope and further enhance the model's applicability and reliability in clinical environments.

\section*{Data Availability Statement}
The dataset analyzed in this study can be found at \url{https://github.com/akabircs/Skin-Lesions-Classification}.





\bibliographystyle{elsarticle-num-names} 
\bibliography{reference}

\begin{thebibliography}{58}
\expandafter\ifx\csname natexlab\endcsname\relax\def\natexlab#1{#1}\fi
\providecommand{\url}[1]{\texttt{#1}}
\providecommand{\href}[2]{#2}
\providecommand{\path}[1]{#1}
\providecommand{\DOIprefix}{doi:}
\providecommand{\ArXivprefix}{arXiv:}
\providecommand{\URLprefix}{URL: }
\providecommand{\Pubmedprefix}{pmid:}
\providecommand{\doi}[1]{\href{http://dx.doi.org/#1}{\path{#1}}}
\providecommand{\Pubmed}[1]{\href{pmid:#1}{\path{#1}}}
\providecommand{\bibinfo}[2]{#2}
\ifx\xfnm\relax \def\xfnm[#1]{\unskip,\space#1}\fi
\bibitem[{Ski(2024)}]{Skin_Disorder}
\bibinfo{title}{Skin disorder}, \bibinfo{howpublished}{\url{https://www.britannica.com/science/human-skin-disease}}, \bibinfo{year}{2024}. \bibinfo{note}{Accessed on 18 May 2024}.
\bibitem[{Abn(2024)}]{Abnormal_behaviour_of_skin}
\bibinfo{title}{Abnormal behaviour of skin}, \bibinfo{howpublished}{\url{https://my.clevelandclinic.org/health/diseases/15818-skin-cancer}}, \bibinfo{year}{2024}. \bibinfo{note}{Accessed on 18 July 2023}.
\bibitem[{Sta(2024)}]{Statistics}
\bibinfo{title}{Statistics for skin disease}, \bibinfo{howpublished}{\url{https://jamanetwork.com/journals/jamadermatology/fullarticle/2790344}}, \bibinfo{year}{2024}. \bibinfo{note}{Accessed on 18 May 2023}.
\bibitem[{Choudhary et~al.(2022)Choudhary, Singhai, and Yadav}]{choudhary2022skin}
\bibinfo{author}{P.~Choudhary}, \bibinfo{author}{J.~Singhai}, \bibinfo{author}{J.~Yadav},
\newblock \bibinfo{title}{Skin lesion detection based on deep neural networks},
\newblock \bibinfo{journal}{Chemometrics and Intelligent Laboratory Systems} \bibinfo{volume}{230} (\bibinfo{year}{2022}) \bibinfo{pages}{104659}.
\bibitem[{Al-Masni et~al.(2020)Al-Masni, Kim, and Kim}]{al2020multiple}
\bibinfo{author}{M.~A. Al-Masni}, \bibinfo{author}{D.-H. Kim}, \bibinfo{author}{T.-S. Kim},
\newblock \bibinfo{title}{Multiple skin lesions diagnostics via integrated deep convolutional networks for segmentation and classification},
\newblock \bibinfo{journal}{Computer methods and programs in biomedicine} \bibinfo{volume}{190} (\bibinfo{year}{2020}) \bibinfo{pages}{105351}.
\bibitem[{Cau(2024)}]{Cause}
\bibinfo{title}{Cause of skin disease}, \bibinfo{howpublished}{\url{https://www.healthline.com/health/skin-disorders\#causes}}, \bibinfo{year}{2024}. \bibinfo{note}{Accessed on 18 May 2023}.
\bibitem[{Debelee(2023)}]{debelee2023skin}
\bibinfo{author}{T.~G. Debelee},
\newblock \bibinfo{title}{Skin lesion classification and detection using machine learning techniques: A systematic review},
\newblock \bibinfo{journal}{Diagnostics} \bibinfo{volume}{13} (\bibinfo{year}{2023}) \bibinfo{pages}{3147}.
\bibitem[{Arora et~al.(2023)Arora, Dubey, Jaffery, and Rocha}]{arora2023comparative}
\bibinfo{author}{G.~Arora}, \bibinfo{author}{A.~K. Dubey}, \bibinfo{author}{Z.~A. Jaffery}, \bibinfo{author}{A.~Rocha},
\newblock \bibinfo{title}{A comparative study of fourteen deep learning networks for multi skin lesion classification (mslc) on unbalanced data},
\newblock \bibinfo{journal}{Neural Computing and Applications} \bibinfo{volume}{35} (\bibinfo{year}{2023}) \bibinfo{pages}{7989--8015}.
\bibitem[{Kassem et~al.(2021)Kassem, Hosny, Dama{\v{s}}evi{\v{c}}ius, and Eltoukhy}]{kassem2021machine}
\bibinfo{author}{M.~A. Kassem}, \bibinfo{author}{K.~M. Hosny}, \bibinfo{author}{R.~Dama{\v{s}}evi{\v{c}}ius}, \bibinfo{author}{M.~M. Eltoukhy},
\newblock \bibinfo{title}{Machine learning and deep learning methods for skin lesion classification and diagnosis: a systematic review},
\newblock \bibinfo{journal}{Diagnostics} \bibinfo{volume}{11} (\bibinfo{year}{2021}) \bibinfo{pages}{1390}.
\bibitem[{Rafay and Hussain(2023)}]{rafay2023efficientskindis}
\bibinfo{author}{A.~Rafay}, \bibinfo{author}{W.~Hussain},
\newblock \bibinfo{title}{Efficientskindis: An efficientnet-based classification model for a large manually curated dataset of 31 skin diseases},
\newblock \bibinfo{journal}{Biomedical Signal Processing and Control} \bibinfo{volume}{85} (\bibinfo{year}{2023}) \bibinfo{pages}{104869}.
\bibitem[{Ski(2024{\natexlab{a}})}]{Skinnnn}
\bibinfo{title}{Skin disease2}, \bibinfo{howpublished}{\url{https://www.kaggle.com/datasets/arsanyfawzy/skin-diseases}}, \bibinfo{year}{2024}{\natexlab{a}}. \bibinfo{note}{Accessed on 24 May 2024}.
\bibitem[{Ski(2024{\natexlab{b}})}]{Skin_Lesions_Classification_Dataset}
\bibinfo{title}{Skin lesions classification}, \bibinfo{howpublished}{\url{https://www.kaggle.com/datasets/ahmedxc4/skin-ds}}, \bibinfo{year}{2024}{\natexlab{b}}. \bibinfo{note}{Accessed on 18 May 2024}.
\bibitem[{Dat(2024)}]{Dataset_23}
\bibinfo{title}{Dataset skin}, \bibinfo{howpublished}{\url{https://www.kaggle.com/datasets/onurinan1/dateset-23-skin}}, \bibinfo{year}{2024}. \bibinfo{note}{Accessed on 18 May 2024}.
\bibitem[{Ski(2024)}]{SkinDiseases}
\bibinfo{title}{Skindiseasess}, \bibinfo{howpublished}{\url{https://www.kaggle.com/datasets/ascanipek/skin-diseases}}, \bibinfo{year}{2024}. \bibinfo{note}{Accessed on 24 May 2024}.
\bibitem[{Prasanna~Kumar et~al.(2023)Prasanna~Kumar, Venkatraman, Jawahar, Harish, Bharathraj, and Mukesh}]{10421742}
\bibinfo{author}{R.~Prasanna~Kumar}, \bibinfo{author}{K.~Venkatraman}, \bibinfo{author}{C.~Jawahar}, \bibinfo{author}{B.~Harish}, \bibinfo{author}{S.~Bharathraj}, \bibinfo{author}{K.~Mukesh},
\newblock \bibinfo{title}{Attention-guided residual network for skin lesion classification using deep reinforcement learning},
\newblock in: \bibinfo{booktitle}{2023 International Conference on Integrated Intelligence and Communication Systems (ICIICS)}, \bibinfo{year}{2023}, pp. \bibinfo{pages}{1--7}. \DOIprefix\doi{10.1109/ICIICS59993.2023.10421742}.
\bibitem[{Codella et~al.(2018)Codella, Gutman, Celebi, Helba, Marchetti, Dusza, Kalloo, Liopyris, Mishra, Kittler et~al.}]{codella2018skin}
\bibinfo{author}{N.~C. Codella}, \bibinfo{author}{D.~Gutman}, \bibinfo{author}{M.~E. Celebi}, \bibinfo{author}{B.~Helba}, \bibinfo{author}{M.~A. Marchetti}, \bibinfo{author}{S.~W. Dusza}, \bibinfo{author}{A.~Kalloo}, \bibinfo{author}{K.~Liopyris}, \bibinfo{author}{N.~Mishra}, \bibinfo{author}{H.~Kittler}, et~al.,
\newblock \bibinfo{title}{Skin lesion analysis toward melanoma detection: A challenge at the 2017 international symposium on biomedical imaging (isbi), hosted by the international skin imaging collaboration (isic)},
\newblock \bibinfo{organization}{IEEE}, \bibinfo{year}{2018}, pp. \bibinfo{pages}{168--172}.
\bibitem[{Choudhary et~al.(2022)Choudhary, Singhai, and Yadav}]{CHOUDHARY2022104659}
\bibinfo{author}{P.~Choudhary}, \bibinfo{author}{J.~Singhai}, \bibinfo{author}{J.~Yadav},
\newblock \bibinfo{title}{Skin lesion detection based on deep neural networks},
\newblock \bibinfo{journal}{Chemometrics and Intelligent Laboratory Systems} \bibinfo{volume}{230} (\bibinfo{year}{2022}) \bibinfo{pages}{104659}. \DOIprefix\doi{10.1016/j.chemolab.2022.104659}.
\bibitem[{Codella et~al.(2019)Codella, Rotemberg, Tschandl, Celebi, Dusza, Gutman, Helba, Kalloo, Liopyris, Marchetti et~al.}]{codella2019skin}
\bibinfo{author}{N.~Codella}, \bibinfo{author}{V.~Rotemberg}, \bibinfo{author}{P.~Tschandl}, \bibinfo{author}{M.~E. Celebi}, \bibinfo{author}{S.~Dusza}, \bibinfo{author}{D.~Gutman}, \bibinfo{author}{B.~Helba}, \bibinfo{author}{A.~Kalloo}, \bibinfo{author}{K.~Liopyris}, \bibinfo{author}{M.~Marchetti}, et~al.,
\newblock \bibinfo{title}{Skin lesion analysis toward melanoma detection 2018: A challenge hosted by the international skin imaging collaboration (isic)},
\newblock \bibinfo{journal}{arXiv preprint arXiv:1902.03368}  (\bibinfo{year}{2019}).
\bibitem[{Young et~al.(2019)Young, Booth, Simpson, Dutton, and Shrapnel}]{young2019deep}
\bibinfo{author}{K.~Young}, \bibinfo{author}{G.~Booth}, \bibinfo{author}{B.~Simpson}, \bibinfo{author}{R.~Dutton}, \bibinfo{author}{S.~Shrapnel},
\newblock \bibinfo{title}{Deep neural network or dermatologist?},
\newblock \bibinfo{organization}{Springer}, \bibinfo{year}{2019}, pp. \bibinfo{pages}{48--55}.
\bibitem[{Tschandl et~al.(2018)Tschandl, Rosendahl, and Kittler}]{tschandl2018ham10000}
\bibinfo{author}{P.~Tschandl}, \bibinfo{author}{C.~Rosendahl}, \bibinfo{author}{H.~Kittler},
\newblock \bibinfo{title}{The ham10000 dataset, a large collection of multi-source dermatoscopic images of common pigmented skin lesions},
\newblock \bibinfo{journal}{Scientific data} \bibinfo{volume}{5} (\bibinfo{year}{2018}) \bibinfo{pages}{1--9}.
\bibitem[{Chowdhury et~al.(2021)Chowdhury, Bajwa, Chakraborti, Rittscher, and Pal}]{chowdhury2021exploring}
\bibinfo{author}{T.~Chowdhury}, \bibinfo{author}{A.~R. Bajwa}, \bibinfo{author}{T.~Chakraborti}, \bibinfo{author}{J.~Rittscher}, \bibinfo{author}{U.~Pal},
\newblock \bibinfo{title}{Exploring the correlation between deep learned and clinical features in melanoma detection},
\newblock \bibinfo{organization}{Springer}, \bibinfo{year}{2021}, pp. \bibinfo{pages}{3--17}.
\bibitem[{Srinivasu et~al.(2021)Srinivasu, SivaSai, Ijaz, Bhoi, Kim, and Kang}]{srinivasu2021classification}
\bibinfo{author}{P.~N. Srinivasu}, \bibinfo{author}{J.~G. SivaSai}, \bibinfo{author}{M.~F. Ijaz}, \bibinfo{author}{A.~K. Bhoi}, \bibinfo{author}{W.~Kim}, \bibinfo{author}{J.~J. Kang},
\newblock \bibinfo{title}{Classification of skin disease using deep learning neural networks with mobilenet v2 and lstm},
\newblock \bibinfo{journal}{Sensors} \bibinfo{volume}{21} (\bibinfo{year}{2021}) \bibinfo{pages}{2852}.
\bibitem[{Ali et~al.(2023)Ali, Khalid, Alshanbari, Zafar, and Lee}]{bioengineering10121430}
\bibinfo{author}{M.~U. Ali}, \bibinfo{author}{M.~Khalid}, \bibinfo{author}{H.~Alshanbari}, \bibinfo{author}{A.~Zafar}, \bibinfo{author}{S.~W. Lee},
\newblock \bibinfo{title}{Enhancing skin lesion detection: A multistage multiclass convolutional neural network-based framework},
\newblock \bibinfo{journal}{Bioengineering} \bibinfo{volume}{10} (\bibinfo{year}{2023}). \DOIprefix\doi{10.3390/bioengineering10121430}.
\bibitem[{Ayas(2023)}]{ayas2023multiclass}
\bibinfo{author}{S.~Ayas},
\newblock \bibinfo{title}{Multiclass skin lesion classification in dermoscopic images using swin transformer model},
\newblock \bibinfo{journal}{Neural Computing and Applications} \bibinfo{volume}{35} (\bibinfo{year}{2023}) \bibinfo{pages}{6713--6722}.
\bibitem[{Combalia et~al.(2019)Combalia, Codella, Rotemberg, Helba, Vilaplana, Reiter, Carrera, Barreiro, Halpern, Puig et~al.}]{combalia2019bcn20000}
\bibinfo{author}{M.~Combalia}, \bibinfo{author}{N.~C. Codella}, \bibinfo{author}{V.~Rotemberg}, \bibinfo{author}{B.~Helba}, \bibinfo{author}{V.~Vilaplana}, \bibinfo{author}{O.~Reiter}, \bibinfo{author}{C.~Carrera}, \bibinfo{author}{A.~Barreiro}, \bibinfo{author}{A.~C. Halpern}, \bibinfo{author}{S.~Puig}, et~al.,
\newblock \bibinfo{title}{Bcn20000: Dermoscopic lesions in the wild},
\newblock \bibinfo{journal}{arXiv preprint arXiv:1908.02288}  (\bibinfo{year}{2019}).
\bibitem[{Desale and Patil(2024)}]{desale2024efficient}
\bibinfo{author}{R.~Desale}, \bibinfo{author}{P.~Patil},
\newblock \bibinfo{title}{An efficient multi-class classification of skin cancer using optimized vision transformer},
\newblock \bibinfo{journal}{Medical \& Biological Engineering \& Computing} \bibinfo{volume}{62} (\bibinfo{year}{2024}) \bibinfo{pages}{773--789}.
\bibitem[{Saha et~al.(2024)Saha, Joy, and Majumder}]{saha2024yotransvit}
\bibinfo{author}{D.~K. Saha}, \bibinfo{author}{A.~M. Joy}, \bibinfo{author}{A.~Majumder},
\newblock \bibinfo{title}{Yotransvit: A transformer and cnn method for predicting and classifying skin diseases using segmentation techniques},
\newblock \bibinfo{journal}{Informatics in Medicine Unlocked} \bibinfo{volume}{47} (\bibinfo{year}{2024}) \bibinfo{pages}{101495}.
\bibitem[{Kassem et~al.(2020)Kassem, Hosny, and Fouad}]{9121248}
\bibinfo{author}{M.~A. Kassem}, \bibinfo{author}{K.~M. Hosny}, \bibinfo{author}{M.~M. Fouad},
\newblock \bibinfo{title}{Skin lesions classification into eight classes for isic 2019 using deep convolutional neural network and transfer learning},
\newblock \bibinfo{journal}{IEEE Access} \bibinfo{volume}{8} (\bibinfo{year}{2020}) \bibinfo{pages}{114822--114832}. \DOIprefix\doi{10.1109/ACCESS.2020.3003890}.
\bibitem[{Alsahafi et~al.(2023)Alsahafi, Kassem, and Hosny}]{alsahafi2023skin}
\bibinfo{author}{Y.~S. Alsahafi}, \bibinfo{author}{M.~A. Kassem}, \bibinfo{author}{K.~M. Hosny},
\newblock \bibinfo{title}{Skin-net: a novel deep residual network for skin lesions classification using multilevel feature extraction and cross-channel correlation with detection of outlier},
\newblock \bibinfo{journal}{Journal of Big Data} \bibinfo{volume}{10} (\bibinfo{year}{2023}) \bibinfo{pages}{105}.
\bibitem[{Zeng et~al.(2024)Zeng, Ji, Zhang, Chen, Liao, Wang, Lyu, and Zhao}]{zeng2024dsp}
\bibinfo{author}{X.~Zeng}, \bibinfo{author}{Z.~Ji}, \bibinfo{author}{H.~Zhang}, \bibinfo{author}{R.~Chen}, \bibinfo{author}{Q.~Liao}, \bibinfo{author}{J.~Wang}, \bibinfo{author}{T.~Lyu}, \bibinfo{author}{L.~Zhao},
\newblock \bibinfo{title}{Dsp-kd: dual-stage progressive knowledge distillation for skin disease classification},
\newblock \bibinfo{journal}{Bioengineering} \bibinfo{volume}{11} (\bibinfo{year}{2024}) \bibinfo{pages}{70}.
\bibitem[{Nakai et~al.(2022)Nakai, Chen, and Han}]{nakai2022enhanced}
\bibinfo{author}{K.~Nakai}, \bibinfo{author}{Y.-W. Chen}, \bibinfo{author}{X.-H. Han},
\newblock \bibinfo{title}{Enhanced deep bottleneck transformer model for skin lesion classification},
\newblock \bibinfo{journal}{Biomedical Signal Processing and Control} \bibinfo{volume}{78} (\bibinfo{year}{2022}) \bibinfo{pages}{103997}.
\bibitem[{de~A.~Rodrigues et~al.(2020)de~A.~Rodrigues, Ivo, Satapathy, Wang, Hemanth, and Filho}]{RODRIGUES20208}
\bibinfo{author}{D.~de~A.~Rodrigues}, \bibinfo{author}{R.~F. Ivo}, \bibinfo{author}{S.~C. Satapathy}, \bibinfo{author}{S.~Wang}, \bibinfo{author}{J.~Hemanth}, \bibinfo{author}{P.~P.~R. Filho},
\newblock \bibinfo{title}{A new approach for classification skin lesion based on transfer learning, deep learning, and iot system},
\newblock \bibinfo{journal}{Pattern Recognition Letters} \bibinfo{volume}{136} (\bibinfo{year}{2020}) \bibinfo{pages}{8--15}. \DOIprefix\doi{10.1016/j.patrec.2020.05.019}.
\bibitem[{Mendon{\c{c}}a et~al.(2015)Mendon{\c{c}}a, Celebi, Mendonca, and Marques}]{mendoncca2015ph2}
\bibinfo{author}{T.~Mendon{\c{c}}a}, \bibinfo{author}{M.~Celebi}, \bibinfo{author}{T.~Mendonca}, \bibinfo{author}{J.~Marques},
\newblock \bibinfo{title}{Ph2: A public database for the analysis of dermoscopic images},
\newblock \bibinfo{journal}{Dermoscopy image analysis} \bibinfo{volume}{2} (\bibinfo{year}{2015}).
\bibitem[{Rezaee and Zadeh(2024)}]{rezaee2024self}
\bibinfo{author}{K.~Rezaee}, \bibinfo{author}{H.~G. Zadeh},
\newblock \bibinfo{title}{Self-attention transformer unit-based deep learning framework for skin lesions classification in smart healthcare},
\newblock \bibinfo{journal}{Discover Applied Sciences} \bibinfo{volume}{6} (\bibinfo{year}{2024}) \bibinfo{pages}{3}.
\bibitem[{Ahmad et~al.(2024)Ahmad, Amin, Lali, Abbas, and Sharif}]{ahmad2024novel}
\bibinfo{author}{I.~Ahmad}, \bibinfo{author}{J.~Amin}, \bibinfo{author}{M.~I. Lali}, \bibinfo{author}{F.~Abbas}, \bibinfo{author}{M.~I. Sharif},
\newblock \bibinfo{title}{A novel deeplabv3+ and vision-based transformer model for segmentation and classification of skin lesions},
\newblock \bibinfo{journal}{Biomedical Signal Processing and Control} \bibinfo{volume}{92} (\bibinfo{year}{2024}) \bibinfo{pages}{106084}.
\bibitem[{Rotemberg et~al.(2021)Rotemberg, Kurtansky, Betz-Stablein, Caffery, Chousakos, Codella, Combalia, Dusza, Guitera, Gutman et~al.}]{rotemberg2021patient}
\bibinfo{author}{V.~Rotemberg}, \bibinfo{author}{N.~Kurtansky}, \bibinfo{author}{B.~Betz-Stablein}, \bibinfo{author}{L.~Caffery}, \bibinfo{author}{E.~Chousakos}, \bibinfo{author}{N.~Codella}, \bibinfo{author}{M.~Combalia}, \bibinfo{author}{S.~Dusza}, \bibinfo{author}{P.~Guitera}, \bibinfo{author}{D.~Gutman}, et~al.,
\newblock \bibinfo{title}{A patient-centric dataset of images and metadata for identifying melanomas using clinical context},
\newblock \bibinfo{journal}{Scientific data} \bibinfo{volume}{8} (\bibinfo{year}{2021}) \bibinfo{pages}{34}.
\bibitem[{Nagadevi et~al.(2024)Nagadevi, Suman, and Lakshmi}]{nagadevi2024enhanced}
\bibinfo{author}{D.~Nagadevi}, \bibinfo{author}{K.~Suman}, \bibinfo{author}{P.~S. Lakshmi},
\newblock \bibinfo{title}{An enhanced skin lesion detection and classification model using hybrid convolution-based ensemble learning model},
\newblock \bibinfo{journal}{Research on Biomedical Engineering}  (\bibinfo{year}{2024}) \bibinfo{pages}{1--26}.
\bibitem[{Khan et~al.(2024)Khan, Muhammad, Sharif, Akram, and Kadry}]{khan2024intelligent}
\bibinfo{author}{M.~A. Khan}, \bibinfo{author}{K.~Muhammad}, \bibinfo{author}{M.~Sharif}, \bibinfo{author}{T.~Akram}, \bibinfo{author}{S.~Kadry},
\newblock \bibinfo{title}{Intelligent fusion-assisted skin lesion localization and classification for smart healthcare},
\newblock \bibinfo{journal}{Neural Computing and Applications} \bibinfo{volume}{36} (\bibinfo{year}{2024}) \bibinfo{pages}{37--52}.
\bibitem[{Gutman et~al.(2016)Gutman, Codella, Celebi, Helba, Marchetti, Mishra, and Halpern}]{gutman2016skin}
\bibinfo{author}{D.~Gutman}, \bibinfo{author}{N.~C. Codella}, \bibinfo{author}{E.~Celebi}, \bibinfo{author}{B.~Helba}, \bibinfo{author}{M.~Marchetti}, \bibinfo{author}{N.~Mishra}, \bibinfo{author}{A.~Halpern},
\newblock \bibinfo{title}{Skin lesion analysis toward melanoma detection: A challenge at the international symposium on biomedical imaging (isbi) 2016, hosted by the international skin imaging collaboration (isic)},
\newblock \bibinfo{journal}{arXiv preprint arXiv:1605.01397}  (\bibinfo{year}{2016}).
\bibitem[{Atl(2024)}]{Atlas}
\bibinfo{title}{Atlas dermatology}, \bibinfo{howpublished}{\url{https://www.atlasdermatologico.com.br/}}, \bibinfo{year}{2024}. \bibinfo{note}{Accessed on 24 May 2024}.
\bibitem[{Sadik et~al.(2023)Sadik, Majumder, Biswas, Ahammad, and Rahman}]{sadik2023depth}
\bibinfo{author}{R.~Sadik}, \bibinfo{author}{A.~Majumder}, \bibinfo{author}{A.~A. Biswas}, \bibinfo{author}{B.~Ahammad}, \bibinfo{author}{M.~M. Rahman},
\newblock \bibinfo{title}{An in-depth analysis of convolutional neural network architectures with transfer learning for skin disease diagnosis},
\newblock \bibinfo{journal}{Healthcare Analytics} \bibinfo{volume}{3} (\bibinfo{year}{2023}) \bibinfo{pages}{100143}.
\bibitem[{Der(2024)}]{Dermnet}
\bibinfo{title}{Dermnet}, \bibinfo{howpublished}{\url{https://dermnet.com/}}, \bibinfo{year}{2024}. \bibinfo{note}{Accessed on 24 May 2024}.
\bibitem[{Ali et~al.(2024)Ali, Ahmed, Jahan, Paul, Sani, Noor, Asma, and Hasan}]{ali2024web}
\bibinfo{author}{S.~N. Ali}, \bibinfo{author}{M.~T. Ahmed}, \bibinfo{author}{T.~Jahan}, \bibinfo{author}{J.~Paul}, \bibinfo{author}{S.~S. Sani}, \bibinfo{author}{N.~Noor}, \bibinfo{author}{A.~N. Asma}, \bibinfo{author}{T.~Hasan},
\newblock \bibinfo{title}{A web-based mpox skin lesion detection system using state-of-the-art deep learning models considering racial diversity},
\newblock \bibinfo{journal}{Biomedical Signal Processing and Control} \bibinfo{volume}{98} (\bibinfo{year}{2024}) \bibinfo{pages}{106742}.
\bibitem[{der(2024)}]{dermn}
\bibinfo{title}{Dermnet dataset}, \bibinfo{howpublished}{\url{https://dermnet.com/}}, \bibinfo{year}{2024}. \bibinfo{note}{Accessed on 18 May 2023}.
\bibitem[{Wang et~al.(2020)Wang, Wu, Zhu, Li, Zuo, and Hu}]{wang2020eca}
\bibinfo{author}{Q.~Wang}, \bibinfo{author}{B.~Wu}, \bibinfo{author}{P.~Zhu}, \bibinfo{author}{P.~Li}, \bibinfo{author}{W.~Zuo}, \bibinfo{author}{Q.~Hu},
\newblock \bibinfo{title}{Eca-net: Efficient channel attention for deep convolutional neural networks},
\newblock in: \bibinfo{booktitle}{Proceedings of the IEEE/CVF conference on computer vision and pattern recognition}, \bibinfo{year}{2020}, pp. \bibinfo{pages}{11534--11542}.
\bibitem[{Woo et~al.(2018)Woo, Park, Lee, and Kweon}]{woo2018cbam}
\bibinfo{author}{S.~Woo}, \bibinfo{author}{J.~Park}, \bibinfo{author}{J.-Y. Lee}, \bibinfo{author}{I.~S. Kweon},
\newblock \bibinfo{title}{Cbam: Convolutional block attention module},
\newblock in: \bibinfo{booktitle}{Proceedings of the European conference on computer vision (ECCV)}, \bibinfo{year}{2018}, pp. \bibinfo{pages}{3--19}.
\bibitem[{Yin et~al.(2023)Yin, Chen, and Zhang}]{rs15092406}
\bibinfo{author}{M.~Yin}, \bibinfo{author}{Z.~Chen}, \bibinfo{author}{C.~Zhang},
\newblock \bibinfo{title}{A cnn-transformer network combining cbam for change detection in high-resolution remote sensing images},
\newblock \bibinfo{journal}{Remote Sensing} \bibinfo{volume}{15} (\bibinfo{year}{2023}). \DOIprefix\doi{10.3390/rs15092406}.
\bibitem[{van Hooij et~al.(2017)van Hooij, Tjon Kon~Fat, van~den Eeden, Wilson, Batista~da Silva, Salgado, Spencer, Corstjens, and Geluk}]{van2017field}
\bibinfo{author}{A.~van Hooij}, \bibinfo{author}{E.~M. Tjon Kon~Fat}, \bibinfo{author}{S.~J. van~den Eeden}, \bibinfo{author}{L.~Wilson}, \bibinfo{author}{M.~Batista~da Silva}, \bibinfo{author}{C.~G. Salgado}, \bibinfo{author}{J.~S. Spencer}, \bibinfo{author}{P.~L. Corstjens}, \bibinfo{author}{A.~Geluk},
\newblock \bibinfo{title}{Field-friendly serological tests for determination of m. leprae-specific antibodies},
\newblock \bibinfo{journal}{Scientific reports} \bibinfo{volume}{7} (\bibinfo{year}{2017}) \bibinfo{pages}{8868}.
\bibitem[{Alshahrani et~al.(2024)Alshahrani, Al-Jabbar, Senan, Ahmed, and Mohammed~Saif}]{alshahrani2024analysis}
\bibinfo{author}{M.~Alshahrani}, \bibinfo{author}{M.~Al-Jabbar}, \bibinfo{author}{E.~M. Senan}, \bibinfo{author}{I.~A. Ahmed}, \bibinfo{author}{J.~A. Mohammed~Saif},
\newblock \bibinfo{title}{Analysis of dermoscopy images of multi-class for early detection of skin lesions by hybrid systems based on integrating features of cnn models},
\newblock \bibinfo{journal}{Plos one} \bibinfo{volume}{19} (\bibinfo{year}{2024}) \bibinfo{pages}{1--36}. \DOIprefix\doi{10.1371/journal.pone.0298305}.
\bibitem[{Mekala et~al.(2024)Mekala, Pahde, Baur, Chandrashekar, Diep, Wenzel, Wisotzky, Ümit Yolcu, Lapuschkin, Ma, Eisert, Lindvall, Porter, and Samek}]{mekala2024synthetic}
\bibinfo{author}{R.~R. Mekala}, \bibinfo{author}{F.~Pahde}, \bibinfo{author}{S.~Baur}, \bibinfo{author}{S.~Chandrashekar}, \bibinfo{author}{M.~Diep}, \bibinfo{author}{M.~Wenzel}, \bibinfo{author}{E.~L. Wisotzky}, \bibinfo{author}{G.~Ümit Yolcu}, \bibinfo{author}{S.~Lapuschkin}, \bibinfo{author}{J.~Ma}, \bibinfo{author}{P.~Eisert}, \bibinfo{author}{M.~Lindvall}, \bibinfo{author}{A.~Porter}, \bibinfo{author}{W.~Samek}, \bibinfo{title}{Synthetic generation of dermatoscopic images with gan and closed-form factorization}, \bibinfo{year}{2024}. \href{http://arxiv.org/abs/2410.05114}{{\tt arXiv:2410.05114}}.
\bibitem[{Yunusa et~al.(2024)Yunusa, Qin, Chukkol, Yusuf, Bello, and Lawan}]{yunusa2024exploring}
\bibinfo{author}{H.~Yunusa}, \bibinfo{author}{S.~Qin}, \bibinfo{author}{A.~H.~A. Chukkol}, \bibinfo{author}{A.~A. Yusuf}, \bibinfo{author}{I.~Bello}, \bibinfo{author}{A.~Lawan}, \bibinfo{title}{Exploring the synergies of hybrid cnns and vits architectures for computer vision: A survey}, \bibinfo{year}{2024}. \href{http://arxiv.org/abs/2402.02941}{{\tt arXiv:2402.02941}}.
\bibitem[{Luo et~al.(2023)Luo, Zhong, Su, Cheng, Ma, and Hao}]{luo2023artificial}
\bibinfo{author}{N.~Luo}, \bibinfo{author}{X.~Zhong}, \bibinfo{author}{L.~Su}, \bibinfo{author}{Z.~Cheng}, \bibinfo{author}{W.~Ma}, \bibinfo{author}{P.~Hao},
\newblock \bibinfo{title}{Artificial intelligence-assisted dermatology diagnosis: from unimodal to multimodal},
\newblock \bibinfo{journal}{Computers in Biology and Medicine}  (\bibinfo{year}{2023}) \bibinfo{pages}{107413}. \DOIprefix\doi{10.1016/j.compbiomed.2023.107413}.
\bibitem[{Yan et~al.(2024)Yan, Yu, Primiero, Vico-Alonso, Wang, Yang, Tschandl, Hu, Tan, Tang, Ng, Powell, Bonnington, See, Janda, Mar, Kittler, Soyer, and Ge}]{yan2024multimodal}
\bibinfo{author}{S.~Yan}, \bibinfo{author}{Z.~Yu}, \bibinfo{author}{C.~Primiero}, \bibinfo{author}{C.~Vico-Alonso}, \bibinfo{author}{Z.~Wang}, \bibinfo{author}{L.~Yang}, \bibinfo{author}{P.~Tschandl}, \bibinfo{author}{M.~Hu}, \bibinfo{author}{G.~Tan}, \bibinfo{author}{V.~Tang}, \bibinfo{author}{A.~B. Ng}, \bibinfo{author}{D.~Powell}, \bibinfo{author}{P.~Bonnington}, \bibinfo{author}{S.~See}, \bibinfo{author}{M.~Janda}, \bibinfo{author}{V.~Mar}, \bibinfo{author}{H.~Kittler}, \bibinfo{author}{H.~P. Soyer}, \bibinfo{author}{Z.~Ge}, \bibinfo{title}{A general-purpose multimodal foundation model for dermatology}, \bibinfo{year}{2024}. \href{http://arxiv.org/abs/2410.15038}{{\tt arXiv:2410.15038}}.
\bibitem[{Van~der Velden et~al.(2022)Van~der Velden, Kuijf, Gilhuijs, and Viergever}]{van2022explainable}
\bibinfo{author}{B.~H. Van~der Velden}, \bibinfo{author}{H.~J. Kuijf}, \bibinfo{author}{K.~G. Gilhuijs}, \bibinfo{author}{M.~A. Viergever},
\newblock \bibinfo{title}{Explainable artificial intelligence (xai) in deep learning-based medical image analysis},
\newblock \bibinfo{journal}{Medical Image Analysis} \bibinfo{volume}{79} (\bibinfo{year}{2022}) \bibinfo{pages}{102470}. \DOIprefix\doi{10.1016/j.media.2022.102470}.
\bibitem[{Tjoa and Guan(2020)}]{tjoa2020survey}
\bibinfo{author}{E.~Tjoa}, \bibinfo{author}{C.~Guan},
\newblock \bibinfo{title}{A survey on explainable artificial intelligence (xai): Toward medical xai},
\newblock \bibinfo{journal}{IEEE transactions on neural networks and learning systems} \bibinfo{volume}{32} (\bibinfo{year}{2020}) \bibinfo{pages}{4793--4813}. \DOIprefix\doi{10.1109/TNNLS.2020.3027314}.
\bibitem[{Ghnemat et~al.(2023)Ghnemat, Alodibat, and Abu Al-Haija}]{ghnemat2023explainable}
\bibinfo{author}{R.~Ghnemat}, \bibinfo{author}{S.~Alodibat}, \bibinfo{author}{Q.~Abu Al-Haija},
\newblock \bibinfo{title}{Explainable artificial intelligence (xai) for deep learning based medical imaging classification},
\newblock \bibinfo{journal}{Journal of Imaging} \bibinfo{volume}{9} (\bibinfo{year}{2023}) \bibinfo{pages}{177}. \DOIprefix\doi{10.3390/jimaging9090177}.
\bibitem[{Tmamna et~al.(2024)Tmamna, Ayed, Fourati, Gogate, Arslan, Hussain, and Ayed}]{tmamna2024pruning}
\bibinfo{author}{J.~Tmamna}, \bibinfo{author}{E.~B. Ayed}, \bibinfo{author}{R.~Fourati}, \bibinfo{author}{M.~Gogate}, \bibinfo{author}{T.~Arslan}, \bibinfo{author}{A.~Hussain}, \bibinfo{author}{M.~B. Ayed},
\newblock \bibinfo{title}{Pruning deep neural networks for green energy-efficient models: A survey},
\newblock \bibinfo{journal}{Cognitive Computation}  (\bibinfo{year}{2024}) \bibinfo{pages}{1--22}. \DOIprefix\doi{10.1007/s12559-024-10313-0}.
\bibitem[{Mishra and Gupta(2023)}]{mishra2023transforming}
\bibinfo{author}{R.~Mishra}, \bibinfo{author}{H.~Gupta},
\newblock \bibinfo{title}{Transforming large-size to lightweight deep neural networks for iot applications},
\newblock \bibinfo{journal}{ACM Computing Surveys} \bibinfo{volume}{55} (\bibinfo{year}{2023}) \bibinfo{pages}{1--35}. \DOIprefix\doi{10.1145/3570955}.

\end{thebibliography}

\end{document}